%% file: article.tex
\documentclass{jfm}
\usepackage{newtxtext}
\usepackage{newtxmath}
\usepackage{natbib}
\usepackage{hyperref}
\usepackage{array,multirow,makecell}
\usepackage{colortbl}
\usepackage{xcolor}
\usepackage{changes}
\hypersetup{
    colorlinks = true,
    urlcolor   = blue,
    citecolor  = black,
}

\input{mypreamble.tex}

\def\mapacolor{jet} 

\newcolumntype{C}[1]{>{\centering\arraybackslash }m{#1}}

\newcommand{\comJok}[1]{\textcolor{black}{{ #1}}}

\usepackage{ulem}
\usepackage{upgreek}

\input{aaMaketitle.tex}

\begin{document}

\maketitle
\def\thefootnote{*}\footnotetext{These authors contributed equally to this work.}\def\thefootnote{\arabic{footnote}}
\begin{abstract}
\input{aaAbstract.tex} 
\end{abstract}



\input{aaIntro_V2.tex}

\input{aaExperimentalSetup_v2.tex}

\input{aaModelling_v2.tex}

\input{aaResults_v2.tex}

\input{aaConclusion.tex}


\input{aaAcknowledgements.tex}






\input{aaApendices_v2.tex}

\bibliographystyle{jfm}
\bibliography{transport_dynamics_particle_bib}

\end{document}

%% file: mypreamble.tex
\usepackage{array}
\usepackage{textcomp,gensymb}

\usepackage{graphicx}
\usepackage{epstopdf, epsfig}
\usepackage{float}
\usepackage{stmaryrd}
\usepackage{amssymb}
\usepackage{pgfplotstable}
\usepackage{amsmath}
\usepackage{graphicx}
\usepackage{natbib}
\usepackage{pgfplots}
\usepackage{tikz}
\usetikzlibrary{spy}
\usetikzlibrary{shapes}
\usetikzlibrary{arrows}
\usetikzlibrary{arrows.meta}
\usetikzlibrary{patterns}
\usepackage{xcolor}
\usepgfplotslibrary{colormaps}
\usepackage{transparent}
\usepackage{anysize}
\marginsize{2.5cm}{2.5cm}{2.5cm}{2.5cm}
\usepackage{hyperref}
\usepackage{cleveref}
\pgfplotsset{compat=newest}
\usepgflibrary{decorations.markings}
\usepgflibrary{decorations.shapes}

\definecolor{green}{rgb}{0.0,.6,0.0}
\definecolor{migris}{rgb}{.85,.85,.85}%

\tikzset{dashed/.style={dash pattern=on 3pt off 1pt}}
\tikzset{dashed22/.style={dash pattern=on 2pt off 2pt}}
\tikzset{dashdot/.style={dash pattern=on .4pt off 3pt on 4pt off 3pt}}
\tikzset{,
	MyPersp1/.style={scale=1.8,x={(-0.8cm,-0.4cm)},y={(0.8cm,-0.4cm)},z={(0cm,1cm)}},
	MyPoints/.style={fill=white,draw=black,thick}
		}
\pgfplotsset{
	axis on top,
	xtick align=center,
	ytick align=center,
	xlabel near ticks,
	ylabel near ticks,
         legend cell align=left,
	width={0.9\textwidth},
    	height={0.0\textwidth},
	scale only axis,
	legend style={font=\tiny},
	title style={font=\footnotesize},
	every axis/.append style={font=\footnotesize},
	ticklabel style={font=\footnotesize},
	xlabel style={font=\footnotesize},
	ylabel style={font=\footnotesize},
	xticklabel style={font=\footnotesize},
	every axis plot/.append style={line width=.75pt,mark size=3pt},
	}

\newcommand{\D}{\mathcal{D}}

\newcommand{\V}{\mathcal{V}}
\newcommand{\T}{\boldsymbol{T}}
\newcommand{\Ca}{{\rm Ca}}
\newcommand{\La}{{\rm La}}
\newcommand{\Oh}{{\rm Oh}}

\renewcommand{\Re}{{\rm Re}}

\newcommand{\We}{{\rm We}}

\newcommand{\Vb}{V_{0, 0}}
\newcommand{\Vp}{V_{0, \infty}}

\renewcommand{\exp}[1]{e^{#1}}

\newcommand{\vzero}{\boldsymbol{0}}
\newcommand{\vpsi}{\boldsymbol{\psi}}
\newcommand{\x}{\times}

\newcommand{\vn}{\boldsymbol{n}}

\newcommand{\ve}{\boldsymbol{e}}

\newcommand{\vT}{\boldsymbol{T}}
\newcommand{\dbrackets}[1]{\llbracket  #1 \rrbracket}
\newcommand{\vx}{\boldsymbol{x}}

\newcommand{\veps}{\boldsymbol{\varepsilon}}
\newcommand{\vf}{\boldsymbol{f}}

\newcommand{\id}{\mathcal I}

\renewcommand{\O}{\mathcal{O}}

\newcommand{\figref}[1]{fig.~\ref{#1}}

\newcommand{\tabref}[1]{table~\ref{#1}}

\newcommand{\secref}[1]{sec.~\ref{#1}}

\newcommand{\appref}[1]{app.~\ref{#1}}

\newcommand{\grad}{\boldsymbol{\nabla}}

\newcommand{\dd}{{\rm{d}}}

%% file: aaMaketitle.tex
\title{Beads, bubbles and drops in microchannels: \\
stability of centered position and equilibrium velocity}

\author{Jean Cappello$^{1,*}$, Javier Rivero-Rodr\'iguez$^{1,2,*}$, Youen Vitry$^{1,3}$, Adrien Dewandre$^{1,3}$, Benjamin Sobac$^{1,4}$  \corresp{\email{benjamin.sobac@cnrs.fr}}and Benoit Scheid$^{1}$}

\affiliation {$^{1}$Transfers, Interfaces and Processes (TIPs), Universit\'e libre de Bruxelles, 1050 Brussels, Belgium.\\
$^{2}$ Escuela T\'ecnica Superior de Ingenieros Industriales, Universidad de M\'alaga, Plaza El Ejido s/n, 29013 M\'alaga, Spain. \\
$^{3}$ Secoya Technologies, 1348 Louvain-la-Neuve, Belgium.\\
$^{4}$ Laboratoire des Fluides Complexes et leurs R\'eservoirs (LFCR), UMR 5150 CNRS, Universit\'e de Pau et des Pays de l'Adour, Anglet, France.\\}

%% file: aaAbstract.tex
Understand and predict the dynamics of dispersed micro-objects in microfluidics is crucial in numerous natural, industrial and technological situations. In this paper, we experimentally characterized the equilibrium velocity $V$ and lateral position $\varepsilon$ of various dispersed micro-objects such as beads, bubbles and drops, in a cylindrical microchannel over an unprecedent wide range of parameters. By systematically varying the dimensionless object size ($d \in [0.1; 1]$), the viscosity ratio ($\lambda \in [10^{-2}; \infty[$), the density ratio ($\varphi \in [10^{-3}; 2]$), the Reynolds number ($\Re \in [10^{-2}; 10^2]$), and the capillary number ($\Ca \in [10^{-3}; 0.3]$), we offer a general study exploring various dynamics from the nonderformable viscous regime to the deformable visco-inertio-capillary regime, thus enabling to highlight the sole and combined roles of inertia and capillary effects on lateral migration. The experiments are compared and well-agree with a steady 3D Navier-Stokes model for incompressible two-phase fluids including both the effects of inertia and possible interfacial deformations. This model enables to rationalize the experiments and to provide an exhaustive parametric analysis on the influence of the main parameters of the problem, mainly on two aspects: the stability of the centered position and the velocity of the dispersed object. Interestingly, we propose a useful correlation for the object velocity $V$ as functions of the $d$, $\varepsilon$ and $\lambda$, obtained in the $\text{Re}=\text{Ca}=0$ limit, but actually valid for a larger range of values of $\text{Re}$ and $\text{Ca}$ in the linear regimes.  

%% file: aaIntro_V2.tex
\section{Introduction}
\label{sec:intro}

Mastering the dynamics of dispersed micro-objects, such as beads, drops or bubbles transported by an external flow, is crucial in many situations, including the control and optimization of two-phase flows through capillaries or porous materials for the food, pharmaceutical or cosmetic industries where emulsions are widely used \citep{Muschiolik2007, Park2021}, to the improvement of heat and mass transfer or the intensification of heterogeneous reactions for energy or chemical applications \citep{Song2006}, the enhanced recovery of residual oils in the petroleum industry \citep{Green1998}, or even for the biomedical and biological fields where these objects represent model systems for the study of bio-micro-objects, like for example deformable cells \citep{Hur2011a, Chen2014}. 
Nowadays, the emergence of microfluidics that eases manipulation of small objects with the help of a continuous phase, rises new challenges such as object focusing and separation. Different strategies have been developed for continuous flow separation \citep{Pamme2007}. While these techniques appear very different, most of them share the same fundamental principles: the hydrodynamic forces leading to a migration of the objects are modulated by either geometry such as obstacles, constrictions, expansions and surface texture, or external forces acting on the object such as gravitational, electrical, magnetic, centrifugal, optical or acoustical forces. The sources of hydrodynamic forces are twofold: viscous and inertial. The Reynolds number is the dimensionless number that compares viscous to inertial forces and is defined as $\Re = \rho_c  J d_h / \mu_c$ with $d_h$ the hydraulic diameter of the channel, $J$ the superficial velocity, $\rho_c$ the density of the continuous phase and $\mu_c$ its dynamic viscosity. Microfluidics is generally associated with negligible inertia because of the small characteristic size of the channels. In that case, the hydrodynamic forces are restricted to viscous forces and the linear Stokes equations are used to model the flow \citep{Camille_HOward}. Nevertheless, an estimation of the Reynolds number for common situations, for instance considering water of viscosity $\mu_c = 1$ mPa$\cdot$s and of density $\rho_c = 1000$ kg/m$^3$ flowing in a channel of diameter $d_h = 100$ $\upmu$m with a velocity of 10 mm/s, leads to a value $\sim 1$. Thus, the inertial forces are not necessary negligible and the non-linear Navier-Stokes equations are often needed to model the flow, especially when an object is transported by the flow  \citep{DiCarlo2007}.

The nonlinear term of the Navier-Stokes equation which is associated to inertia is of major importance because it breaks the linearity of the Stokes equation and leads to lateral (i.e. cross-stream) motion of dispersed objects evolving in flows with shear gradients \citep{Bretherton1962}. \cite{Segre1962} observed for the first time such inertial migration for small neutrally buoyant rigid spheres transported in Poiseuille flow. They noticed that the beads migrate radially to equilibrium positions located at a distance of about $0.3 d_h$ from the circular channel centerline when their diameter $d_d$ is small compared to $d_h$. As a consequence, a randomly distributed suspension of beads focuses onto a narrow equilibrium annulus as it moves downstream the channel. These observations, unexplained at that time, have drawn the interest of the scientific community. They lead to series of experimental studies on inertial migration in various situations:  the case of non-rotating sphere has been studied by \cite{Oliver1962}, non-neutrally buoyant objects in vertical flow were investigated in various studies \citep{Repetti1964, Jeffrey1965, Karnis1966, Aoki1979} and observations of inertial migration have been extended to different flow geometries such as plane Poiseuille flows \citep{Tachibana1973} or more recently flows in rectangular cross-section duct \citep{DiCarlo2007, DiCarlo2009, Hur2011, Masaeli2012}. A comprehensive analysis of inertial migration has been obtained by the mean of analytical models based on the technique of matched asymptotic expansion. In the case of small but finite Reynolds numbers and small bead size compared to the channel diameter, \cite{Cox1968} derived a scaling for the migration force. Afterwards, \cite{Ho1974} provided the spatial evolution of the inertial force along the lateral position in the channel and by this mean, showed that this force changes sign at a distance close to $0.3d_h$ from the channel centerline. This study highlights that the lateral migration of beads to an equilibrium position results from the interplay between a wall-induced force pointing toward the channel center and a shear-gradient-induced force oriented toward the channel wall. \comJok{The former force develops when a bead is close enough to a wall and arises from the asymmetries of the streamlines and the velocities on either sides of the bead, causing a pressure imbalance with a higher pressure on the side near the wall and generating a force pointing toward the channel centerline.} On the contrary, the latter force is due to the curvature of the velocity profile and can be understood as follows:  because of the non-uniform velocity gradient, the relative velocity of the flow to the bead is higher at the side of larger shear (i.e. close to the wall in a Poiseuille flow) resulting in a smaller pressure at this side. Therefore, the bead moves towards the region of large shear where the magnitude of the relative velocity is the highest. This seminal study has been extended to the cases of larger channel's Reynolds number \citep{Schonberg1989, Asmolov1999} and two dimensional Poiseuille flows \citep{ Hogg1994}, but remain valid only for small bead's Reynolds numbers ($\Re_d = (d_d/d_h)^2 \Re$), i.e. in the case of small dispersed objects compared to the channel dimension. According to the investigations of \cite{Matas2003} and \cite{DiCarlo2009}, some aspects of inertial migration, as the impact of the finite-size of the dispersed object on its equilibrium position when $d_d/d_h$ gets close to unity, cannot be predicted using these assumptions. 

In parallel, experiments on deformable dispersed objects at vanishing Reynolds number, using drops transported in a Poiseuille flow, have shown that the coupling between deformation and the external flow results in a deformation-induced migration force \citep{Goldsmith1962}. In this situation, the deformability of the dispersed object is quantified by the capillary number $\Ca = \mu_c J /\gamma$ that compares viscous to capillary forces, with $\gamma$ the interfacial tension. Similarly to \cite{Ho1974}, \cite{Chan1979} used a matched asymptotic expansion method to derive an analytic expression of this deformation-induced migration force. Their study reveals that the orientation of the force strongly depends on the viscosity ratio $ \lambda = \mu_d/\mu_c$, with  $\mu_d$ the viscosity of the dispersed phase. In a cylindrical Poiseuille flow, the force points toward the wall for values of $\lambda$ between 0.7 and 11.5, while it points toward the channel centerline for $\lambda <0.7$ and $\lambda >11.5$. However, their theory is accompanied with inherent limiting hypothesis such as small drop deformations, small drop diameters compared to the channel dimension, and $\lambda < 1/\Ca$. Moreover, it is also important to note that the wall effects that push deformed drops toward the channel centerline were neglected \citep{Kennedy1994}. \cite{Coulliette1998} used numerical simulations based on boundary integral methods to study the case of deformation-induced migration of drops in a Stokes flow (i.e. inertialess flow) with a viscosity ratio $\lambda=1$. By this mean, the authors investigated the case of large deformations and studied the impact of the drop diameter $d_d$ on the capillary migration processes. They showed that, in all cases, after an initial period of rapid deformation, the drops migrate radially toward the centerline. 

The first study dealing with deformable dispersed objects at finite Reynolds number gathering inertial and deformation induced migration forces was led by \cite{Mortazavi2000}. To investigate the dynamics of drops in Poiseuille flow for few values of capillary number, Reynolds number, viscosity ratio and drop diameter, they used two dimensional, supplemented with few three dimensional, numerical simulations based on finite element methods coupled with locally adaptive moving mesh. They have shown that large drops with $d_d \sim d_h$ always move toward the channel centerline. For smaller drops and intermediate Reynolds numbers (i.e. $\Re$ = [5-50]), the competition between inertial and deformation induced forces leads to the migration toward an off-centered equilibrium position. This equilibrium position gets slightly closer to the wall when increasing the Reynolds number while keeping constant the Weber number $\We=\Re\cdot \Ca$, which compares inertial to capillary forces. On the contrary, increasing the drop viscosity or the drop deformation (by increasing the Weber number) has the opposite effect. They also reported that, for large Reynolds number and/or small viscosity ratio, the equilibrium position is reached after transient oscillations around the steady state. Moreover, above a critical Reynolds number and/or below a critical viscosity ratio, oscillations are not damped anymore and no equilibrium position is observed. \cite{Chen2014} carried out 3D transient simulations and experiments of deformable drops in rectangular cross section channels of a width larger than the height. In this flow geometry, they have shown that drops migrate to the centerline in the height direction and to two equilibrium positions in the width direction. 
Moreover, in agreement with the study of \cite{Mortazavi2000}, the authors reported that, increasing the drop deformation by increasing the Weber number, displaces the equilibrium positions closer to the channel centerline. More recently, also using numerical simulations based on the finite element method, \cite{Rivero2018} solved the full steady Navier-Stokes equations to provide a detailed study on the dynamics of dispersed objects depending on the inertial and capillary migration forces for the case of bubbles. The authors realized a systematic exploration of the influence of the bubble size, Reynolds and capillary numbers, on the velocity and the lateral position for these objects, and this for two boundary conditions at the bubbles' interface: stress-free and non-slip, in order to consider the effect of the interface rigidity in two extreme cases namely for a perfectly clean and deformable bubble and for a rigid bubble when involving surfactants or dusts for instance (equiv. to the rigid bead case), respectively.

Although these works give insights on the dynamics of dispersed objects transported by an external flow in a microchannel subject to inertial and/or capillary forces, they focused on specific dispersed objects, some limiting cases, or a few set of parameters. In the present study, we aim to offer a general vision of this problem by performing, both experimentally and numerically, a systematic study of the influence of all parameters of the problem: Reynolds and capillary numbers, dispersed object diameter, density ratio and viscosity ratio, on the equilibrium velocity and position of the dispersed objects, and this over a wide range of parameters. With this intention, we explore the dynamics of various dispersed objects, such as beads, bubbles and drops, in various regimes,  which enables to highlight the sole and combined roles of inertia and capillary effects on lateral migration and its consequence on the object velocity. 

The structure of the article is as follows. We first describe, in \secref{sec:Experimental_techniques}, the experimental set-up and methods we used to characterize the equilibrium velocity and lateral position of these various dispersed objects (beads, drops and bubbles) in a cylindrical microchannel.
In \secref{sec:Numerical_simulations}, we describe a steady 3D Navier-Stokes model for incompressible two-phase fluids including both the effects of inertia and potential interfacial deformations of the dispersed objects. Two reduced versions of the model are then proposed to specifically and easily compute the equilibrium velocity of the dispersed objects or the stability of their equilibrium centered positions. In \secref{sec:results}, we first compare the experimental results with the numerically-determined ones and then use the numerical models to extend our understanding on the dynamics of dispersed objects over a larger range of parameters. Moreover, we propose a useful correlation for the equilibrium velocity of the dispersed objects as functions of its diameter, position and viscosity ratio and we discuss the influence of the Reynolds number, capillary number, viscosity ratio and density ratio on the stability of their equilibrium centered positions. Finally, conclusions are presented in \secref{sec:Conclusion}.

%% file: aaExperimentalSetup_v2.tex
\section{Experimental setup and methods}
\label{sec:Experimental_techniques}

\begin{figure}
  \begin{center}
    \includegraphics[width=1\columnwidth]{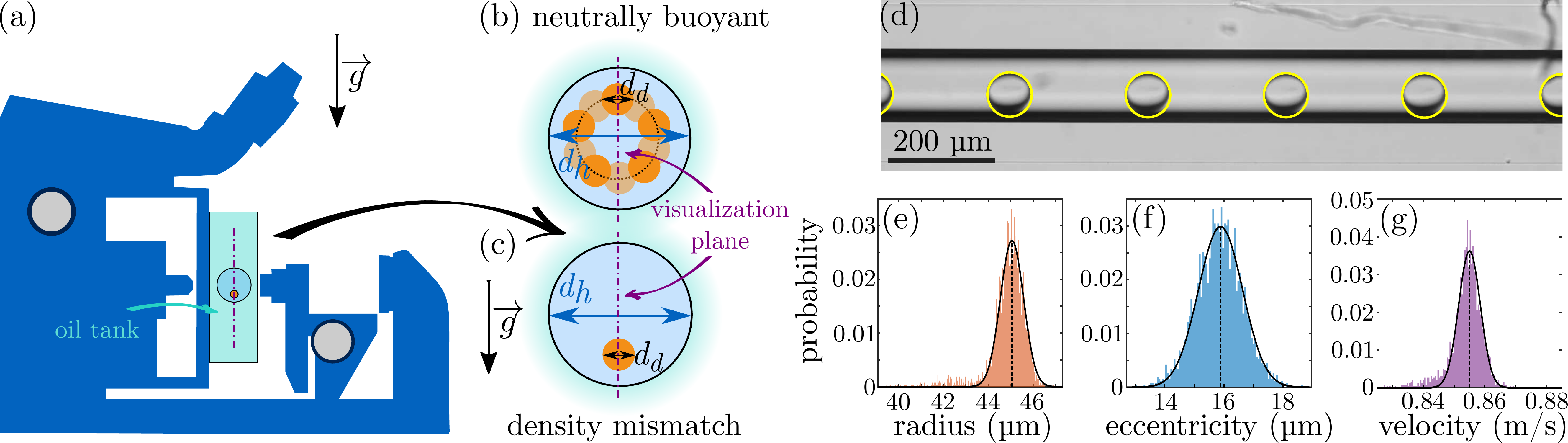}
        \caption{ (a) Schematic illustration of the experimental setup involving an inverted microscope flipped at $90^\circ$ to observe the dynamics of dispersed objects transported by an external flow within a microcapillary. (b-c) Cross-section of the microcapillary. If migration occurs, while a neutrally buoyant object would migrate toward an annulus of equilibrium, a density mismatch between the two phases results in a migration occurring  excusively along the gravitational axis. The latter scheme corresponds to the case where the object is denser than the continuous phase. (d) Typical experimental images showing a drop of FC-770 transported from left to right in water at different times ($d_h = 153$ $\pm$ $3\upmu$m, $\Re = 87 $, $\Ca= 1.3 \times 10^{-3}$, see \tabref{tab:liquids} for additional information). A downward migration is observed here due to inertial effects. (e-g) Distributions of the drop radius, eccentricity, and velocity resulting from the image analysis of the experiment presented in (d), in which the vertical dotted lines correspond to the averaged values.}
  \label{fig:fig_material_methods}
  \end{center}
\end{figure}

The experiments consist of characterising the stationnary dynamics of a dispersed micro-object, such as a bead, a bubble or a drop, transported by an external flow within a circular microchannel. A schematic of the experimental set-up is provided in \figref{fig:fig_material_methods}.

\subsection{Experimental apparatus and image analysis}
\label{subsec:Experimental_apparatus}

Two-phase flows are generated and flow within a horizontal glass microcapillary  (Postnova Analytics). The flow rate of the continuous liquid phase is imposed using a high precision syringe pump (Nemesys, Cetoni), and a flow meter (Coriolis M12, Bronkhorst) is used in series to verify it reaches its setpoint and measure its value accurately. The flow rate of the dispersed phase is controlled using either a pressure controller (MFCS-EX, Fluigent) for experiments involving bubbles or a second syringe pump (Nemesys, Cetoni) coupled with a flow meter (Flow unit, Fluigent) when drops are involved. Note that microcapillaries of different inner diameters $d_h$ have been used depending on the type of dispersed micro-object involved in the experiments:  $d_h = 52 \pm 1$ $\upmu$m or $59 \pm 1$ $\upmu$m with beads and $d_h = 153 \pm 3$ $\upmu$m or $188 \pm 1$ $\upmu$m with drops and bubbles. These values have been measured using an optical profilometer (Keyence VK-X200 series). The dynamics of the micro-objects are observed using an inverted microscope (Nikon Eclipse-Ti) equipped with a 10X TU Plan Fluor objective and images are recorded with a high-speed camera (IDT Y3 Motionpro) at a rate of 3030 frames/s. To ensure a precise observation of the dispersed micro-objects, we first limit image deformations caused by optical refraction effects by placing the microcapillary within a glass tank filled with light mineral oil (Sigma Aldrich), whose refractive index ($\rm{n_{\rm{oil}}}$ = 1.467) is very close to that of glass ($\rm{n_{\rm glass}} =1.470$). Second, we take advantage of gravity by turning the microscope by 90$^{\circ}$ to force the dispersed micro-objects, whose density slightly mismatch the density of the continuous phase, to always stay in the visualization plane parallel to the gravitational axis. Moreover, this trick makes unique the equilibrium position as visible in \figref{fig:fig_material_methods}~(b) and (c). The dynamics of the transported dispersed objects are recorded at a distance of few tens of centimeters from the capillary inlet in order to ensure a quasi-static regime to be achieved. A typical experimental visualisation is shown in \figref{fig:fig_material_methods}(d). Note that this image actually corresponds to the superposition of raw images recorded at six consecutive times and hence shows the same moving object (a drop here) at six different positions in the field of view of the camera. 

Then, a classical image processing, based on binarization and standard Hough's transform (using ImageJ \citep{Schneider2012} and Matlab routines), is performed to fit the dispersed object shape overtime and extract its radius $d_d/2$, as well as its equilibrium velocity $V$ and eccentricity $\varepsilon_d$ (i.e. the transverse distance of the object center from the microchannel centerline). Note that, since in a recorded movie we are able to follow the dynamics of numerous dispersed objects and because experiments are repeated several times, a statistical measurement of these quantities is obtained, from which we derive their mean values (see \figref{fig:fig_material_methods}~(e)-(g)). Typical standard deviation on the eccentricity distribution corresponds to the size of one pixel (1.085 $\upmu$m) and thus on the resolution of our images. The radius distribution is even narrower, the standard deviation corresponding to half a pixel, highlighting the high monodispersity of the dispersed objects generated. The standard deviation on the velocity distribution corresponds to variations of the distance traveled by the object between two consecutive images and is of the order of one to two pixels divided by the time interval between consecutive frames ($0.7$ mm/s).

\subsection{Continuous and dispersed phases}
\label{subsec:Particles_and_fluid}

In order to vary the viscosity ratio of the dispersed phase to the continuous phase $ \lambda = \mu_d/\mu_c$, we consider different micro-objects for the dispersed phase, such as beads ($\lambda \to \infty$), bubbles ($\lambda \to 0$) and drops (intermediate $\lambda$), and various liquids for the continuous phase. 

As nondeformable objects, we used rigid, monodispersed and spherical polystyrene beads (Microbeads AS) suspended in a solution of water containing 7$\%$ of NaOH salt in order to almost match the density of the continuous phase with the one of the polystyrene beads. To additionally vary the size of this dispersed object, we experimented beads of various diameters from $d_d=10 ~\upmu$m to $50~\upmu$m.

As deformable objects, we considered bubbles and drops generated using the Raydrop device (Secoya Technologies), a drop generator based on a non-embedded \textit{co-flow-focusing} technology, ensuring the generation of drops or bubbles at high frequency with a very good reproducibility~\citep{Dewandre2020}. The versatility of this tool enables the use of different couples of fluids and an excellent control of the drops or bubbles sizes. Note that, to vary the size of these dispersed objects, we use different strategies depending on their nature: For drops, we impose a constant flow rate for the dispersed phase and vary the one of the continuous phase, while, for bubbles, we use a constant flow rate for the continuous phase and we vary the inlet pressure of the dispersed phase. 

Table~\ref{tab:liquids} summarizes all beads and fluids employed for the continuous and dispersed phases together with the corresponding ratios of viscosity $\lambda$ and density $\varphi = \rho_d/\rho_c$, and the ranges of dimensionless diameters $d=d_d/d_h$,  Reynolds number $\Re$ and capillary number $\Ca$ encountered in our experiments. For convenience, our results are sometimes represented as functions of the Laplace number $\La=\Re / \Ca$, that relates the inertial and capillary forces to the viscous forces, and of the Weber number $\We=\Re \cdot \Ca$, which describes the competition between inertial forces and capillary forces. 

\begin{table*}
      \caption{Summary of the different situations experimentally investigated in this work.}
\setlength\tabcolsep{0pt}
\begin{tabular}{|C{1.7cm}|C{1.7cm}|C{1.8cm}|C{1.8cm}|C{1.8cm}|C{1.8cm}|C{2cm}|C{1.4cm}|C{1.8cm}|}
  \hline
\rowcolor[gray]{.9}
{\rm{continuous phase}} &  \rm{dispersed phase} &\rm{$ \lambda = \mu_d/\mu_c$}  &  $\varphi=\rho_d/\rho_c$& $d=d_d/d_h$ &$\Re$  & $\Ca$     & $\La$ & $\We$\\
  \hline
  \rm{water + 7\%NaOH} & polystyrene beads &  $\infty$ & $\approx1$  & 0.17 $-$ 0.77 & 4.1 $-$ 24 & 0  & $\infty$  & 0\\ 
  \hline
  \rm{water} & \rm{FC-40} & 4.1 & 1.85 &  0.55 $-$ 0.89 & 9.8 $-$ 78 & 0.001 $-$ 0.010  & 7795& 0.01 $-$ 0.78\\
  \hline
  \rm{water} & \rm{FC-770} & 1.4 & 1.75  & 0.63 $-$ 0.92 & 15 $-$ 87 & 0.002 $-$ 0.013 &  6738& 0.03 $-$ 1.12\\
  \hline
  \rm{FC-40} & \rm{water} & 2.4$\times10^{-1}$ & 0.54  & 0.54 $-$ 0.96 & 6.8 $-$ 38 & 0.008 $-$ 0.041  &  947& 0.05 $-$ 1.56  \\
  \hline
  \rm{mineral oil light} & \rm{ethanol} & 3.9$\times10^{-2}$ & 0.94 & 0.19 $-$ 0.59 & 0.03  $-$ 0.18 & 0.091 $-$ 0.33  & 0.355 &  0.003 $-$ 0.04\\
  \hline
  \rm{water} & \rm{air} & 1.8$\times 10^{-2}$ & 1.18$\times 10^{-3}$ & 0.14 $-$ 0.96 & 17 $-$ 35 & 0.002 $-$ 0.003 & 11213 & 0.02 $-$ 0.11\\ 
  \hline
  \end{tabular}
    \label{tab:liquids}
   \end{table*}

%% file: aaModelling_v2.tex
\section{Modelling}
\label{sec:Numerical_simulations}

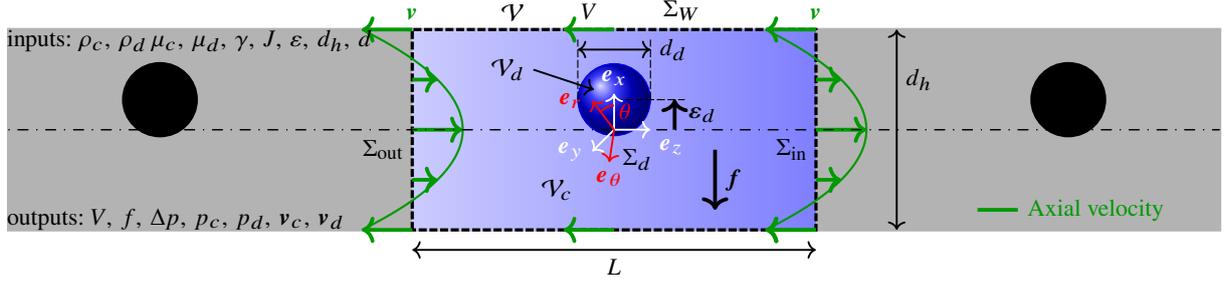
\begin{figure}
\input{./texfigures/Sketch.tex}
\caption{Sketch of the modelled segment of a train of equally spaced dispersed objects in a circular microchannel.}
\label{Sketch1}
\end{figure}

The experimental study is complemented by numerical simulations that model the dynamics of dispersed micro-objects (i.e., beads, bubbles, or drops) transported by an external flow in a cylindrical channel for Reynolds number values ranging from low to intermediate. More specifically, we model the dynamics of a steady train of equally spaced dispersed objects in a cylindrical microchannel. Thus, the present modeling can actually be considered as a generalization of the one presented in \cite{Rivero2018}, dedicated to the bubbles and beads cases (i.e., in the limit cases $\lambda \to 0$ and $\lambda \to \infty$, respectively), to drops (i.e., for all $\lambda$) by considering the internal flow within the dispersed objects. As previously mentioned, different equilibrium positions are possible depending on the size of the dispersed object and on the balance of the involved forces such as viscous, inertial, capillary and body ones, as due to gravity or magnetic fields. To model this situation, we consider a volume $\V$ containing one dispersed object of volume $\V_d$ and of equivalent diameter $d_d = ({6 \V_d/\pi})^{1/3}$, delimited by the walls of the channel, $\Sigma_W$, two cross-sections of the channel, $\Sigma_{\rm in}$ and $\Sigma_{\rm out}$, and the dispersed object surface, $\Sigma_d$, as schematised in Fig.~\ref{Sketch1}. The continuous phase has a density $\rho_c$ and a viscosity $\mu_c$, whereas the dispersed phase has a density $\rho_d$ and a viscosity $\mu_d$. It is assumed that the system is isothermal and the evolution of the dispersed object can be considered as quasi-steady in the absence of either vortex shedding or turbulence. We include interfacial tension $\gamma$ and a uniform body force $\vf$ exerted on the continuous phase in the transverse direction. Gravity is neglected as in the experiments the Froude number that compares the gravitational to the inertial forces, $\text{Fr}=\sqrt{J^2 \rho_c/\left(|\rho_c-\rho_d|gd_h\right)}$ where $g$ in the gravitational acceleration, is always large ($\rm{Fr}\gg1$) due to the small channel size and the large flow velocities. The two phases flow inside a cylindrical channel of hydraulic diameter $d_h$ with a mean velocity $J$, producing a pressure drop due to the Poiseuille flow modified by the presence of a dispersed object, $\Delta p$, along a segment of length $L$, which is the spacial periodicity of the train. Note that the length $L$ is taken large enough to avoid interactions between consecutive objects. The dispersed object travels with a velocity $V$ at a transverse equilibrium position $\veps$ measured from the centre of the channel which is determined by balancing the forces acting on the surface of the dispersed object in the transverse direction. Periodic boundary conditions are considered without loss of generality between the $\Sigma_{\rm in}$ and $\Sigma_{\rm out}$ cross-sections. An upstream velocity $V$ is imposed at the wall of the channel such that the frame of reference is moving with the dispersed object, whose velocity is determined by balancing the forces acting on the dispersed object surface in the streamwise direction. We make use of either Cartesian or cylindrical coordinates depending on the needs. Note that, although in the present study we consider only small dispersed objects with $d_d/d_h < 1$, there is no size limitation in the described model.

In what follows, we first present a general model, based on a steady three-dimensionnal isothermal two-phase flow modelling including the effect of inertia and the deformability of the interphase, and composed by equations \eqref{NS}--\eqref{global}. Then, the general model is conveniently simplified yielding to two reduced and independent models, enabling to specifically compute: (i) the velocity of the dispersed object in the inertialess and nondeformable limit and (ii) the stability of the centered position using a linear perturbation and expansion in its lateral position around the axisymmetric solution. These two reduced models offer the advantages of a gain both in physical understanding and in computation time, the latter enabling to provide an exhaustive parametric analysis. Note that the computed results from the two reduced models have been compared to the ones from the general model for validation.

It is worth mentioning that the equations of the models are presented in dimensionless form. To do so, the characteristic length, velocity and pressure are taken as the diameter of the channel $d_h$, the superficial velocity $J$, and the viscous stress $\mu_c J / d_h$. The dimensionless numbers of the problem are
\begin{align}
\Ca=\frac{\mu_c J }{\gamma} \,,
\quad 
\Re=\frac{\rho_c J d_h }{\mu_c} \,,
\quad 
\lambda=\frac{\mu_d}{\mu_c}  \,,
\quad 
\varphi=\frac{\rho_d}{\rho_c}  \,,
\quad 
d = \frac{d_d}{d_h} \,,
\quad 
\varepsilon = \frac{\varepsilon_d}{d_h}\,.
\end{align}
\noindent In addition, the Laplace number $\La=\Re / \Ca$ and the Weber number $\We=\Re \cdot \Ca$ will also be used for representing the results.

\subsection{Equations of the general model} 
\label{SecEqs}

The flow of both phases in the modeled segment of the channel can be analyzed by solving, in the reference frame attached to the dispersed object, the steady and dimensionless Navier-Stokes equations for incompressible fluids
\refstepcounter{equation}
  $$
  \grad \cdot \boldsymbol{v}_i = 0\,, \quad
  \varphi_i \Re \, \left(\boldsymbol{v}_i \cdot \grad \right) \boldsymbol{v}_i = \grad \cdot \T_{\!i} \,, \qquad \mbox{in } \V_i  \,,
  \eqno{(\theequation{\mathit{a},\mathit{b}})}
  \label{NS}
  $$

with $\T_{\!i} = -\grad p_i + \lambda_i [ \grad \boldsymbol{v}_i + (\grad \boldsymbol{v}_i)^T ]$, $p_i$ and $\boldsymbol{v}_i = (u_i, v_i, w_i)$ being the dimensionless stress tensor, pressure and velocity vector, respectively. The subscript $i$ may refer to continuous $c$ or dispersed $d$ phases with $\lambda_c=1$, $\lambda_d=\lambda$,  $\varphi_c=1$ and $\varphi_d=\varphi$. For the sake of compactness, the subscript $i$ will be omitted when referring to any of both phases and explicited when referring to only one of them. 

Since the difference between the variables in both phases appears in the equations at the level of the interface, we introduce the double brackets operator defined as $\dbrackets{\star} = \star_c - \star_d$. The impermeability condition, continuity of velocity and stress jump together with Young--Laplace equations write

\refstepcounter{equation}
$$
\boldsymbol{v}_c  \cdot \vn=\vzero \,, \qquad 
\dbrackets{\boldsymbol{v}}=0 \,, \qquad
\dbrackets{\T} \cdot \vn= \mathcal{D}_s \Ca^{-1} + \vf \cdot (\vx-\veps) \vn  \,,  \qquad \mbox{at } \Sigma_{d} \,,
  \eqno{(\theequation{\mathit{a},\mathit{b},\mathit{c}})}\label{velcontYL}
$$

where $\vx$ is the position vector, $\vn$ is the outer normal to the continuous domain and $\D_s \star = \grad \cdot \id_s \star $ is the intrinsic surface derivative previously introduced in \cite{Rivero2018}, with $\id_s=\id - \vn \vn$ the surface identity tensor, which is independent of the orientation of $\vn$.  

The velocity field and the reduced pressure gradient are periodic along a distance $L$, producing a pressure drop $\Delta p$,
\begin{align}
\label{periodicity}
\boldsymbol{v} \left(\vx\right)=\boldsymbol{v} \left(\vx \!+\! L \ve_z \right) \,, \quad
\partial_z \boldsymbol{v} \left(\vx\right)= \partial_z \boldsymbol{v} \left(\vx \!+\! L \ve_z\right) \,, \quad 
 p \left(\vx\right) = p \left(\vx \!+\! L \ve_z\right) + \Delta p \,,
\end{align}
which must be imposed at any position in $\Sigma_{\rm out}$. In the reference frame attached to the dispersed object moving at the equilibrium velocity $V$, the velocity of the liquid at the wall writes
\begin{align}
\label{wall}
\boldsymbol{v}_c =-V \ve_z \qquad \mbox{at } \Sigma_W \,. 
\end{align}

Both domains have impermeable boundaries, as shown in (\ref{velcontYL}a,b) and (\ref{wall}), thus requiring to impose a pressure reference at one point for each phase. One pressure represents the absolute pressure reference which is set to 0, i.e. $p_c=0$, at one point arbitrarily chosen in $\vx \in \mathcal{V}_c$, whereas the other one remains to be determined, i.e. $p_d=p_{\rm{ref}}$, at one point arbitrarily chosen $\vx \in \mathcal{V}_d$. Note that $p_{\rm{ref}}$ is an integration constant which is determined as follows by a volume integral constraint.

Finally, the volume of the drop, centroid position, average flow rate through any cross section $\Sigma_{\rm cross}$ and null drag exerted on the object are also imposed,
\refstepcounter{equation}
$$
  \int_{\mathcal{V}_d}  \dd \V = \V_d \,, \quad
  \int_{\mathcal{V}_d} \vx \dd \V = \V_d \veps \,, \quad
  \int_{\Sigma_{\rm cross}} \left(\boldsymbol{v}_c \cdot \ve_z + V - 1\right)  \dd \Sigma =0 \,, \quad   
  \vf \cdot \ve_z =0 \,,
  \eqno{(\theequation{\mathit{a},\mathit{b},\mathit{c},\mathit{d}})}
  \label{global}
$$
which determine the values of the variables $p_{\rm ref}$, $\vf$, $\Delta p$ and $V$, respectively.

Respecting the symmetry, the Cartesian coordinate system can be oriented with the vector $\ve_x$ aligned with the lateral eccentricity $\veps$ and $\vf$. Therefore, these vectors can be written as $\veps=\varepsilon \ve_x$ and migration force $\vf=f \ve_x$, where (\ref{global}d) has already been considered.\\

From the solution of the general model, composed by the system of equations \eqref{NS}-\eqref{global}, it can be observed that migration forces are induced when either inertia, quantified by the Reynolds number $\Re$, or deformability of the interphase induced by the viscous forces, quantified by the capillary number $\Ca$, are taken into account. Asymptotic expansion of this system of equations in terms of $\Re$ and $\Ca$ leads to the following expansion of $V$ and $f$
\begin{subequations}
\begin{align}
\label{TaylorExpension}
V \left(\lambda, d,\varepsilon,\Re,\Ca \right) &= V_0 \left(\lambda, d,\varepsilon \right) + \mathcal{O}\left(\Re^2, \Re \, \Ca, \Ca^2\right) \, , \\
f \left(\lambda,d,\varepsilon,\Re,\Ca \right) &= f_{1,\Re} \left(\lambda, d,\varepsilon \right) \Re + f_{1,\Ca} \left(\lambda, d,\varepsilon \right) \Ca +  \mathcal{O}\left(\Re^3, \Re^2\, \Ca, \Re \,\Ca^2, \Ca^3 \right)  \, ,
\end{align}
\end{subequations}
where vanishing terms have been removed, according to the results in \cite{Rivero2018}, based on the reversibility of the flow. It is observed that for sufficiently small values of $\Re$ and $\Ca$, the velocity of the dispersed object is independent of these numbers, whereas the migration force is proportional to these numbers, vanishing only when both numbers vanish, $\Re=\Ca=0$, for any arbitrary lateral position of the dispersed object. 

\begin{figure}
  \begin{center}
\input{./texfigures/Regimes.tex}
        \caption{The considered flow regimes involving inertial and capillary migrations. The bricks and dots represent linear and nonlinear problems respectively.
       }
  \label{fig:regimes}
  \end{center}
\end{figure}
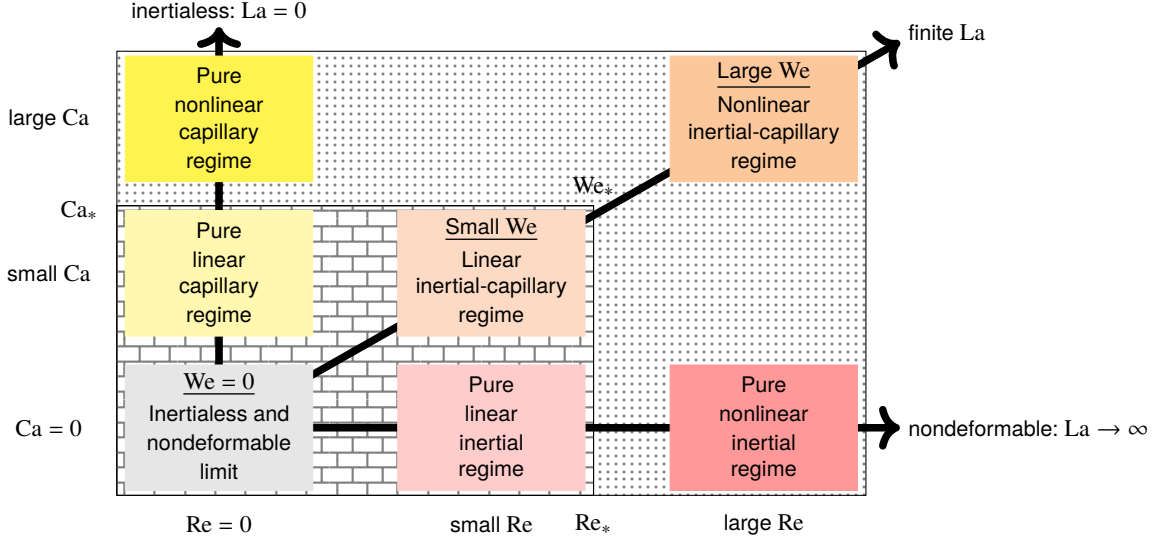

Then, several regimes can be classified with respect to the dimensionless numbers as sketched in Fig.~\ref{fig:regimes}. For $\Re=\Ca=0$, the system is in the inertialess and nondeformable limit. For non-zero but small values, the migration force is proportional to $\Re$ and/or $\Ca$, and we refer to as the linear regime. For sufficiently large values, nonlinearities arise. Analogous regimes can be considered for $\La=0$ or $\La \rightarrow \infty$, which represent inertialess or nondeformable systems, respectively, and the nonlinearity of the regime is gauged by $\Ca$ or $\Re$.

Solving the general model to derive the equilibrium velocity and lateral position of dispersed objects is time and resource consuming because of the numbers of parameters ($d$, $\lambda$, $\Ca$, $\Re$ and $\varphi$) and the tridimensionality of the problem. Therefore, in the following, the general model is not used as is, but rather aptly developed and reduced in two different limits, namely, the inertialess and nondeformable limit, mainly used for predicting the equilibrium  velocity of the dispersed objects, and the limit case of axisymmetric solutions, to address the question of the stability of their centered position.


\subsection{Inertialess and/or nondeformable limit(s)} \label{SecEqs2}

The system of equations \eqref{NS}-\eqref{global} allows two non-exclusive limits, namely, inertialess for $\Re=0$ and nondeformable for $\Ca=0$. The inertialess limit can be obtained by substituting $\Re=0$ in the equations, whereas the nondeformable limit requires more modifications than simply substituting $\Ca=0$ in the system. In the latter limit, the variations of the curvature with respect to the undeformed dispersed object for large interfacial tensions are inversely proportional to interfacial tension, leading to a finite pressure $p_s$ irrespective of the value of Ca, provided it is sufficiently small. Thus, the interfacial tension term in (\ref{velcontYL}$\mathit{c}$) is of the form $p_s \vn$ leading to 
\begin{align}
\label{ps}
\dbrackets{\T} \cdot \vn= -p_s \vn + \vf \cdot (\vx-\veps) \vn  \,.
\end{align}
Furthermore, since the interphase deformation vanishes for $\Ca = 0$, the equations (\ref{global}$\mathit{a}$,$\mathit{b}$) do not longer hold, the volume and position of the object being \textit{a priori} imposed. Instead, it must be considered that the overall surface pressure forces applied on a closed surface must vanish, namely
\begin{align}
\label{ps0}
\int_{\Sigma_{d}} p_s \vn \dd \Sigma &=0 \,.
\end{align}
This model simplification is analogous to that rigorously developed in \cite{Rivero2018} in terms of asymptotic expansion in $\Ca$ around $\Ca=0$. 

In the inertialess and nondeformable limit, the general model is simplified by combining all modifications explained above for the two non-exclusive limits. Thus, this simplified model reduces to the system of equations \eqref{NS}-(\ref{velcontYL}$\mathit{b}$), \eqref{ps}, \eqref{periodicity}-\eqref{wall}, \eqref{ps0}, (\ref{global}$\mathit{c},\mathit{d}$), in which $\Re=0$ and $\Ca=0$. It is worth mentioning that, despite the vanishing force $\vf=\vzero$ in this inertialess and nondeformable limit, the prediction from this simplified model for the dispersed object velocity remain valid over a relatively larger range of values of $\Re$ and $\Ca$ in the linear regimes, as observed in \eqref{TaylorExpension} since the first order correction vanishes and as shown later in \secref{sec:parametric_analysis_num_stability}.

\subsection{Linear stability of axisymmetric solutions}\label{SecEqs3}

The equilibrium eccentricity of a dispersed object is analyzed here via the determination of the stability threshold of its centered position (i.e. $\varepsilon=0$), hence considering a neutrally buoyant object (i.e. $f=0$). To do this, the general model \eqref{NS}-\eqref{global} is perturbed up to first order in eccentricity $\varepsilon$ around $\varepsilon=0$. Thus, $f(\varepsilon)= f_0 + f_1 \varepsilon + \mathcal{O}(\varepsilon^2)$, matching the Taylor expansion $ f(\varepsilon) = f\vert_{\varepsilon=0} + \partial f / \partial \varepsilon \vert_{\varepsilon=0} \,\varepsilon + \mathcal{O}(\varepsilon^2)$, should reveal from the sign of $f_1 = \partial f / \partial \varepsilon \vert_{\varepsilon=0}$, the stable or unstable character of the centered position. If $f_1<0$, the body force acts in the opposite direction of the lateral displacement and the centered position is stable, while if $f_1>0$, the centered position is unstable and the body force leads to the lateral migration of the dispersed object. Compared to the previous analysis of the position stability based on the prediction of the pitchwork bifurcation when $\partial f / \partial \varepsilon = 0$ (see \cite{Rivero2018} in the case of bubbles and beads), this alternative method has the advantage of reducing both the dimensionality of the problem from three-dimensional to axisymmetric two-dimensional and the parametric space, thus requiring much less computational effort and enabling extended parametric analysis on the equilibrium eccentricity.


To proceed, we perturbe the axisymmetric geometry which undergoes a displacement of the dispersed object interphase $\boldsymbol{\delta}= \delta \vn$. Then, we seek an expansion of the variables in terms of $\varepsilon$ and an analytical $\theta$-dependence in cylindrical coordinates :
\begin{subequations}
\label{exp}
\begin{align}
\label{expv}
p_i \left(r, \theta, z\right)&=p_{i0}(r,z) + \varepsilon \exp{i\theta} \, p_{i1}\left(r,z\right) + \O\left(\varepsilon^2\right)   \qquad {\mbox{in }}\V_{i0} \,,
\\
\boldsymbol{v}_i\left(r, \theta, z\right)&=\boldsymbol{v}_{i0}\left(r,z,\theta\right) + \varepsilon \exp{i\theta} \, \boldsymbol{v}_{i1}\left(r,z,\theta\right) + \O\left(\varepsilon^2\right)   \qquad {\mbox{in }} \V_{i0} \,,
\\
\label{expdelta}
\delta\left(r, \theta, z\right) &= \varepsilon \exp{i\theta} \, \delta_1\left(r,z\right) + \O\left(\varepsilon^2\right)  \qquad \mbox{at } \Sigma_{d0} \,,
\\
\eta  &= \eta_0 + \varepsilon \, \eta_1 + \O\left(\varepsilon^2\right)  \,,
\end{align}
\end{subequations}
where $\eta$ is any of the global variables $f$, $V$ or $\Delta p$. The subscript $0$ refers to the unperturbed axisymmetric geometry and the subscript $1$ refers to the perturbation. Although  the vectors $\boldsymbol{v}_{i0}$ and $\boldsymbol{v}_{i1}$ depend on $\theta$, their components in cylindrical coordinates do not,
\begin{align}
\boldsymbol{v}_{i0}= u_{i0} \ve_z + v_{i0} \ve_r\,,\qquad \boldsymbol{v}_{i1} =  u_{i1} \ve_z + v_{i1} \ve_r + w_{i1} \ve_\theta \,,
\end{align}
i.e., $\ve_r$ and $\ve_\theta$ are $\theta$-dependent whereas the components of the vectors, $u_{i0}$, $v_{i0}$, $u_{i1}$, $v_{i1}$ and $w_{i1}$, are not. Note that in the sought solution, the $\theta$-dependence is analytical, and hence, every variable and the geometry exclusively depend on the position in the $z$-$r$ plane. Thus, the variables in the volume $\V_{i0}$ or the surface $\Sigma_{d0}$, reduces to the variables in their intersection with the $r$-$z$ plane, namely $S$ and $\Gamma$, respectively. Conversely, the revolution around the $z$-axis of the two-dimensional geometries $S$ and $\Gamma$ leads to the unperturbed axisymmetric tridimensional geometries.

In axisymmetric geometries, the differential operators appearing in the previous equations, $\grad$ and $\D_s$, can be splitted into the $r$-$z$ components and the $\theta$ component as 
\begin{align}
\label{defdifop}
\grad \star = \grad_{rz}  \star + \frac{\ve_\theta}{r} \partial_\theta \star  \,, \qquad
\mathcal{D}_{s} \star = \mathcal{D}_{s,{rz}} \star + \frac{\ve_\theta}{r} \cdot \partial_\theta \id_s \star  \,,
\end{align}
where $\grad_{rz} \star = \id_{rz} \cdot \grad \star$ and $\mathcal{D}_{s,{rz}} \star = \grad_{rz} \cdot \id_s \star$. Notice that the vector $\D_{s,{rz}} 1 = \grad_{rz} \cdot \id_s $ represents the curvature of the planar curve $\Gamma$, i.e., the axial curvature of the surface.




Introducing the expansion \eqref{exp} into the governing system of equations \eqref{NS}-\eqref{global} leads to the equations governing the zeroth and first order. In doing so, the continuity equation (\ref{NS}$\mathit{a}$) multiplied by a factor $r$ writes
\begin{subequations}
\label{cont01}
\begin{align}
0&= \grad_{{rz}} \cdot \left( r \boldsymbol{v}_{0} \right) \,, \\
0&=\grad_{{rz}} \cdot \left( r \boldsymbol{v}_{1} \right) + i w_{1}  \,,
\label{cont1split}
\end{align}
\end{subequations}
at $S$ whereas the momentum equation (\ref{NS}$\mathit{b}$) multiplied by a factor $r$ writes
\begin{subequations}
\label{moment01}
\begin{align}
r \varphi \Re \, \boldsymbol{v}_0 \cdot \grad_{rz} \boldsymbol{v}_0 &= \grad_{{rz}} \cdot \left( r \T_{0} \right) + \ve_z \times \vT_{0\theta} \,, \\
r \varphi \Re  \left(\boldsymbol{v}_0 \cdot \grad_{rz} \boldsymbol{v}_1+ \boldsymbol{v}_1 \cdot \grad_{rz} \boldsymbol{v}_0 + w_1 v_0 \ve_\theta \right) &=\grad_{{rz}} \cdot \left(r \T_{1}\right)+ \ve_z \times \vT_{1\theta}  + i \vT_{1\theta}   \,,
\label{NS1split}
\end{align}
\end{subequations}
at $S$ where $\vT_\theta = \T \cdot \ve_{\theta}$ and the stress tensors, $\T = \T_0 + \varepsilon  \exp{i\theta}  \T_1$, are given in the \appref{intsteps} by their components (\ref{Trz}-\ref{Ttheta}) in cylindrical coordinates. In the first-order expressions \eqref{cont1split} and \eqref{NS1split}, the factor $\exp{i\theta}$ has been cancelled out, as it will be done in further equations for the first-order terms. To derive \eqref{cont01} and \eqref{moment01}, it is convenient to use the alternative expressions of the differential operators \eqref{gradtheta}.

The boundary conditions \eqref{velcontYL} are for zeroth-order
\begin{subequations}
\begin{align}
r \vn \cdot \boldsymbol{v}_{c0}  &= 0  \,,
\\
\dbrackets{\boldsymbol{v}_0}  &= \vzero  \,,
\\
 r\vn \cdot  \dbrackets{\T_0} &= \Ca^{-1} \left( \mathcal{D}_{s,{rz}} r - \ve_r  \right) \,,
\end{align}
\end{subequations}
and for first-order
\begin{subequations}
\begin{align}
r \vn\cdot \boldsymbol{v}_{c1}  &= \mathcal{D}_{s,{rz}} \cdot \left( r \delta_1 \boldsymbol{v}_{c0} \right)    \,,
\\
\dbrackets{\boldsymbol{v}_1} + \delta_1 \vn \cdot   \grad_{{rz}} \dbrackets{\boldsymbol{v}_0} &= \vzero  \,,
\\
r\vn\cdot \dbrackets{  \T_1 } &= \mathcal{D}_{s,{rz}} \cdot \left( r \delta_1 \dbrackets{\T_0} \right) + i \delta_1 \dbrackets{\vT_{0\theta}}
 + \delta_1 \ve_{z}\times \dbrackets{\vT_{0\theta}}  -  r \delta_1 \varphi \Re \,  \dbrackets{\boldsymbol{v}_0 \cdot \grad_{{rz}} \boldsymbol{v}_0}  +
 \\ \nonumber
 &  \quad + r^2 f_1 \vn +  \Ca^{-1}  \mathcal{D}_{s,{rz}} \cdot \left( r \Psi_1 \right) + \Ca^{-1} \partial_\theta \vpsi_{\theta1}  
 \,,
\end{align}
\end{subequations}
at $\Gamma$ where the last two terms correspond to the perturbation of the interfacial tension given by \eqref{pertST}, whose details and those of the perturbation of the flux terms are given in \appref{intsteps}. In addition, the first-order uniform body force is written in cylindrical coordinates as $\vf_1 =f_1 \exp{i\theta} ( \ve_r + i \ve_\theta )= f_1 (\ve_x + i \ve_y ) $, being independent of $\theta$, as well as fulfilling (\ref{global}$\mathit{d}$). The appearance of $\exp{i\theta}$ as a common factor in the first-order terms allows it to be cancelled out.

The equations (\ref{global}$\mathit{a}$,$\mathit{b}$) are written after the expansion \eqref{expdelta}, using the Reynolds transport theorem \eqref{pertvolpos} and carrying out the $\theta$-integrals \eqref{thetaintegral}, 

\refstepcounter{equation}
$$
2 \pi \int_{S}  r \dd \Sigma  = \V_d
\,, \qquad
2 \pi \int_{S}  r z \dd \Sigma  = 0
\,, \qquad
\ \pi   \int_{\Gamma}  r \delta_1  \dd \Gamma  \exp{i\theta} (\ve_r + i \ve_\theta) = \V \veps_1 \,, 
 \eqno{(\theequation{\mathit{a},\mathit{b},\mathit{c}})}
\label{integralesthetahechas}
$$
where the vectorial character of \eqref{integralesthetahechas} can be removed by considering that $\veps_1 =  \exp{i\theta} (\ve_r + i \ve_\theta)$, hence $\veps_1 $ and $\vf_1$ are parallel. These equations determine the volume and the position in the longitudinal direction for the zeroth-order and the position in the transverse direction for the first-order, whereas the others hidden equations related to the missing order are automatically fulfilled.



The perturbation of equations (\ref{global}$\mathit{c}$) vanish and the perturbation $\Delta p_1$ also does.  The boundary conditions at the wall \eqref{wall} write
\begin{align}
\boldsymbol{v}_{c0} = - V \ve_z \,, \qquad \boldsymbol{v}_{c1}=\vzero \,,
\end{align}
for which (\ref{global}$\mathit{c}$) is automatically fulfilled for the first-order.

Periodicity \eqref{periodicity} must also be imposed for $\boldsymbol{v}$ and $p$. Concerning the pressure references, they should not be imposed for the first-order since, in fact, the pressure vanishes at the axis, ensuring the regularity of $p_{{\rm ref}1} \exp{i\theta}$ at $r=0$ and serving as a reference itself. 
\begin{subequations}
\label{periodicity01}
\begin{align}
&\boldsymbol{v}_0 (\vx)=\boldsymbol{v}_0 (\vx \!+\! L \ve_z) \,, \quad
\partial_z \boldsymbol{v}_0 (\vx)= \partial_z \boldsymbol{v}_0 (\vx \!+\! L \ve_z) \,, \quad 
 p_0 (\vx) = p_0 (\vx \!+\! L \ve_z) + \Delta p \,,  
 \\
&\boldsymbol{v}_1 (\vx)=\boldsymbol{v}_1 (\vx \!+\! L \ve_z) \,, \quad
\partial_z \boldsymbol{v}_1 (\vx)= \partial_z \boldsymbol{v}_1 (\vx \!+\! L \ve_z) \,, \quad 
 p_1 (\vx) = p_1 (\vx \!+\! L \ve_z)  \,. 
\end{align}
\end{subequations}

In summary, this reduced model concerning the stability of the centered position of a dispersed object is composed by the system of equations \eqref{cont01}-\eqref{periodicity01}. It is worth specifying that this model is obtained through a linearisation in $\varepsilon$, rather than in $\Re$ and $\Ca$, hence it should not be confused with a linear model in the sense of the flow regimes presented in \figref{fig:regimes}, since the model is actually valid for all values of $\Re$ and $\Ca$. Consequently, in addition to predict the stability of the centered position of a dispersed object, this reduced model is also able to evidence when non-linearities in the sense of flow regimes arise, as later shown in \secref{sec:parametric_analysis_num_stability}.\\

The three systems of PDEs have been solved using the finite element method with the help of Comsol Multiphysics. The equations have been implemented in the general purpose {\it Weak Form PDE} and {\it Weak Form Boundary PDE} modules, using first-order elements for pressure and quadratic for any other variable. Independence of the mesh has been checked and reduced models have been validated by comparison with the general model. When the system of equations needs to be solved in a deformable domain, such as the system in \secref{SecEqs} or \secref{SecEqs3}, the Arbitrary Lagrangian-Eulerian (ALE) method implemented in the {\it Moving Mesh} module and the Differential Boundary Arbitrary Lagrangian-Eulerian (DBALE) method proposed by \cite{rivero2021alternative}, have been used to allow mesh deformation, starting from the undeformed mesh corresponding to a spherical dispersed object of the same volume. Nondeformable mesh is used otherwise, such as in \secref{SecEqs2}.

%% file: texfigures/Sketch.tex
\begin{center}

\begin{tikzpicture}[x={(.166\textwidth,0cm)},y={(0cm,.166\textwidth)}]

\filldraw[color=black!30, very thick] (-3.0,0.5) rectangle (3.0,-0.5);

\shadedraw[color=black, fill=blue!20, very thick, dashed,left color=blue!20,right color=blue!50] (-1,-0.5)rectangle (1,0.5);


\draw[color=black, dashdot,line width=.15mm] (-3,0)-- (3,0) ;

\filldraw[color=blue, fill=blue,shading=ball] (0,.15) circle (.18);

\filldraw[color=black, fill=black, very thick] (-2.25,.15) circle (.18);
\filldraw[color=black, fill=black, very thick] (2.25,.15) circle (.18);

\draw[color=green, line width=.5mm,->] (.0,.5)-- (-.25,0.5) node[black, midway, above] {$V$} ;
\draw[color=green, line width=.5mm,->] (.0,-.5)-- (-.25,-0.5) ;

\draw[color=black, dashed,line width=.15mm] (0,0.15)-- (0.35,0.15);
\draw[color=black, line width=.5mm,->] (0.3,0)-- (0.3,0.15) node[midway, right] {$\boldsymbol{ \, \varepsilon}_d$} ;

\draw[color=black, line width=.5mm,->] (.5,-.1)-- (0.5,-.4) node[midway, right] {$\boldsymbol{f}$} ;

\draw[color=black, line width=.25mm,<->] (-1.,-.6)-- (1.,-.6) node[midway, below] {$L$} ;
\draw[color=black, line width=.25mm,<->] (1.4,-.5)-- (1.4,.5) node[near end, right] {$d_h$} ;
\draw[color=black, line width=.25mm,<->] (-.18,.4)-- (.18,.4) node[anchor=west] {$d_d$} ;
\draw[color=black, dashed,line width=.15mm] (-.18,0.15)-- (-0.18,0.45);
\draw[color=black, dashed,line width=.15mm] (+.18,0.15)-- (+0.18,0.45);



\draw[color=green, line width=.5mm,<-] ({0.75-2},-.5)--({1-2},-.5); \node[below,black,anchor=east] at (-1,-.1) {$\Sigma_{\rm out}$};
\draw[color=green, line width=.5mm,->] ({1-2},-.25)-- ({1.125-2},-.25);
\draw[color=green, line width=.5mm,->] ({1-2},0)-- ({1.25-2},0);
\draw[color=green, line width=.5mm,->] ({1-2},.25)-- ({1.125-2},.25);
\draw[color=green, line width=.5mm,<-] ({.75-2},.5)-- ({1-2},.5) node[above, green] {$\boldsymbol{v}$};

\draw [green, thick,  domain=0:1, samples=40] 
 plot ({-2+1.25-.5*\x^2}, {+.5*\x} );
 \draw [green, thick,  domain=0:1, samples=40] 
 plot ({-2+1.25-.5*\x^2}, {-.5*\x} );



\draw[color=black, line width=.25mm,<-] (-0.1,.2)--(-0.4,.3) node[anchor=east]  {$\mathcal V_d$}; 
\draw[color=green, line width=.5mm,<-] (0.75,-.5)--(1,-.5); \node[below,black,anchor=west] at (-0.4,-.3) {$\mathcal V_c$};
\draw[color=green, line width=.5mm,<-] (0.75,-.5)--(1,-.5); \node[below,black,anchor=west] at (-0.,-.15) {$\Sigma_{d}$};
\draw[color=green, line width=.5mm,<-] (0.75,-.5)--(1,-.5); \node[below,black,anchor=south west] at (0.2,.5) {$\Sigma_{W}$};
\draw[color=green, line width=.5mm,<-] (0.75,-.5)--(1,-.5); \node[below,black,anchor=east] at (1,-.1) {$\Sigma_{\rm in}$};
\node[below,black,anchor=south west] at (-0.6,.5) {$\V$};

\draw[color=green, line width=.5mm,->] (1,-.25)-- (1.125,-.25);
\draw[color=green, line width=.5mm,->] (1,0)-- (1.25,0);
\draw[color=green, line width=.5mm,->] (1,.25)-- (1.125,.25);
\draw[color=green, line width=.5mm,<-] (.75,.5)-- (1,.5) node[above, green] {$\boldsymbol{v}$};

\draw [green, thick,  domain=0:1, samples=40] 
 plot ({1.25-.5*\x^2}, {+.5*\x} );
\draw [green, thick,  domain=0:1, samples=40] 
plot ({1.25-.5*\x^2}, {-.5*\x} );



\draw [green, line width=.5mm] (1.8,-.4)-- (2,-.4) node[right] {Axial velocity};

\node[anchor=north west] at (-3.05,+.54) {inputs: $\rho_{c},\, \rho_d \, \mu_{c}, \, \mu_{d}, \, \gamma, \, J ,\, \varepsilon,\, d_h,\, d$};
\node[anchor=south west] at (-3.05,-.54) {outputs: $V ,\, f ,\, \Delta p ,\, p_{c},\, p_{d},\, \boldsymbol{v}_{c},\, \boldsymbol{v}_{d} $};


\draw[color=black, dashdot,line width=.15mm] (-3,0)-- (3,0) ;
\draw[color=white,line width=.25mm, ->] (0,0) -- (.18,.0) node[anchor=north west] {\footnotesize{$\ve_z$}};
\draw[color=white,line width=.25mm, ->] (0,0) -- (.0,.18) node[anchor=south] {\footnotesize{$\ve_x$}};
\draw[color=white, line width=.25mm, ->] (0,0) -- (-.108,-.108) node[anchor=east] {\footnotesize{$\ve_y$}};

\draw[color=red,line width=.25mm, ->] (0,0) -- (-.108,.144) node[anchor=east] {\footnotesize{$\ve_r$}};
\draw[color=red, line width=.25mm, ->] (0,0) -- (-.024,-.162) node[anchor=north] {\footnotesize{$\ve_\theta$}};
\draw [color=red, line width=.25mm, -]    (.0,.12) to[out=160,in=-160]  (-.06,.09) node[anchor=west] {\footnotesize{$\, \, \theta$}};

\end{tikzpicture}
\end{center}

%% file: texfigures/Regimes.tex
\begin{tikzpicture}[y=.4\textwidth, x=.7\textwidth,font=\sffamily]
	\def\x{.2}
	\def\y{.52}
	\def\z{.84}
	\def\f{.95}
	\def\fx{1.}
	\def\fy{1.}
	
	
    	
	\draw[pattern=dots, pattern color = gray] (\x-.08-.04,\x-.1-.04) rectangle (\z+.08+.04,\z+.1+.04);
	\label{patNLin}
	
	\draw[fill=white, pattern color = gray] (\x-.08-.04,\x-.1-.04) rectangle (\y+.08+.04,\y+.1+.04);
	\draw[pattern=bricks, pattern color = gray] (\x-.08-.04,\x-.1-.04) rectangle (\y+.08+.04,\y+.1+.04);
	
	\label{patLinear}

	\node[] at (\x,0) {$\Re=0$};
	\node[] at (\y,0) {small $\Re$};
	\node[] at (\z,0) {large $\Re$};
	\node[] at (\x+\y-.08,0) {$\Re_*$};

	\node[] at (0,\x) {$\Ca=0$};
	\node[] at (0,\y) {small $\Ca$};
	\node[] at (0,\z) {large $\Ca$};
	\node[] at (0,\y+.08+.05) {\, \, \, \, \, \, \,$\Ca_*$};

	\node[] at (\x+\y-.08,\y+.08+.1) {$\We_*$};
	
	\draw[|->,line width=.1cm] (\x,\x) -- (\fx,\fy); 
	\draw[|->,line width=.1cm] (\x,\x) -- (\x,\fy+0.03);
	\draw[|->,line width=.1cm] (\x,\x) -- (\fx,\x);
	
	\node[anchor=west] at (\fx,\fy+0.02) {finite $\La$};
	\node[anchor=south] at (\x,\fy+0.03) {inertialess: $\La = 0$};
	\node[anchor=west] at (\fx,\x) {nondeformable: $\La \rightarrow \infty$};

	\draw[fill=black!10,black!10] (\x-.08-0.03,\x-.1-0.03) rectangle (\x+.08+0.03,\x+.1+0.03);
	
	\draw[fill=red!20,red!20] (\y-.08-0.03,\x-.1-0.03) rectangle (\y+.08+0.03,\x+.1+0.03);
	\draw[fill=red!40,red!40] (\z-.08-0.03,\x-.1-0.03) rectangle (\z+.08+0.03,\x+.1+0.03);

	\draw[fill=yellow!40,yellow!40] (\x-.08-0.03,\y-.1-0.03) rectangle (\x+.08+0.03,\y+.1+0.03);
	\draw[fill=yellow!80,yellow!80] (\x-.08-0.03,\z-.1-0.03) rectangle (\x+.08+0.03,\z+.1+0.03);	

	\draw[fill=yellow!40!red!40,yellow!20!red!20] (\y-.08-0.03,\y-.1-0.03) rectangle (\y+.08+0.03,\y+.1+0.03);
	\draw[fill=yellow!80!red!80,yellow!40!red!40] (\z-.08-0.03,\z-.1-0.03) rectangle (\z+.08+0.03,\z+.1+0.03);
	
	\node[] at (\x,\x+.06+0.03) {\underline{$\We=0$}};
	\node[] at (\x,\x+0.03) {Inertialess and};
	\node[] at (\x,\x-.06+0.03) {nondeformable};
	\node[] at (\x,\x-.12+0.03) {limit};
	
	\node[] at (\y,\x+.06+0.03) {Pure};
	\node[] at (\y,\x+0+0.03) {linear};
	\node[] at (\y,\x-.06+0.03) {inertial};
	\node[] at (\y,\x-.12+0.03) {regime};

	\node[] at (\z,\x+.06+0.03) {Pure};
	\node[] at (\z,\x+0+0.03) {nonlinear};
	\node[] at (\z,\x-.06+0.03) {inertial};
	\node[] at (\z,\x-.12+0.03) {regime};
	
	\node[] at (\x,\y+.06+.03) {Pure};
	\node[] at (\x,\y+0+.03) {linear};
	\node[] at (\x,\y-.06+.03) {capillary};
	\node[] at (\x,\y-.12+.03) {regime};

	\node[] at (\x,\z+.06+.03) {Pure};
	\node[] at (\x,\z+0+.03) {nonlinear};
	\node[] at (\x,\z-.06+.03) {capillary};
	\node[] at (\x,\z-.12+.03) {regime};

	\node[] at (\z,\z+.06+.03) {\underline{Large $\We$}};
	\node[] at (\z,\z+0+.03) {Nonlinear};
	\node[] at (\z,\z-.06+.03) {inertial-capillary};
	\node[] at (\z,\z-.12+.03) {regime};
	
	\node[] at (\y,\y+.06+.03) {\underline{Small $\We$}};
	\node[] at (\y,\y+0+.03) {Linear};
	\node[] at (\y,\y-.06+.03) {inertial-capillary};
	\node[] at (\y,\y-.12+.03) {regime};
	

\end{tikzpicture}

%% file: aaResults_v2.tex
\section{Results and discussion}
\label{sec:results}

\subsection{Experimental observations and numerical validation}

\subsubsection{Beads: nondeformable case}
\label{subsec:rigid_particles}

\begin{figure}
\centering
\input{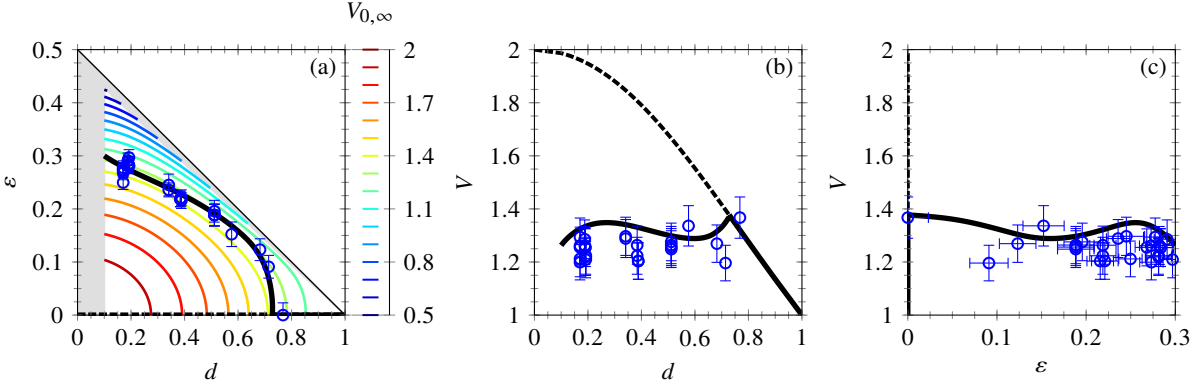}
\input{./texfigures/Fig4/Fig4b.tex}
\input{./texfigures/Fig4/Fig4c.tex}
 \caption{(a) Equilibrium eccentricity $\varepsilon$ vs. the diameter $d$ for neutrally buoyant beads ($f=0$). The experimental results (blue circle) are compared with numerical ones (black line) computed with the general model considering $\varphi=1$, $\Ca=0$, $\lambda\to \infty$ and small but finite values of $\Re$. The numerically-determined stable (solid line) and unstable (dashed line) equilibrium positions are shown. In addition, the numerically-determined $V$ computed in the $\Re=\Ca=0$ limit with $\lambda\to \infty$, i.e. $V_0(\lambda\to\infty,d,\varepsilon,)=\Vp(d,\varepsilon)$, is provided by the color code. Gray area corresponds to unexplored regions. (b-c) Equilibrium velocity $V$ vs. $d$ and $\varepsilon$, respectively. The same symbols and color code are used. However, the numerical results show $\Vp$ using for the equilibrium lateral position, the stable (solid line) and unstable (dashed line) results numerically-determined in (a).}
  \label{fig:fig_rigid_beads}
\end{figure}

Our investigation begins with the analysis of the dynamics of beads. This situation corresponds to the nondeformable rigid limit of the problem, i.e. when $\Ca=0$ (or $\La \rightarrow \infty$) and $\lambda \rightarrow \infty$, since no internal flow motion and deformation of the dispersed object may occur. In this limit, capillary effects are negligible and if a lateral migration is observed, it results solely from the effect of inertia.

Figures \ref{fig:fig_rigid_beads}(a), (b) and (c) report for beads the equilibrium eccentricity $\varepsilon$ as a function of their diameter $d$, as well as the equilibrium velocity $V$ as functions of their diameter $d$ and their equilibrium eccentricity $\varepsilon$, respectively. The experimental results (blue circles), obtained with the conditions $\varphi \approx1$, $\Re= 4.1-24$ and $\Ca=0$ (see \tabref{tab:liquids} for complementary information), are compared and well-agree with the numerical predictions of two different models (black lines) depending on whether the equilibrium lateral position or the equilibrium velocity is considered.

{In \figref{fig:fig_rigid_beads}(a), the experimental results concerning $\varepsilon$ vs. $d$ are compared with the numerical ones computed with the general model in the nonderformable limit (i.e. $\Ca=0$), for neutral buoyant beads (i.e. $f=0$ and $\lambda=10^{5}$) and for finite but small values of $\Re$, in accordance with the above-mentioned experimental conditions. Note that these numerical results reproduce exactly those already computed for beads in~\cite{Rivero2018}, this enabling to suppose that this modeling is also valid in the linear regime up to a value of $\Re=32$.}
It is observed that, for large beads with $d$ larger than a critical diameter $d_c=0.73$, it exists only one lateral equilibrium position corresponding to the centered position $\varepsilon=0$ (solid line). However, for smaller beads with $d < d_c$, the centered position loses its stability through a perfect pitchfork bifurcation and two branches of stability appear, one corresponding to an unstable equilibrium for centered positions $\varepsilon=0$ (dashed lines), and the other corresponding to a stable equilibrium for off-centered positions $\varepsilon \neq 0$ (solid line). In the latter case, a decrease of $d$ results in an increase of $\varepsilon$ due to inertial migration. It is worth to notice that the experimental results well-agree with the numerical prediction of the stable equilibrium position whatever $d$ in the large range of experimentally explored sizes, namely $d = 0.17 - 0.77$. Note also that for small beads of $d \approx 0.1$, the equilibrium eccentricity is close to $0.3 $, recovering both the seminal observation of \cite{Segre1962} and the analytical result of \cite{Ho1974}.

In figs~\ref{fig:fig_rigid_beads}(b) and (c), the experimental results concerning $V$ as functions of $d$ and $\varepsilon$ are compared with the numerical ones computed with the reduced 
model in the nonderformable and inertialess limit (i.e. $\Ca=\Re=0$), i.e. $V_0(\lambda\rightarrow \infty, d,\varepsilon) = \Vp (d,\varepsilon)$, using for the equilibrium lateral position, the stable (solid line) and unstable (dashed line) results numerically-determined in \figref{fig:fig_rigid_beads}~(a). It is observed that, for centered beads ($\varepsilon=0$), $\Vp$ monotonically decreases from the maximum value of the Poiseuille flow ($\Vp=2$) when $d\rightarrow0$, to the mean flow value ($\Vp=1$) when $d\rightarrow1$. However, as previously mentioned, when $d<d_c$, centered positions become unstable and inertial effects result in stable off-centered positions leading to intricate variations of $\Vp$ as functions of $d$ and $\varepsilon$. As shown in \figref{fig:fig_rigid_beads}(a) with the numerical prediction in the $\Re=\Ca=0$ limit of $\Vp(d,\varepsilon)$ and discussed in detail later in \secref{sec:parametric_analysis_num_velocity}, $\Vp$ decreases with the increases of $d$ and $\varepsilon$. Thus, the variations of $\Vp$ observed in figs~\ref{fig:fig_rigid_beads}(b) and (c) are direct consequences of  the stable equilibrium positions observed in \figref{fig:fig_rigid_beads}(a) for which $\varepsilon$ increases when $d$ decreases due to the inertial migration when $d<d_c$. It is worth to observe that the experimental results fairly-agree with the numerical prediction of the equilibrium velocity $\Vp$ in the large range of the experimentally explored sizes $d = 0.17 - 0.77$, and note that, because of the inertial migration, despite little variations, the equilibrium velocity of beads $V$ only slightly varies around a value $\Vp\approx1.3$ in this range of $d$. 

\subsubsection{Drops and bubbles: deformable cases}
\label{sec:velocities}

Increasing the complexity of the problem, our investigation continues with the analysis of the dynamics of bubbles and drops. With these dispersed objects, both internal flow motion and object deformation may occur, resulting in the additional influence of two dimensionless numbers on their dynamics: the viscosity ratio $\lambda$ and the capillary number $\Ca$. Thus, in this situation involving deformable objects, capillary effects can additionally participate in the problem and if a lateral migration is observed, it can be due to the sole or combined effect of inertia or/and capillarity.

\begin{figure}
  \begin{center}
    \includegraphics[height=0.8\textheight]{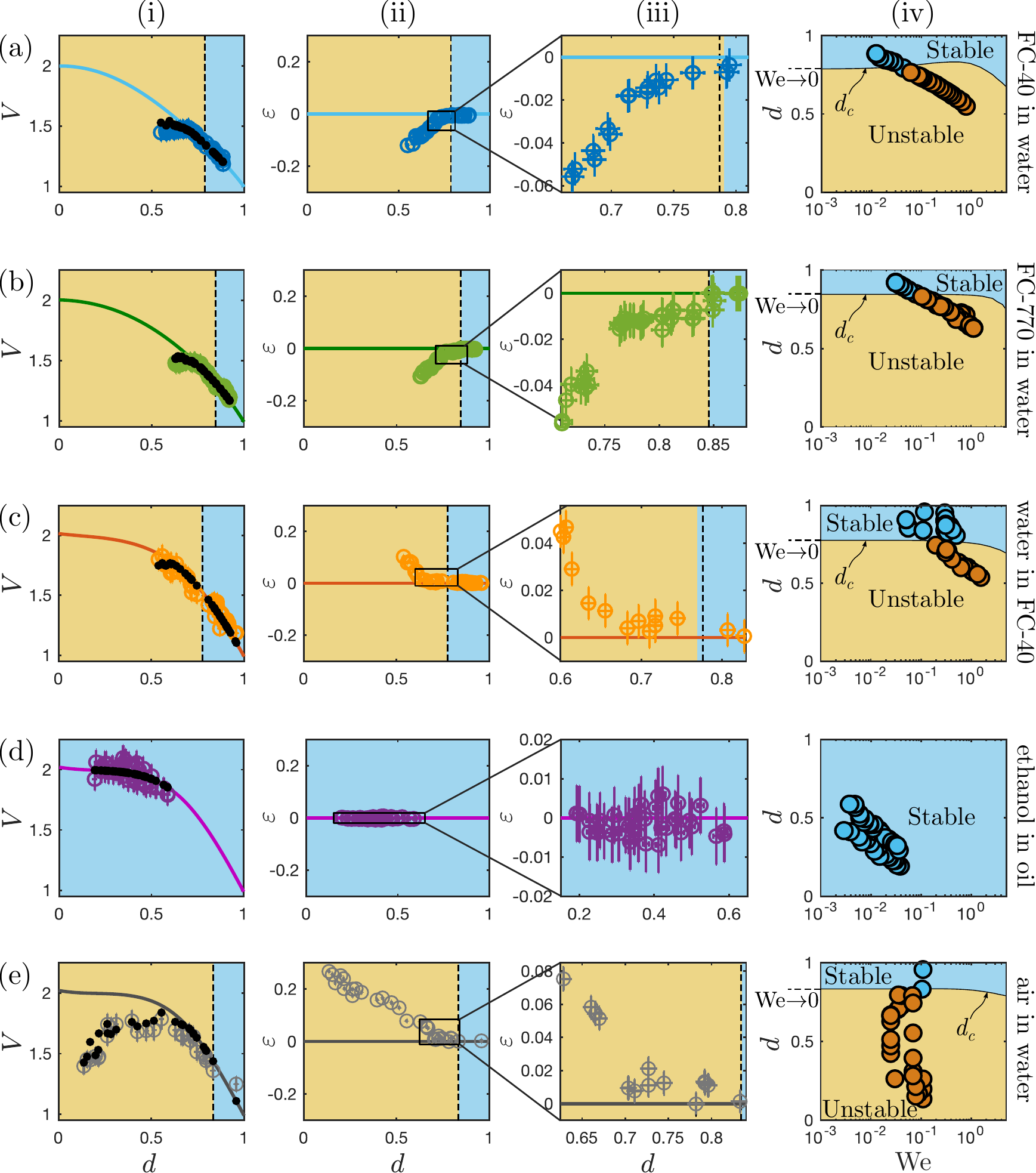}
        \caption{ Equilibrium (i) velocity $V$ and (ii-iii) eccentricity $\varepsilon$ as a function of the diameter $d$ of two types of deformable dispersed objects: drops (a-d) and bubble (e). The experimental results (colored open circles) obtained in the conditions summarized in \tabref{tab:liquids} are compared with numerical ones computed in the $\Re=\Ca=0$ limit, i.e. $V_0$, for centered objects $\varepsilon=0$ (solid lines). In column (i), the experimental results are also compared with numerical predictions computed from the same model but considering for $\varepsilon$ the experimental eccentricities from column (ii) (black dots). Column (iii) corresponds to a zoom of column (ii) at the vicinity of the centered/off-centered threshold, when defined. (iv) Stability maps of the centered position of the dispersed objects as functions of $d$ and the Weber number $\We$. The maps, in which centered positions appear as stable in blue and unstable in orange, compare experimental results (filled circles) and numerical ones obtained from the linear stability analysis of axisymmetric solution (background color). The same color code is reported in columns (i) to (iii) and the transition from stable to unstable centered position is determined from column (iv) for $\We$ corresponding to the experiments. The dotted lines show the transition at small 
 $\We$ (i.e. in the linear regime).}
  \label{fig:stability_centered_position}
  \end{center}
\end{figure}

Figure~\ref{fig:stability_centered_position} presents for five couples of fluids, corresponding to situations involving either drops (rows (a) to (d), intermediate $\lambda$) or bubbles (row (e), $\lambda \rightarrow 0$), the variations of the equilibrium velocity $V$ (column (i)) and eccentricity $\varepsilon$ (columns (ii) and (iii)) as a function of their diameter $d$. In addition, the stability of their centered position as a function of $d$ and $\We$ is plotted in column (iv). Note that the values of all dimensionless numbers of the problem concerning these couples of fluids are also provided in \tabref{tab:liquids}.

In columns (i), (ii) and (iii), the experimental results (colored opened circles) concerning $V$ and $\varepsilon$ are first compared with the numerical ones computed in the $\Ca=\Re=0$ limit for centered objects $\varepsilon=0$ (solid lines). Two types of behaviors are observed: while for the case (d), the comparison is perfect for all the experimentally explored range of $d$, for the cases (a), (b), (c) and (e), the comparison remains only excellent for large $d$ and gets poorer when $d$ decreases as deviations are observed for smaller $d$. Naturally, the observation done for $V$ in column (i) is directly linked with the one done for $\varepsilon$ in column (ii) in which it is seen that, except the case (d), for which the drops remain centered whatever $d$, for the cases (a), (b), (c) and (e), it exists a critical diameter $d_c$ (whose value varies for the different couples of fluids considered) above which the drops and bubbles are centered and below which a decrease of $d$ results in an increase of the absolute value of $\varepsilon$. To validate this interpretation and reveal the influence of $\varepsilon \neq 0$ on $V$, the numerical prediction of $V_0$ (computed in the $\Re=\Ca=0$ limit) using the experimentally measured eccentricities for the lateral position, are additionally plotted in column (i) as black points and show a good agreement. This confirms, as mentioned in \secref{SecEqs2}, that even if derived in the limit $\Re=\Ca=0$, the prediction of the equilibrium velocity $V$ is actually valid for a larger range of $\Ca$ and $\Re$ (see \tabref{tab:liquids}). Note also that, the equilibrium eccentricity can be either positive or negative depending on the value of the density ratio $\varphi$ compared to 1 ($\varphi<1$ or $\varphi>1$, respectively), with the same consequence on $V$, the velocity profile of a Poiseuille flow being axisymmetric. 

In addition, column (iv) shows numerically-determined phase diagrams highlighting the stable (in blue) or unstable (in orange) nature of the centered position for these drops or bubbles. The numerical results (background color) are computed with the linear stability around the centered position presented in \secref{SecEqs3}. For comparison, the experimental results have been added, a filled circle being blue when the dispersed object is centered or red when it is off-centered, and a very good agreement with the model is found. The predicted threshold, which corresponds to $f_1=0$, enables to obtain $d_c$ as a function of $\We$. When such a stability threshold exists (i.e. except in case (d)), for small values of $\We$, $d_c$ is independent of $\We$, which is a signature of the linear regime, while for larger values of $\We$, different behaviors depending on the couple of fluid considered are observed. In cases (b), (c) and (e), $d_c$ monotonously decreases with increasing $\We$, while in case (a), $d_c$ increases first to a maximum before decreasing. For the present conditions of the experiments, the thresholds for the stability of the centered positions are always within or very close to the linear regime, meaning that the numerically-determined $d_c$ for vanishing $\We$ are good predictions of the experimentally observed thresholds. 

The color code for the numerically-determined phase diagrams used in column (iv), as well as the predictions of $d_c$ for vanishing $\We$ (dashed line), are reported in columns (i), (ii) and (iii). It is observed that the predicted phases and transitions between centered and off-centered positions for these drops and bubbles are in very good agreement with the experiments.  Moreover, as seen in column (iii)  by zooming in on $\varepsilon$ as a function of $d$ in the vicinity of $d_c$, when $d<d_c$, the increase of $|\varepsilon|$ with decreasing $d$ is very small at first, before getting larger away from the transition (typically when $d \lesssim d_c -0.1$). A consequence of such a behavior is directly visible on $V$ in column (i), the experimental data matching on the predictions for the velocity of centered objects (solid curves) even for $d$ lower than $d_c$ before starting to deviate more markedly for $d\lesssim d_c-0.1$. 

Finally, the two types of behaviors observed experimentally can be understood and interpreted through the analysis of their Laplace number $\La$. The cases (a), (b), (c) and (e) are all characterized by large values of $\La$ ($\La\gtrsim10^3$). In these situations, the inertial effects dominate over capillary ones leading to the destabilization of the centered position when $d<d_c$ because of inertial migration forces pushing the dispersed objects away from the microchannel centerline, similarly to the bead case studied in \secref{subsec:rigid_particles}. On the contrary, the case (d) is characterized by a small value of $\La$ since $\La<1$. In this situation, the capillary effects dominate over the inertial ones and the deformation-induced migration forces stabilize the centered position whatever $d$. Interestingly, while the cases (d) and (e) correspond both to the bubble limit ($\lambda \ll 1$), their behaviors are strongly different because of their respective $\La$, and consequently because of different dominating forces. 

\begin{figure}
  \begin{center}
    \includegraphics[width=0.8\textwidth]{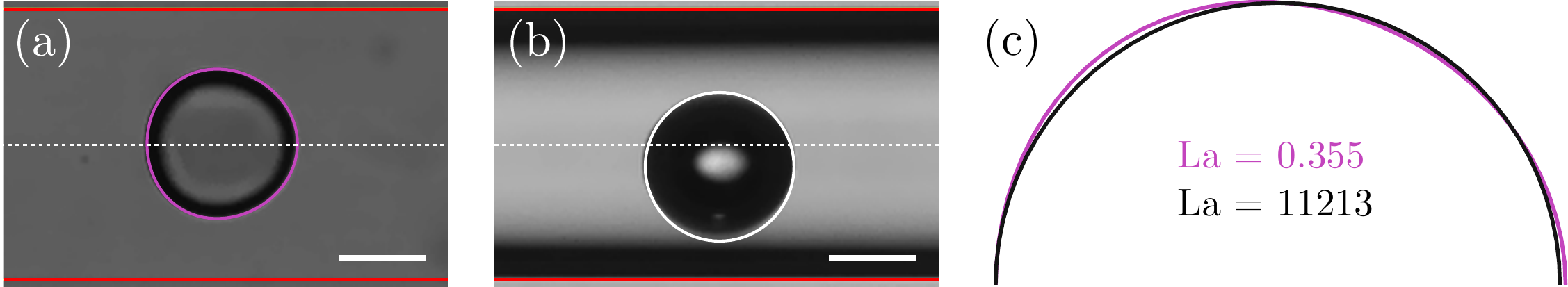}
        \caption{(a-b) Experimental images of an ethanol drop in mineral oil and an air bubble in water of identical size $d=0.56$, respectively. The two situations, corresponding to cases (d) and (e) in \figref{fig:stability_centered_position}, are obtained in the bubble limit since $\lambda \sim 10^{-2}$ for the following values of Laplace numbers: (a) $\La = 0.355$ and 
(b) $\La = 11213$. Complementary information are reported in \tabref{tab:liquids}. The scale bar is 50 $\upmu$m. The full red lines and the white dashed lines evidence the wall and the centerline of the capillary, respectively.  Numerically-determined shapes of the dispersed objects computed for the experimental conditions of (a) and (b) are superimposed on these pictures (magenta and black lines, respectively) and compared in (c) to highlight the influence of the regime on the object shape.}
  \label{fig:bubble_drop_comp}
  \end{center}
\end{figure}

{Figure \ref{fig:bubble_drop_comp}(a) and (b) show experimental pictures of an ethanol drop in mineral oil and an air bubble in water, of identical size $d=0.56$, extracted from the experiments corresponding to cases (d) and (e) in \figref{fig:stability_centered_position}, respectively. While the drop is deformed in the inertialess limit when $\La <1$ in (a), the bubble remains spherical in the nondeformable limit when $\La \gg 1$ in (b), although both interfaces are deformable by nature. This deformation of the dispersed object shape when $\La < 1$ compared to the axisymmetric case when $\La \gg 1$ is evidenced in \figref{fig:bubble_drop_comp}(c) via a comparison of the numerically-determined shapes corresponding to cases (a) and (b). Besides, a very satisfying comparison is also shown by the superimposition of these computed object shapes to the experimental ones in (a) and (b).}

\subsection{Numerical extension of the parametric analysis}
\label{sec:parametric_analysis_num}

Now that the numerically solved model has been validated by comparison with experiments, we propose in this section to use the model to extend our understanding about the velocity of a dispersed object, but also the stability of its centered position, over a larger range of parameters, and beyond the experimentally attainable ranges.

\subsubsection{Dispersed object velocity}
\label{sec:parametric_analysis_num_velocity}

\begin{figure}
\centering
\input{./texfigures/Velocity2/Vcentro.tex}
\input{./texfigures/Velocity2/Vpared.tex}
\caption{Equilibrium velocity of a dispersed object $V_0$ as functions of its size $d$ and of the viscosity ratio $\lambda$ for (a) a centered object [$\varepsilon=0$] and (b) an off-centered object near the wall [$\varepsilon = 0.45 (1-d)]$, computed in the $\Re=\Ca=0$ limit.}
\label{fig:velocity_evo_numerics}
\end{figure}
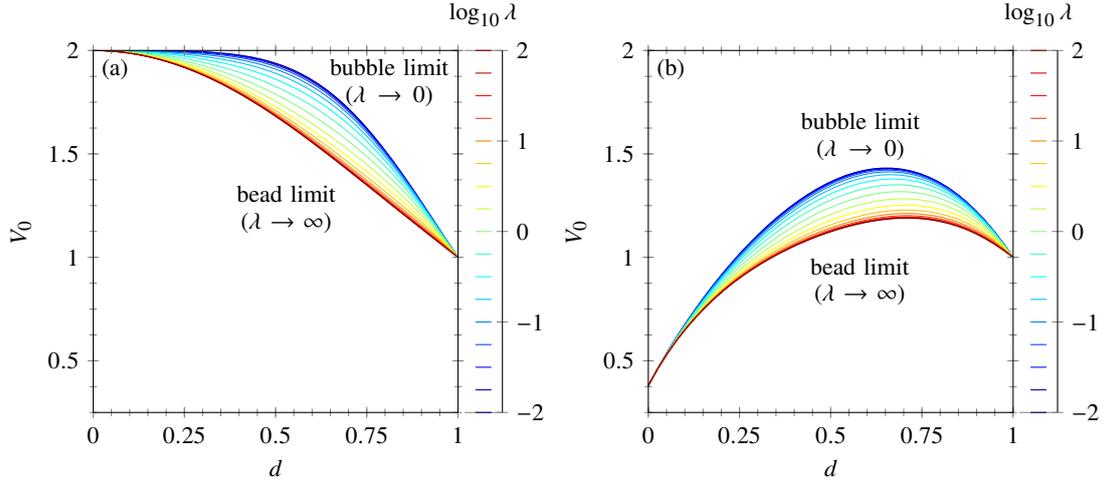

Figure~\ref{fig:velocity_evo_numerics} shows the velocity of a dispersed object as a function of its size for various values of the viscosity ratio $\lambda$ and at two given positions in the microchannel: centered ($\varepsilon=0$) and near the wall ($\varepsilon=0.45 (1-d)$), computed in the $\Re=\Ca=0$ limit. Note that to consider a physically sound situation close to the wall, the imposed eccentricity varies with the object size.

For a centered object in \figref{fig:velocity_evo_numerics}(a), the velocity decreases monotonically with its size, from the local value of the Poiseuille flow velocity when $d \rightarrow 0$, which is $V_0=2$ at the center, to its mean value $V_0 = 1$ when $d \rightarrow 1$. Between these two limit sizes, the velocity is also influenced by $\lambda$. The velocity increases monotonically with decreasing $\lambda$, from a minimum value when $\lambda\rightarrow\infty$ corresponding to the bead limit, to a maximum value when $\lambda\rightarrow 0$ corresponding to the bubble limit. Such a variation of $V_0$ with $\lambda$ results directly from the effect of the viscosity of the dispersed object on its internal flow, which ultimately leads to a change at the interphase from a no-slip situation for the bead limit to a stress-free situation for the bubble limit.

Similarly to the centered case, the velocity of the off-centered object in \figref{fig:velocity_evo_numerics}(b) increases monotonically with decreasing $\lambda$, from the bead limit value for $\lambda \rightarrow \infty$ to the bubble limit value for $\lambda\rightarrow0$, and varies from the local value of the Poiseuille flow velocity when $d \rightarrow 0$, which corresponds to $V_0 = 0.38$ when $\varepsilon=0.45$, to its mean value $V_0 = 1$ when $d \rightarrow 1$. 
However, between these two limit sizes, the variation of the velocity of the off-centered object is no longer monotonous and evidences a maximum value for a diameter between 0.65 and 0.7 depending on $\lambda$. 

In the end, it should be remembered that, whatever the given position and size of the dispersed object, bubbles are transported faster than drops, which are both faster than beads. However, for a given object of size $d$, its velocity depends on the eccentricity and is maximum for a centered position.\\

Inspired by the works of \cite{hadamard1911mouvement} and \cite{rybczynski1911uber}, we propose to generalize the analytical expression they derived for the stationary velocity of a liquid sphere
moving in another liquid as a function of the viscosity ratio $\lambda$. In our confined situation involving a circular microchannel, we account for both the dependences of the size $d$ and the position $\varepsilon$ of the dispersed object, still in the inertialess and nondeformable limit. The general expression for the velocity reads
\begin{align}
\label{Vdropecc}
V_0 (\lambda,d,\varepsilon) = \frac{\lambda_\star(d,\varepsilon) \Vb(d,\varepsilon) + \lambda \Vp (d,\varepsilon)}{\lambda_\star(d,\varepsilon) + \lambda} \,,
\end{align}

\noindent where the functions $\Vb$, $\Vp$ and $\lambda_\star$ are numerically determined from the computation of $V_0(\lambda,d,\varepsilon)$, in the $\Re=\Ca=0$ limit, as $\Vb (d,\varepsilon) = V_0(\lambda\rightarrow 0, d, \varepsilon)$  (bubble limit),  $\Vp (d,\varepsilon) = V_0(\lambda\rightarrow \infty, d,\varepsilon)$ (bead limit) and $\lambda_\star = \lambda (V_0- \Vp)/(\Vb- V_0)$, respectively. Note that $\lambda_\star$ can be determined for any arbitrary and intermediate value of $\lambda$ providing the same results. Since the dependencies of the limit velocities $\Vb$ and $\Vp$ with $d$ and $\varepsilon$ can be observed in \figref{fig:velocity_evo_numerics} and have been already discussed in depth in \cite{Rivero2018}, we only provide the results for $\lambda_\star$ in \appref{app:velocity}. Moreover, to enable an easy use of the equation \eqref{Vdropecc} predicting the velocity of a dispersed object in a circular microchannel, we provide the polynomial fittings of $\Vb(d,\varepsilon)$, $\Vp(d,\varepsilon)$ and $\lambda_\star(d,\varepsilon)$ in the same \appref{app:velocity}. Finally, it is worth noting that in the absence of confinement and shear, i.e. $d \rightarrow 0$ and $\varepsilon=0$, the original Hadamard-Rybczynski equation is recovered since $\lim_{d \rightarrow 0} \lambda_\star(d,0) = 2/3$. 

\subsubsection{Stability of the centered position and eccentricity}
\label{sec:parametric_analysis_num_stability}

\begin{figure}
\centering
\def\ydesp{-.38\textwidth}
\input{./texfigures/Nonlinear/mu_03_def_paperWe.tex}
\input{./texfigures/Nonlinear/mu_3_def_paperWe.tex} 
\input{./texfigures/Nonlinear/WeLa_nonlin_paper3.tex}
\input{./texfigures/Nonlinear/ReCa_nonlin_paper3.tex}
\caption{(a-b) Stability threshold of the centered equilibrium position $d_c$ of a dispersed object as functions of its size $d$ and the Weber number $\We$, for several values of the Laplace number $\La$, two given viscosity ratios: (a) $\lambda=0.3$ and (b) $\lambda=3$ and $\varphi=1$ using the stability analysis around the centered position. The color lines correspond to critical diameters $d_c$ above which the centered position is stable and below which it is unstable. The color dots correspond to critical Weber numbers $\We_*$ below which the regime is linear and above which it is non-linear in the sense of \figref{fig:regimes}. (c-d) Threshold between linear and nonlinear regimes plotted in a (c) $\La$-$\We$ plane and (d) $\Re$-$\Ca$ plane. These results being directly extracted from (a) and (b), they are obtained for the same conditions.}
\label{fig:nonlinear_regime}
\end{figure}
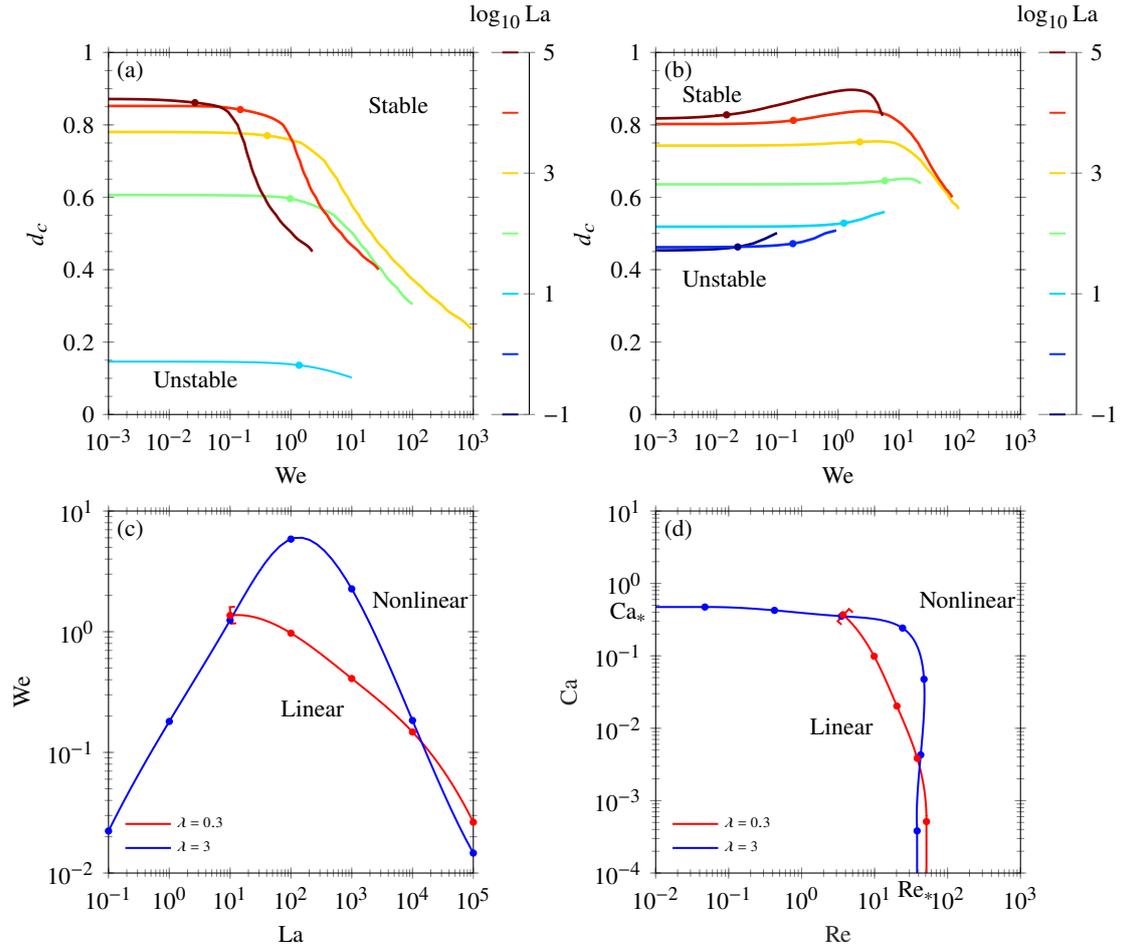

Figures \ref{fig:nonlinear_regime}(a) and (b) show stability maps for the centered position of dispersed objects as a function of $\We$, for various values of $\La$ and for two given viscosity ratios: $\lambda=0.3$ in (a) and $\lambda=3$ in (b), computed for $\varphi=1$ with the stability analysis presented in \secref{SecEqs3}. The threshold corresponds to the critical diameter $d_c$, above which the centered position is stable and below which it is unstable, thus resulting in a lateral migration with $\varepsilon \neq 0$. For small values of $\We$, $d_c$ is (quasi)-independent of $\We$ whatever the values of $\La$ and $\lambda$, which is a typical signature of the linear regime (in the sense of \figref{fig:regimes}). In the linear regime, it is observed that a decrease of $\La$ results in a decrease of $d_c$ for the two $\lambda$, highlighting the stabilizing role of capillarity. Note also that for $\lambda=0.3$, no curves appear for $\La <10^{1}$, the centered position for such a drop being always stable whatever $d$. For larger values of $\We$, when non-linearities arise, the behaviors between the two viscosity ratios are strongly different (as already evidenced in \secref{sec:velocities}). While $d_c$ monotonously decreases when $\We$ increases for $\lambda=0.3$, it first increases until it reaches a maximum before also decreasing for $\lambda=3$. This figure shows that the best strategy for stabilizing dispersed objects at their centered positions consists in increasing $\Ca$, since it leads simultaneously to an increase of $\We$ and a decrease of $\La$, both causing the stabilization.

In parallel to show that $\La$ and $\lambda$ strikingly affect the stability threshold of the centered equilibrium position $d_c$, figs~\ref{fig:nonlinear_regime}(a) and (b) also evidence and enable to characterize the influence of these two parameters on the transition between the linear and non-linear regimes. This transition is represented by color dots in \figref{fig:nonlinear_regime}(c) that corresponds to the transition Weber numbers $\We_*$ above which $d_c(\We,\La)$ differs by more than $0.01$ from its value in the linear regime (i.e. for $\lim_{\We \rightarrow 0} d_c(\We,\La)$). Two markedly different behaviors are observable depending on the value of $\lambda$. While $\We_*$, above which the regime is non-linear and below which the regime is linear, monotonously decreases when $\La$ increases for $\lambda=0.3$, it first increases until a maximum and then decreases for $\lambda=3$. Note that the curve for $\lambda=0.3$ can not be plotted below $\La = 10^{1}$ since for smaller values of $\La$, $\We_*$ can not be defined, the centered position being stable for any drop diameter. In addition, this transition between the linear and non-linear regimes are plotted in (d) as functions of $\Re$ and $\Ca$. It is interesting to observe that, while for $\lambda=0.3$, non-linearities arise for $\Re>[5-50]$ depending on $\Ca$ (with the transition Reynolds number $\Re_*$ decreasing when $\Ca$ increases), for $\lambda=3$, the transitions are stiffer and non-linearities develop when $\Re>[40-50]$ or/and $\Ca>[0.3-0.5]$. Interestingly, for these two values of $\lambda$ and for $\varphi=1$, figs~\ref{fig:nonlinear_regime}(c) and (d) highlight when non-linearities become important in this problem, as well as the domain of validity of the linear regime as sketched previously in \figref{fig:regimes}.

Note that, as a complement of \figref{fig:nonlinear_regime}, the reader can find results concerning these two transitions for the bubble and bead limits ($\lambda\to 0$ and $\lambda\to \infty$, respectively) in \cite{Rivero2018}.

\begin{figure}
\centering
\input{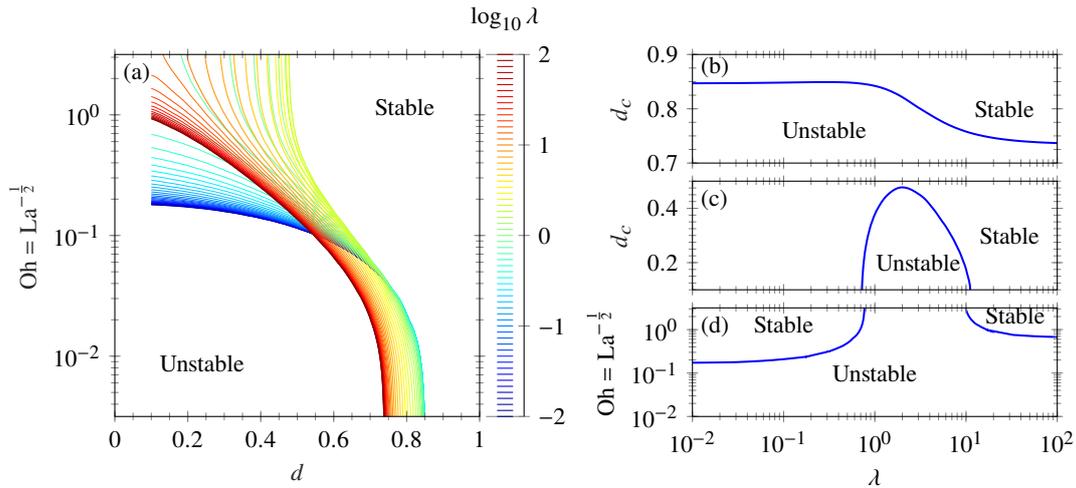}
\input{./texfigures/stability_linear/limited01Lin_paper.tex}
\input{./texfigures/La0Lainf_linear/La0_linear_paper.tex}
\input{./texfigures/La0Lainf_linear/Linf_linear_paper.tex}
\caption{(a) Stability map for the centered position of a dispersed object as functions of its size $d$ and of the Ohnesorge number $\Oh =\La^{-\frac12}$, for several values of $\lambda$, for $\varphi=0$ and $\We\to0$, i.e. in the linear regime. The panel presents the influence of $\lambda$ on the stable-unstable threshold as well as its various limits: (b)  nondeformable limit computed for $\La=10^{5}$, (c) inertialess limit computed for $\La = 0$, and (d) small objects computed for $d=0.1$.}
\label{fig:stability_linear_fig}
\end{figure}

In order to provide an exhaustive parametric analysis on the stability of the centered position of dispersed objects, we now provide numerical results computed 
in the linear regime (i.e. for $\We\to0$). Figure \ref{fig:stability_linear_fig}(a) shows the stability diagram of the centered position as functions of its size $d$ and of the Ohnesorge number $\Oh =\La^{-\frac12}$, for various values of $\lambda$ and for $\varphi=0$ (i.e. neglecting inertia of the internal flow). The influence of these parameters can be more easily described and grasped by regarding the different limits of this figure. In \figref{fig:stability_linear_fig}(b), the nondeformable limit obtained for $\La^{-\frac12} \to 0$ (equiv. to $\La\to\infty$) is shown. In this limit, the threshold of stability for the centered position exists for all values of $\lambda$. It is observed that $d_c$ is constant when $\lambda\lesssim0.04$ and then slowly decreases with increasing values of $\lambda$ until reaching a second plateau value for large $\lambda$. In this nondeformable limit, only large dispersed objects remains centered since as soon as $d<d_c=[0.73-0.85]$, inertial forces result in lateral migration leading to off-centered positions. At the opposite, \figref{fig:stability_linear_fig}(c) presents the inertialess limit when  $\La^{- \frac12} \to \infty$ (equiv. to $\La \to 0$). In this limit, capillary forces result in a stabilization of the centered position for objects of any size when $\lambda \lesssim 0.7$ and $\lambda \gtrsim 10$, a result explaining the experimental case (d) in \figref{fig:stability_centered_position} where $\lambda=0.04$. However, for intermediate values of $\lambda$, the same forces results in capillary-induced lateral migration and thus to off-centered positions for small drops with a diameter $d\lesssim0.475$. These results well-agree with the analytical investigation of \cite{Chan1979} that shown that the deformation-induced migration force is oriented toward the channel centerline when $\lambda<0.7$ or $\lambda>11.5$ but toward the channel walls when $0.7<\lambda<11.5$. Finally, \figref{fig:stability_linear_fig}(d) provides the stability map for small objects of size $d=0.1$. As seen above, it is generally the stability of the centered position of small dispersed objects that is rarely achieved. It is confirmed by this figure which evidences that for this object size, its centered position is always unstable for $\La^{-\frac12}\lesssim0.19$ (equiv. $\La \gtrsim 27.7$). For smaller values of $\La$, the capillary effects can lead to a stabilization of the centered position, but only when $ \lambda \lesssim 0.7$ and $\lambda \gtrsim 10$.

\begin{figure}
\input{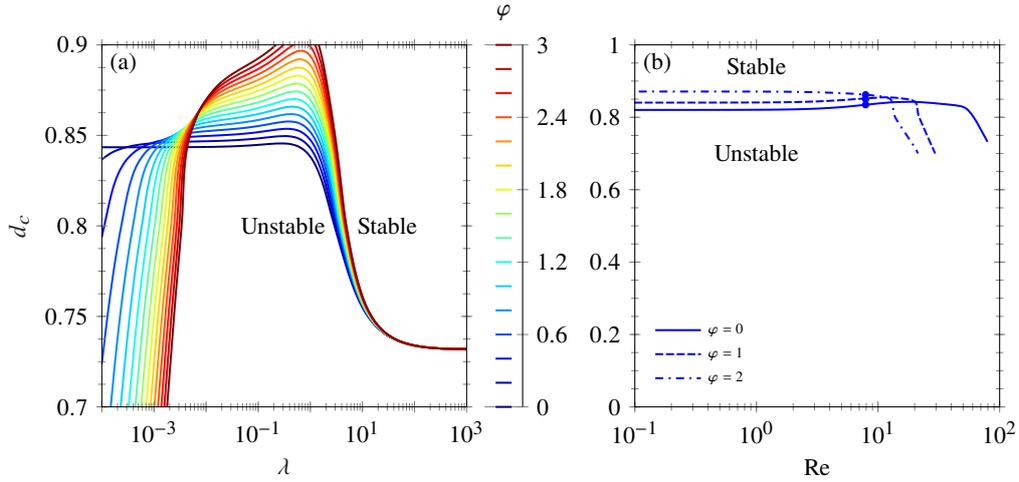}
\input{./texfigures/rhorat_def/rhoratNL_mu01blue_1red_paper.tex}
\caption{(a-b) Stability threshold of the centered position $d_c$ of a dispersed object as functions of (a) the viscosity ratio $\lambda$ and (b) the Reynolds number $\Re$, for several values of the density ratio $\varphi$, for $\La=3162$ (nondeformable limit) and for, in (a) $\We\to0$ (linear regime) and in (b) $\lambda=1$. Similarly to \figref{fig:nonlinear_regime}(a-b), the dots correspond to transition Reynolds numbers $\Re_*$ below which the regime is linear and above which it is non-linear. }
\label{fig:impact_of_phi}
\end{figure}

In order to provide a complete view of the problem, we end up exploring the influence of the density ratio $\varphi$ on the stability of the centered position. As $\varphi$ is related to the ratio of inertia between the inner and the outer flows of the dispersed object, its influence is negligible in the pure capillary regime and should be maximum in the pure inertial regime. Figure \ref{fig:impact_of_phi}(a) provides the stability map concerning the centered position of a dispersed object as function of $\lambda$ and $d$, for various values of $\varphi$ and $\La \approx 10^3$, i.e. in the nondeformable limit. Note that the curve corresponding to $\varphi = 0$ is actually the one already shown in Fig.~\ref{fig:stability_linear_fig}(b). 
{While for large values of $\lambda$ in the bead limit (typically when $\lambda \gtrsim 20$), $\varphi$ has no impact on the stability threshold $d_c$, for lower values of $\lambda$, $\varphi$ appears to affect this threshold, but in a different way according to the values of $\lambda$. For low values of $\lambda$ in the bubble limit (typically when $\lambda \lesssim 10^{-3}$), the inertia of the flow in the dispersed object plays an important stabilizing role as evidenced by the strong decrease of $d_c$ with increasing values of $\varphi$. On the contrary, for intermediate values of $\lambda$ (typically when $10^{-3} \lesssim \lambda \lesssim 20$), the stability threshold $d_c$ increases as a function of $\varphi$ and reaches a maximum value for $\lambda\approx 1$, thus showing the destabilizing character 
of the inner flow inertia in this range of $\lambda$.}
Finally, the influence of $\varphi$ on the stability of the centered position of a dispersed object is extended in \figref{fig:impact_of_phi}(b) by looking at its impact on the linear to non-linear regime transitions. These results are obtained by relaxing the assumption $\We\to0$ and by considering $\La \approx 10^3$ and $\lambda=1$, two conditions for which the destabilizing influence of $\varphi$ is the largest. As expected in this nondeformable limit, for small values of $\Re$, $d_c$ is independent of $\Re$ in the pure linear inertial regime. Then, an increase of $\Re$ leads first to small variations of $d_c$, which can either increase or decrease as a function of $\varphi$, when non-linearities related to inertial effects appear, before $d_c$ suddenly and sharply decreases when these non-linearities increase. It is observed that while the transition Reynolds number $\Re_*$ above which $d_c$ differs more that 0.01 from its value in the pure linear inertial regime does not depend on $\varphi$, the threshold at which the significant decrease in $d_c$ occurs is shortened with increasing relative influence of internal inertia.  Thus, while in the linear regime, the larger $\varphi$, the larger $d_c$, the situation is reversed in the strong non-linear regime.  Surprisingly, while a decrease of $\varphi$ tends to stabilize the centered position of a dispersed object in the linear regime, although this effect is quite relative, it has an opposite effect in the strong non-linear regime.

%% file: texfigures/Fig4/Fig4b.tex
%
%
\begin{tikzpicture}[baseline]

\begin{axis}[%
width=0.22\textwidth,
height=0.22\textwidth,
scale only axis,
xmin=0,
xmax=1,
xlabel={$d$},
ymin=1,
ymax=2,
ylabel={$V$},
axis background/.style={fill=white},
legend style={legend cell align=left, align=left, draw=white!15!blue},
xtick={0,.2,...,1},
ytick={1,1.2,1.4,1.6,1.8,2.0},
minor xtick={0,.05,...,2},
minor ytick={0,.05,...,2},
]

\node[anchor=north east,xshift=0.0cm,yshift=0.0cm] at (rel axis cs:1,1) {(b)};


\addplot[line width=2.0pt, black] table[row sep=crcr] {%
0.1	1.26263686741345\\
0.10625	1.27064060530636\\
0.1125	1.27799170557478\\
0.11875	1.28476355150943\\
0.125	1.29101464424818\\
0.13125	1.29679225965647\\
0.134595608723889	1.2996420629689\\
0.1375	1.30214615552716\\
0.14375	1.30711413544212\\
0.15	1.31170936995004\\
0.15625	1.31595419673708\\
0.1625	1.31986771921792\\
0.16875	1.32346643837122\\
0.175	1.32676474979515\\
0.18125	1.32977533953252\\
0.181588602050153	1.32992139391598\\
0.1875	1.33255224971575\\
0.19375	1.33506709618248\\
0.2	1.33732660278271\\
0.20625	1.33933925394747\\
0.2125	1.34111296309995\\
0.213314809997446	1.34131088499527\\
0.21875	1.34270038739084\\
0.225	1.34407168947407\\
0.23125	1.34522698246756\\
0.2375	1.34617305848347\\
0.240018271795586	1.34646874341636\\
0.24375	1.34694982497746\\
0.25	1.34755448312525\\
0.25625	1.34797156522226\\
0.2625	1.34820815200808\\
0.263930230124845	1.34822175350309\\
0.26875	1.34831629753855\\
0.275	1.34827315144902\\
0.28125	1.34807355739077\\
0.286070228687893	1.3478059658446\\
0.2875	1.3477394575974\\
0.29375	1.34731209101303\\
0.3	1.3467561753199\\
0.30625	1.34608054262024\\
0.307110615221162	1.34597282781422\\
0.3125	1.3453402771748\\
0.31875	1.34449461339279\\
0.325	1.34353795156748\\
0.327446279860101	1.3431252693596\\
0.33125	1.34251381555191\\
0.3375	1.34141099347918\\
0.34375	1.34020717302158\\
0.347239728334087	1.3394864027488\\
0.35	1.33893920727698\\
0.35625	1.3376199573522\\
0.3625	1.33621311533167\\
0.366688567725065	1.33522120045827\\
0.36875	1.33475067144721\\
0.375	1.33326541370322\\
0.38125	1.33171018889831\\
0.386037572023274	1.330476221531\\
0.3875	1.3301120031846\\
0.39375	1.32852288953493\\
0.4	1.32688659631816\\
0.405610651309255	1.32538542483395\\
0.40625	1.32521989800422\\
0.4125	1.32359412986768\\
0.41875	1.32193422975472\\
0.425	1.32024295092003\\
0.425772126200775	1.32003436225912\\
0.43125	1.31859867472767\\
0.4375	1.31693933545342\\
0.44375	1.31525846107131\\
0.446789332935434	1.31444175854198\\
0.45	1.31360369690855\\
0.45625	1.311975497635\\
0.4625	1.31033910569553\\
0.46875	1.30870037492592\\
0.469015752882443	1.30863230278706\\
0.475	1.30713957266528\\
0.48125	1.30559118434525\\
0.4875	1.30405962375431\\
0.49353586663194	1.30260599206405\\
0.49375	1.30255570063856\\
0.5	1.30114995639113\\
0.50625	1.29978417569717\\
0.5125	1.29846337186763\\
0.51875	1.29719196682143\\
0.523274104718552	1.29631622468657\\
0.525	1.29598919569084\\
0.53125	1.29487832249496\\
0.5375	1.29382838811958\\
0.54375	1.29284562298377\\
0.55	1.29193688143027\\
0.55625	1.29110968418136\\
0.5625	1.29037225926919\\
0.56875	1.28973357943014\\
0.575	1.28920339320777\\
0.58125	1.2887922460479\\
0.5875	1.28851148644077\\
0.59375	1.28837325061406\\
0.6	1.28839041734669\\
0.60625	1.28857588917262\\
0.6125	1.2889405482958\\
0.618448343001017	1.2894665534682\\
0.61875	1.28949814651505\\
0.625	1.29032770361199\\
0.63125	1.2913796402435\\
0.632905632545086	1.29170802845417\\
0.6375	1.29274411466703\\
0.642503050089074	1.29405018842549\\
0.64375	1.29441675382374\\
0.64973978351487	1.29637686762359\\
0.65	1.29647193232534\\
0.655513269166901	1.29865549159144\\
0.65625	1.29897838048216\\
0.660316397568635	1.3008738924354\\
0.6625	1.30198951166757\\
0.664442156598611	1.30302969495711\\
0.668057876542805	1.30512337069652\\
0.66875	1.30555145532721\\
0.671257741614998	1.30715162227352\\
0.674170183879427	1.30912162559054\\
0.675	1.30970840742106\\
0.676830111876286	1.31102809654637\\
0.679305738930733	1.31287682351915\\
0.68125	1.31437877841512\\
0.681625617280243	1.31467197002074\\
0.683785575786369	1.31640667386265\\
0.685824594707033	1.3180946657172\\
0.6875	1.31952392410435\\
0.687748946756537	1.31973633543388\\
0.689556940163553	1.32132262711703\\
0.691271578704441	1.32286802231397\\
0.692898513133882	1.32437437232636\\
0.69375	1.32518563325975\\
0.694440257554636	1.32583641389804\\
0.695903122489723	1.32725402798983\\
0.697295012240449	1.32863697058844\\
0.698619471200367	1.32998669095804\\
0.699879742350607	1.33130456021124\\
0.7	1.33143462950493\\
0.70108011213776	1.33257802905979\\
0.702223576759118	1.33382052496173\\
0.70331232091368	1.33503480990947\\
0.704348677158871	1.33622202663981\\
0.705335904423855	1.33738293333194\\
0.70625	1.33848568764763\\
0.706277429661541	1.33851774136308\\
0.707181538914413	1.33961330495988\\
0.708046232152779	1.34068446971684\\
0.70887399851191	1.34173183310835\\
0.70966715891829	1.34275596710765\\
0.710427883967173	1.34375741774663\\
0.711158209958763	1.34473670449939\\
0.711860053333737	1.34569431949941\\
0.7125	1.34658270655558\\
0.712535594933334	1.34663002793121\\
0.713192899697518	1.34753256802852\\
0.713827049016644	1.34841485311887\\
0.714439564194956	1.34927725875652\\
0.715031905458807	1.35012012576398\\
0.715605479749058	1.35094375867566\\
0.716161504102082	1.35174846216054\\
0.716700588974549	1.35253465834322\\
0.717223187673322	1.35330278032333\\
0.717729743106569	1.35405323594601\\
0.718220688409994	1.35478640852383\\
0.718696447577738	1.35550265748359\\
0.71875	1.35558574034236\\
0.719163574387482	1.35619248504196\\
0.719617097411886	1.35686492241393\\
0.720056609319255	1.3575215473076\\
0.720482494374827	1.35816263196479\\
0.720895133308965	1.35878842490912\\
0.7212949040266	1.3593991509954\\
0.721682182379339	1.35999501133462\\
0.722057343011757	1.36057618307974\\
0.722420760295546	1.36114281905464\\
0.722772809367856	1.36169504720426\\
0.723113847761835	1.36223299677127\\
0.723444147371796	1.36275689248599\\
0.723763935080766	1.3632669989001\\
0.724073409710269	1.36376359831801\\
0.724372741842655	1.36424699239389\\
0.724662073402537	1.36471750397157\\
0.724941516977479	1.36517547920361\\
0.725	1.36527533713489\\
0.725215229379669	1.36561517286681\\
0.725480205663191	1.36604159236195\\
0.725735340674485	1.36645685542131\\
0.725980612174969	1.36686143422315\\
0.726215959441372	1.36725584204798\\
0.726441280112222	1.36764063848657\\
0.726656426389277	1.36801643551123\\
0.726861200479729	1.36838390457544\\
0.727055349138201	1.36874378494471\\
0.727238635739745	1.36909676725632\\
0.727411091567602	1.36944309265983\\
0.727572816040678	1.36978287967342\\
0.727723903588712	1.37011624277577\\
0.727864443762333	1.37044329250695\\
0.727994521337079	1.37076413556426\\
0.728114216411641	1.37107887489334\\
0.728223604500495	1.37138760977463\\
0.728322756621147	1.37169043590537\\
0.728411739376162	1.37198744547739\\
0.728490615030142	1.37227872725065\\
0.728559441581819	1.37256436662294\\
0.72861827283141	1.37284444569566\\
0.728667158443364	1.37311904333585\\
0.728706144004642	1.37338823523468\\
0.728735271078632	1.37365209396244\\
0.728754577254826	1.37391068902009\\
0.728764096194343	1.37416408688761\\
0.728763857671404	1.37441235106905\\
0.728753887610838	1.37465554213454\\
0.72873420812169	1.3748937177592\\
0.728704837527011	1.37512693275916\\
0.728665790389885	1.37535523912455\\
0.728617077535749	1.37557868604981\\
0.728558706071054	1.37579731996105\\
0.72849067939831	1.37601118454075\\
0.728412997227546	1.3762203207498\\
0.728325655584213	1.37642476684677\\
0.728228646813556	1.37662455840466\\
0.728121959581469	1.37681972832498\\
0.728005578871833	1.37701030684932\\
0.73125	1.37245740098674\\
0.7375	1.36366074709118\\
0.74375	1.35485607710923\\
0.75	1.34604456988612\\
0.75625	1.33722740426711\\
0.7625	1.32840575909745\\
0.76875	1.31958081322238\\
0.775	1.31075374548717\\
0.78125	1.30192573473704\\
0.7875	1.29309795981726\\
0.79375	1.28427159957308\\
0.8	1.27544783284974\\
0.80625	1.26662783849248\\
0.8125	1.25781279534657\\
0.81875	1.24900388225725\\
0.825	1.24020227806977\\
0.83125	1.23140916162938\\
0.8375	1.22262571178133\\
0.84375	1.21385310737086\\
0.85	1.20509252724323\\
0.85625	1.19634515024368\\
0.8625	1.18761215521747\\
0.86875	1.17889472100984\\
0.875	1.17019402646604\\
0.88125	1.16151125043133\\
0.8875	1.15284757175095\\
0.89375	1.14420416927014\\
0.9	1.13558222183417\\
1 1\\
};

\addplot [color=black, dashed, line width=1.5pt]
  table[row sep=crcr]{%
0	2\\
0.1	1.9866088569792\\
0.10625	1.98490510487271\\
0.1125	1.98309496543699\\
0.11875	1.9811787669022\\
0.125	1.97915683749851\\
0.13125	1.97702950545611\\
0.1375	1.97479709900516\\
0.14375	1.97245994637585\\
0.15	1.97001837579833\\
0.15625	1.96747271550279\\
0.1625	1.96482329371939\\
0.16875	1.96207043867832\\
0.175	1.95921447860975\\
0.18125	1.95625574174384\\
0.1875	1.95319455631078\\
0.19375	1.95003125054074\\
0.2	1.94676615266388\\
0.20625	1.94339959091039\\
0.2125	1.93993189351043\\
0.21875	1.93636338869419\\
0.225	1.93269440469182\\
0.23125	1.92892526973352\\
0.2375	1.92505631204945\\
0.24375	1.92108785986978\\
0.25	1.91702024142469\\
0.25625	1.91285378494435\\
0.2625	1.90858881865893\\
0.26875	1.90422567079862\\
0.275	1.89976466959358\\
0.28125	1.89520614327398\\
0.2875	1.89055042007\\
0.29375	1.88579782821181\\
0.3	1.88094869592959\\
0.30625	1.8760034386209\\
0.3125	1.87096282035288\\
0.31875	1.86582769236004\\
0.325	1.86059890587691\\
0.33125	1.85527731213801\\
0.3375	1.84986376237787\\
0.34375	1.84435910783101\\
0.35	1.83876419973195\\
0.35625	1.83307988931522\\
0.3625	1.82730702781534\\
0.36875	1.82144646646684\\
0.375	1.81549905650423\\
0.38125	1.80946564916203\\
0.3875	1.80334709567479\\
0.39375	1.79714424727701\\
0.4	1.79085795520322\\
0.40625	1.7844891358891\\
0.4125	1.77803896657499\\
0.41875	1.77150868970236\\
0.425	1.76489954771272\\
0.43125	1.75821278304754\\
0.4375	1.75144963814832\\
0.44375	1.74461135545654\\
0.45	1.73769917741369\\
0.45625	1.73071434646126\\
0.4625	1.72365810504074\\
0.46875	1.71653169559361\\
0.475	1.70933636056137\\
0.48125	1.7020733423855\\
0.4875	1.6947438835075\\
0.49375	1.68734922636884\\
0.5	1.67989061341102\\
0.50625	1.67236933030688\\
0.5125	1.66478683565461\\
0.51875	1.6571446312838\\
0.525	1.64944421902399\\
0.53125	1.64168710070476\\
0.5375	1.63387477815565\\
0.54375	1.62600875320623\\
0.55	1.61809052768606\\
0.55625	1.6101216034247\\
0.5625	1.60210348225171\\
0.56875	1.59403766599665\\
0.575	1.58592565648908\\
0.58125	1.57776895555856\\
0.5875	1.56956906503466\\
0.59375	1.56132748674693\\
0.6	1.55304572252494\\
0.60625	1.54472526967401\\
0.6125	1.53636760740254\\
0.61875	1.52797421039473\\
0.625	1.51954655333473\\
0.63125	1.51108611090674\\
0.6375	1.50259435779491\\
0.64375	1.49407276868343\\
0.65	1.48552281825648\\
0.65625	1.47694598119823\\
0.6625	1.46834373219285\\
0.66875	1.45971754592452\\
0.675	1.45106889707742\\
0.68125	1.44239926033572\\
0.6875	1.4337101103836\\
0.69375	1.42500292190524\\
0.7	1.4162791695848\\
0.70625	1.40754027879998\\
0.7125	1.39878747770252\\
0.71875	1.39002194513765\\
0.725	1.38124485995065\\
0.73125	1.37245740098674\\
0.7375	1.36366074709118\\
0.74375	1.35485607710923\\
0.75	1.34604456988612\\
0.75625	1.33722740426711\\
0.7625	1.32840575909745\\
0.76875	1.31958081322238\\
0.775	1.31075374548717\\
0.78125	1.30192573473704\\
0.7875	1.29309795981726\\
0.79375	1.28427159957308\\
0.8	1.27544783284974\\
0.80625	1.26662783849248\\
0.8125	1.25781279534657\\
0.81875	1.24900388225725\\
0.825	1.24020227806977\\
0.83125	1.23140916162938\\
0.8375	1.22262571178133\\
0.84375	1.21385310737086\\
0.85	1.20509252724323\\
0.85625	1.19634515024368\\
0.8625	1.18761215521747\\
0.86875	1.17889472100984\\
0.875	1.17019402646604\\
0.88125	1.16151125043133\\
0.8875	1.15284757175095\\
0.89375	1.14420416927014\\
0.9	1.13558222183417\\
1 1\\
};

\addplot[
  blue,
  only marks,
  mark=o,
  mark size=2pt,
  error bars/.cd, 
    y fixed,
    y dir=both, 
    y explicit
] table [x=d, y=V,x error=inc_d,y error=inc_V, col sep=comma] {  
d,inc_d,eps,inc_eps,V,inc_V
0.1705029838,0.0029071267,0.2732,0.013081174,1.1992080419,0.0665687432
0.1705029838,0.0029071267,0.2497,0.013081174,1.2116318753,0.0672583975
0.1705029838,0.0029071267,0.27287,0.013081174,1.2614169591,0.0700219968
0.1705029838,0.0029071267,0.268,0.013081174,1.2550423029,0.0696681359
0.1705029838,0.0029071267,0.27744,0.013081174,1.2938730656,0.0718236543
0.3410059676,0.0058142535,0.23581,0.013081174,1.2883249263,0.0715156738
0.3410059676,0.0058142535,0.2451,0.0204958487,1.2967423613,0.0719829306
0.6820119352,0.011628507,0.1232,0.0204958487,1.2690835456,0.0704475735
0.7142857143,0.012755102,0.091,0.0214657416,1.1960184062,0.0669957704
0.7680491551,0.0147474876,0,0.0230814426,1.3671985406,0.0777907662
0.5760368664,0.0110606157,0.15225,0.0230814426,1.33623067,0.0760287585
0.5115089514,0.0087213802,0.1884,0.0204958487,1.2484126606,0.0693001205
0.5115089514,0.0087213802,0.1884,0.0204958487,1.2561189115,0.069727899
0.5115089514,0.0087213802,0.1884,0.0204958487,1.2638251625,0.0701556776
0.5115089514,0.0087213802,0.19565,0.0204958487,1.2756366127,0.0708113381
0.1886792453,0.0035599858,0.2821,0.0144756766,1.2621992677,0.0715344924
0.1900327807,0.0036112458,0.2835,0.0145795213,1.2540510421,0.0711860299
0.1900327807,0.0036112458,0.2835,0.0145795213,1.2546050215,0.0712174764
0.1920122888,0.0036868719,0.2812,0.0147313913,1.2187598419,0.0693449116
0.1880936706,0.0035379229,0.2908,0.0144307506,1.2857148852,0.0728171589
0.1920122888,0.0036868719,0.28125,0.0147313913,1.2236618422,0.0696238252
0.1920122888,0.0036868719,0.296875,0.0147313913,1.2092658206,0.0688047214
0.3880669415,0.0075297976,0.215789,0.0148864586,1.2034020957,0.0686359175
0.3840245776,0.0073737438,0.21875,0.0147313913,1.2234796733,0.0696134602
0.3880669415,0.0075297976,0.221,0.0148864586,1.2035723804,0.0686456297
0.3840245776,0.0073737438,0.2188,0.0147313913,1.2631842202,0.0718725667
};

\end{axis}
\end{tikzpicture}%

%% file: texfigures/Fig4/Fig4c.tex
%
%
\begin{tikzpicture}[baseline]

\begin{axis}[%
width=0.22\textwidth,
height=0.22\textwidth,
scale only axis,
xmin=0,
xmax=0.3,
xlabel={$\varepsilon$},
ymin=1,
ymax=2,
ylabel={$V$},
axis background/.style={fill=white},
legend style={legend cell align=left, align=left, draw=white!15!black},
xtick={0,0.1,0.2,0.3},
ytick={1,1.2,1.4,1.6,1.8,2.0},
minor xtick={0,.025,...,0.3},
minor ytick={0,.05,...,2},
]

\node[anchor=north east,xshift=0.0cm,yshift=0.0cm] at (rel axis cs:1,1) {(c)};

\addplot[line width=2.0pt, black] table[row sep=crcr] {%
0.300004921600188	1.26263686741345\\
0.297846229076899	1.27064060530636\\
0.295778661158872	1.27799170557478\\
0.293788655593419	1.28476355150943\\
0.291865480947934	1.29101464424818\\
0.290000542458049	1.29679225965647\\
0.289031544742607	1.2996420629689\\
0.288183108582724	1.30214615552716\\
0.286405420547498	1.30711413544212\\
0.284667379320862	1.31170936995004\\
0.282965184228797	1.31595419673708\\
0.281295647759602	1.31986771921792\\
0.279656076977541	1.32346643837122\\
0.278044180214733	1.32676474979515\\
0.276457992732514	1.32977533953252\\
0.276373692784235	1.32992139391598\\
0.274880459081518	1.33255224971575\\
0.273323777366564	1.33506709618248\\
0.271787547840327	1.33732660278271\\
0.270270622905888	1.33933925394747\\
0.268771965709313	1.34111296309995\\
0.268579240649309	1.34131088499527\\
0.267273985716884	1.34270038739084\\
0.265789260745028	1.34407168947407\\
0.264319447440516	1.34522698246756\\
0.262863732087874	1.34617305848347\\
0.262282756589297	1.34646874341636\\
0.261408901455246	1.34694982497746\\
0.259957607603031	1.34755448312525\\
0.258517355918424	1.34797156522226\\
0.257087277451044	1.34820815200808\\
0.256761837694732	1.34822175350309\\
0.255649490269981	1.34831629753855\\
0.254214453567902	1.34827315144902\\
0.252786095382859	1.34807355739077\\
0.251688132269209	1.3478059658446\\
0.2513581160214	1.3477394575974\\
0.249916834024901	1.34731209101303\\
0.248477994625472	1.3467561753199\\
0.247040338930382	1.34608054262024\\
0.246841843327461	1.34597282781422\\
0.245583673834861	1.3453402771748\\
0.244124418897214	1.34449461339279\\
0.242665607066443	1.34353795156748\\
0.242093067620671	1.3431252693596\\
0.241191491759035	1.34251381555191\\
0.239707195849525	1.34141099347918\\
0.238222275291782	1.34020717302158\\
0.237390551922252	1.3394864027488\\
0.236723793345405	1.33893920727698\\
0.235207856908255	1.3376199573522\\
0.233689477737155	1.33621311533167\\
0.232667735177569	1.33522120045827\\
0.232157810844541	1.33475067144721\\
0.23060162536413	1.33326541370322\\
0.229040258174895	1.33171018889831\\
0.227837619756988	1.330476221531\\
0.227465022153635	1.3301120031846\\
0.225857646369321	1.32852288953493\\
0.224241229754037	1.32688659631816\\
0.222779914089363	1.32538542483395\\
0.22261105788527	1.32521989800422\\
0.220940607221563	1.32359412986768\\
0.219259047371776	1.32193422975472\\
0.217566038263083	1.32024295092003\\
0.217354222543535	1.32003436225912\\
0.215832199054304	1.31859867472767\\
0.21408197196662	1.31693933545342\\
0.212318719178783	1.31525846107131\\
0.211451811610421	1.31444175854198\\
0.210524964518419	1.31360369690855\\
0.208701339855218	1.311975497635\\
0.206862322993914	1.31033910569553\\
0.205006817745323	1.30870037492592\\
0.204926732871932	1.30863230278706\\
0.203104655885743	1.30713957266528\\
0.201182570136989	1.30559118434525\\
0.199240266803236	1.30405962375431\\
0.197342958216656	1.30260599206405\\
0.197275031759844	1.30255570063856\\
0.195260475661859	1.30114995639113\\
0.193221003557742	1.29978417569717\\
0.191155328669915	1.29846337186763\\
0.189062205974352	1.29719196682143\\
0.187524600214226	1.29631622468657\\
0.186934423720444	1.29598919569084\\
0.184763005578536	1.29487832249496\\
0.182561049938005	1.29382838811958\\
0.180326721546572	1.29284562298377\\
0.178057961173717	1.29193688143027\\
0.175752463436612	1.29110968418136\\
0.17340765307107	1.29037225926919\\
0.17102065987209	1.28973357943014\\
0.168588292693965	1.28920339320777\\
0.166107013127908	1.2887922460479\\
0.163572909784363	1.28851148644077\\
0.160981674520729	1.28837325061406\\
0.158328582500107	1.28839041734669\\
0.155608601162255	1.28857588917262\\
0.1528167675025	1.2889405482958\\
0.150086921327334	1.2894665534682\\
0.149946366536452	1.28949814651505\\
0.146959082088815	1.29032770361199\\
0.14387652721967	1.2913796402435\\
0.143037746693858	1.29170802845417\\
0.140652106393712	1.29274411466703\\
0.137971479106248	1.29405018842549\\
0.137283410969914	1.29441675382374\\
0.133878758527617	1.29637686762359\\
0.133725789389235	1.29647193232534\\
0.130393610225497	1.29865549159144\\
0.129931569019938	1.29897838048216\\
0.127315006458162	1.3008738924354\\
0.125853863660354	1.30198951166757\\
0.124523418449734	1.30302969495711\\
0.121949830902926	1.30512337069652\\
0.121439807452214	1.30555145532721\\
0.119554313498608	1.30715162227352\\
0.117286006076214	1.30912162559054\\
0.116620906784378	1.30970840742106\\
0.115129272644073	1.31102809654637\\
0.11305725414649	1.31287682351915\\
0.111385341027069	1.31437877841512\\
0.111057938972165	1.31467197002074\\
0.10913103273154	1.31640667386265\\
0.107261446963302	1.3180946657172\\
0.105682541813619	1.31952392410435\\
0.105445717003505	1.31973633543388\\
0.103683131312564	1.32132262711703\\
0.101964796955232	1.32286802231397\\
0.100287829304717	1.32437437232636\\
0.0993852987315252	1.32518563325975\\
0.0986504402855832	1.32583641389804\\
0.0970496675491781	1.32725402798983\\
0.0954821396936865	1.32863697058844\\
0.0939459617117605	1.32998669095804\\
0.0924393840455258	1.33130456021124\\
0.0922909968957284	1.33143462950493\\
0.0909603877518302	1.33257802905979\\
0.0895074053452572	1.33382052496173\\
0.0880791547287512	1.33503480990947\\
0.0866743428876044	1.33622202663981\\
0.0852914432898138	1.33738293333194\\
0.0839683223314198	1.33848568764763\\
0.0839289297666339	1.33851774136308\\
0.0825839566030445	1.33961330495988\\
0.0812566639027911	1.34068446971684\\
0.0799459293148934	1.34173183310835\\
0.078650713003971	1.34275596710765\\
0.0773700497525211	1.34375741774663\\
0.076103041946412	1.34473670449939\\
0.0748488533332283	1.34569431949941\\
0.0736710820553829	1.34658270655558\\
0.0736066084067185	1.34663002793121\\
0.0723739792169545	1.34753256802852\\
0.0711519856644164	1.34841485311887\\
0.0699399973631886	1.34927725875652\\
0.0687374212418697	1.35012012576398\\
0.0675436985595987	1.35094375867566\\
0.0663583358573843	1.35174846216054\\
0.0651809973023401	1.35253465834322\\
0.0640113916966054	1.35330278032333\\
0.0628492368864281	1.35405323594601\\
0.0616942594321087	1.35478640852383\\
0.0605461942908853	1.35550265748359\\
0.0604122484681571	1.35558574034236\\
0.0594034861207729	1.35619248504196\\
0.0582670719440924	1.35686492241393\\
0.0571368834494881	1.3575215473076\\
0.0560126876506695	1.35816263196479\\
0.0548942579605218	1.35878842490912\\
0.0537813739886169	1.3593991509954\\
0.0526738213426603	1.35999501133462\\
0.0515713914333655	1.36057618307974\\
0.0504738812821966	1.36114281905464\\
0.0493810933313506	1.36169504720426\\
0.0482928386618518	1.36223299677127\\
0.047208948280674	1.36275689248599\\
0.0461292647472549	1.3632669989001\\
0.0450536385707451	1.36376359831801\\
0.0439819277176857	1.36424699239389\\
0.0429139971532765	1.36471750397157\\
0.0418497184129969	1.36517547920361\\
0.0416157502907665	1.36527533713489\\
0.0407883643889554	1.36561517286681\\
0.0397303061725733	1.36604159236195\\
0.0386756023501996	1.36645685542131\\
0.0376241464142572	1.36686143422315\\
0.0365758366683792	1.36725584204798\\
0.0355305759229243	1.36764063848657\\
0.0344882712016654	1.36801643551123\\
0.0334488334568769	1.36838390457544\\
0.0324121772898386	1.36874378494471\\
0.0313782116307208	1.36909676725632\\
0.0303468120715756	1.36944309265983\\
0.0293178473362474	1.36978287967342\\
0.0282911881427353	1.37011624277577\\
0.0272667070995944	1.37044329250695\\
0.0262442786053678	1.37076413556426\\
0.0252237787508731	1.37107887489334\\
0.0242050852241747	1.37138760977463\\
0.0231880772180779	1.37169043590537\\
0.022172635339993	1.37198744547739\\
0.0211586415240183	1.37227872725065\\
0.0201459789450994	1.37256436662294\\
0.0191345319351291	1.37284444569566\\
0.0181241859008534	1.37311904333585\\
0.0171148272434571	1.37338823523468\\
0.0161063432797062	1.37365209396244\\
0.0150986221645263	1.37391068902009\\
0.0140915528149033	1.37416408688761\\
0.0130850248349928	1.37441235106905\\
0.0120789284423299	1.37465554213454\\
0.0110731543950326	1.3748937177592\\
0.0100675939198961	1.37512693275916\\
0.00906213864127533	1.37535523912455\\
0.00805668051065746	1.37557868604981\\
0.00705111173682614	1.37579731996105\\
0.006045324716522	1.37601118454075\\
0.00503921196550452	1.3762203207498\\
0.00403266604992185	1.37642476684677\\
0.00302557951789596	1.37662455840466\\
0.00201784483123129	1.37681972832498\\
0.00100935429715531	1.37701030684932\\
0	1.37245740098674\\
0	1.36366074709118\\
0	1.35485607710923\\
0	1.34604456988612\\
0	1.33722740426711\\
0	1.32840575909745\\
0	1.31958081322238\\
0	1.31075374548717\\
0	1.30192573473704\\
0	1.29309795981726\\
0	1.28427159957308\\
0	1.27544783284974\\
0	1.26662783849248\\
0	1.25781279534657\\
0	1.24900388225725\\
0	1.24020227806977\\
0	1.23140916162938\\
0	1.22262571178133\\
0	1.21385310737086\\
0	1.20509252724323\\
0	1.19634515024368\\
0	1.18761215521747\\
0	1.17889472100984\\
0	1.17019402646604\\
0	1.16151125043133\\
0	1.15284757175095\\
0	1.14420416927014\\
0	1.13558222183417\\
0	1\\
};

\addplot [color=black, dashed, line width=1.5pt]
  table[row sep=crcr]{%
0	1.9866088569792\\
0	1.98490510487271\\
0	1.98309496543699\\
0	1.9811787669022\\
0	1.97915683749851\\
0	1.97702950545611\\
0	1.97479709900516\\
0	1.97245994637585\\
0	1.97001837579833\\
0	1.96747271550279\\
0	1.96482329371939\\
0	1.96207043867832\\
0	1.95921447860975\\
0	1.95625574174384\\
0	1.95319455631078\\
0	1.95003125054074\\
0	1.94676615266388\\
0	1.94339959091039\\
0	1.93993189351043\\
0	1.93636338869419\\
0	1.93269440469182\\
0	1.92892526973352\\
0	1.92505631204945\\
0	1.92108785986978\\
0	1.91702024142469\\
0	1.91285378494435\\
0	1.90858881865893\\
0	1.90422567079862\\
0	1.89976466959358\\
0	1.89520614327398\\
0	1.89055042007\\
0	1.88579782821181\\
0	1.88094869592959\\
0	1.8760034386209\\
0	1.87096282035288\\
0	1.86582769236004\\
0	1.86059890587691\\
0	1.85527731213801\\
0	1.84986376237787\\
0	1.84435910783101\\
0	1.83876419973195\\
0	1.83307988931522\\
0	1.82730702781534\\
0	1.82144646646684\\
0	1.81549905650423\\
0	1.80946564916203\\
0	1.80334709567479\\
0	1.79714424727701\\
0	1.79085795520322\\
0	1.7844891358891\\
0	1.77803896657499\\
0	1.77150868970236\\
0	1.76489954771272\\
0	1.75821278304754\\
0	1.75144963814832\\
0	1.74461135545654\\
0	1.73769917741369\\
0	1.73071434646126\\
0	1.72365810504074\\
0	1.71653169559361\\
0	1.70933636056137\\
0	1.7020733423855\\
0	1.6947438835075\\
0	1.68734922636884\\
0	1.67989061341102\\
0	1.67236933030688\\
0	1.66478683565461\\
0	1.6571446312838\\
0	1.64944421902399\\
0	1.64168710070476\\
0	1.63387477815565\\
0	1.62600875320623\\
0	1.61809052768606\\
0	1.6101216034247\\
0	1.60210348225171\\
0	1.59403766599665\\
0	1.58592565648908\\
0	1.57776895555856\\
0	1.56956906503466\\
0	1.56132748674693\\
0	1.55304572252494\\
0	1.54472526967401\\
0	1.53636760740254\\
0	1.52797421039473\\
0	1.51954655333473\\
0	1.51108611090674\\
0	1.50259435779491\\
0	1.49407276868343\\
0	1.48552281825648\\
0	1.47694598119823\\
0	1.46834373219285\\
0	1.45971754592452\\
0	1.45106889707742\\
0	1.44239926033572\\
0	1.4337101103836\\
0	1.42500292190524\\
0	1.4162791695848\\
0	1.40754027879998\\
0	1.39878747770252\\
0	1.39002194513765\\
0	1.38124485995065\\
0	1.37245740098674\\
0	1.36366074709118\\
0	1.35485607710923\\
0	1.34604456988612\\
0	1.33722740426711\\
0	1.32840575909745\\
0	1.31958081322238\\
0	1.31075374548717\\
0	1.30192573473704\\
0	1.29309795981726\\
0	1.28427159957308\\
0	1.27544783284974\\
0	1.26662783849248\\
0	1.25781279534657\\
0	1.24900388225725\\
0	1.24020227806977\\
0	1.23140916162938\\
0	1.22262571178133\\
0	1.21385310737086\\
0	1.20509252724323\\
0	1.19634515024368\\
0	1.18761215521747\\
0	1.17889472100984\\
0	1.17019402646604\\
0	1.16151125043133\\
0	1.15284757175095\\
0	1.14420416927014\\
0	1.13558222183417\\
0	0\\
};

\addplot[
  blue,
  only marks,
  mark=o,
  mark size=2pt,
  error bars/.cd, 
    x fixed,
    x dir=both, 
    x explicit,
    y fixed,
    y dir=both, 
    y explicit,
] table [x=eps, y=V,x error=inc_eps,y error=inc_V, col sep=comma] {
d,inc_d,eps,inc_eps,V,inc_V
0.1705029838,0.0029071267,0.2732,0.013081174,1.1992080419,0.0665687432
0.1705029838,0.0029071267,0.2497,0.013081174,1.2116318753,0.0672583975
0.1705029838,0.0029071267,0.27287,0.013081174,1.2614169591,0.0700219968
0.1705029838,0.0029071267,0.268,0.013081174,1.2550423029,0.0696681359
0.1705029838,0.0029071267,0.27744,0.013081174,1.2938730656,0.0718236543
0.3410059676,0.0058142535,0.23581,0.013081174,1.2883249263,0.0715156738
0.3410059676,0.0058142535,0.2451,0.0204958487,1.2967423613,0.0719829306
0.6820119352,0.011628507,0.1232,0.0204958487,1.2690835456,0.0704475735
0.7142857143,0.012755102,0.091,0.0214657416,1.1960184062,0.0669957704
0.7680491551,0.0147474876,0,0.0230814426,1.3671985406,0.0777907662
0.5760368664,0.0110606157,0.15225,0.0230814426,1.33623067,0.0760287585
0.5115089514,0.0087213802,0.1884,0.0204958487,1.2484126606,0.0693001205
0.5115089514,0.0087213802,0.1884,0.0204958487,1.2561189115,0.069727899
0.5115089514,0.0087213802,0.1884,0.0204958487,1.2638251625,0.0701556776
0.5115089514,0.0087213802,0.19565,0.0204958487,1.2756366127,0.0708113381
0.1886792453,0.0035599858,0.2821,0.0144756766,1.2621992677,0.0715344924
0.1900327807,0.0036112458,0.2835,0.0145795213,1.2540510421,0.0711860299
0.1900327807,0.0036112458,0.2835,0.0145795213,1.2546050215,0.0712174764
0.1920122888,0.0036868719,0.2812,0.0147313913,1.2187598419,0.0693449116
0.1880936706,0.0035379229,0.2908,0.0144307506,1.2857148852,0.0728171589
0.1920122888,0.0036868719,0.28125,0.0147313913,1.2236618422,0.0696238252
0.1920122888,0.0036868719,0.296875,0.0147313913,1.2092658206,0.0688047214
0.3880669415,0.0075297976,0.215789,0.0148864586,1.2034020957,0.0686359175
0.3840245776,0.0073737438,0.21875,0.0147313913,1.2234796733,0.0696134602
0.3880669415,0.0075297976,0.221,0.0148864586,1.2035723804,0.0686456297
0.3840245776,0.0073737438,0.2188,0.0147313913,1.2631842202,0.0718725667
};

\end{axis}
\end{tikzpicture}%

%% file: texfigures/Velocity2/Vcentro.tex
%
%
\begin{tikzpicture}[text centered]

\begin{axis}[%
width=0.3\textwidth,
height=0.3\textwidth,
scale only axis,
point meta min=-2,
point meta max=2,
colormap/\mapacolor,
xmin=0,
xmax=1,
ymin=0.25,
ymax=2,
xlabel=$d$,
ylabel=$V_0$,
axis background/.style={fill=white},
legend style={legend cell align=left, align=left, draw=white!15!black},
colorbar sampled line={thin, scatter,samples=17,scatter/use mapped color={draw=mapped color},only marks,mark=-,},colorbar style={ytick={-2,-1,0,1,2},},,
colorbar,
colorbar style={title={$\log_{10} \lambda$},xshift=-.025\textwidth},
xtick={0,.25,...,1},
minor xtick={0,.05,...,1},
ytick={0,.5,...,2},
minor ytick={0,.125,...,2},
]

\node[anchor=north west,xshift=0.0cm,yshift=0.0cm] at (rel axis cs:0,1) {(a)};
\node[anchor=north west,xshift=0.0cm,yshift=0.0cm, text width=2cm] at (rel axis cs:.58,1) {bubble limit ($\lambda \to 0$)};
\node[anchor=north west,xshift=0.0cm,yshift=0.0cm, text width=1.5cm] at (rel axis cs:.35,.65) {bead limit ($\lambda \to \infty$)};

\addplot[thin, contour prepared, contour prepared format=matlab, contour/labels=false] table[row sep=crcr] {%
-2	40\\
0	2\\
0.0256410256410256	2.0001642720643\\
0.0512820512820513	2.00011026328898\\
0.0769230769230769	1.99991168815729\\
0.102564102564103	1.9996422611525\\
0.128205128205128	1.99937569624305\\
0.153846153846154	1.9991463873708\\
0.179487179487179	1.99885081446947\\
0.205128205128205	1.99838885764688\\
0.230769230769231	1.99766188639172\\
0.256410256410256	1.99657056854182\\
0.282051282051282	1.99501477588108\\
0.307692307692308	1.99289808410256\\
0.333333333333333	1.99007328170633\\
0.358974358974359	1.9863307230033\\
0.384615384615385	1.98144792686647\\
0.41025641025641	1.97520250888028\\
0.435897435897436	1.9673361523873\\
0.461538461538462	1.95753824223264\\
0.487179487179487	1.94549620452444\\
0.512820512820513	1.93089664776331\\
0.538461538461538	1.9134273500143\\
0.564102564102564	1.89277670825571\\
0.58974358974359	1.868632843566\\
0.615384615384615	1.84068528257201\\
0.641025641025641	1.80872406355456\\
0.666666666666667	1.77260986810707\\
0.692307692307692	1.73219563913145\\
0.717948717948718	1.68738768679535\\
0.743589743589744	1.63834191591719\\
0.769230769230769	1.58541132441986\\
0.794871794871795	1.5289232798611\\
0.82051282051282	1.46920214748223\\
0.846153846153846	1.40657413751859\\
0.871794871794872	1.34136430261914\\
0.897435897435897	1.27389205695713\\
0.923076923076923	1.20501232836271\\
0.948717948717949	1.13573229154001\\
0.974358974358974	1.06705912318659\\
1	1\\
-1.75	40\\
0	2\\
0.0256410256410256	2.00015106983734\\
0.0512820512820513	2.0000673326983\\
0.0769230769230769	1.99982135977998\\
0.102564102564103	1.99948572227949\\
0.128205128205128	1.99913299088682\\
0.153846153846154	1.99879700617335\\
0.179487179487179	1.99837577303767\\
0.205128205128205	1.99777069613918\\
0.230769230769231	1.99688470990511\\
0.256410256410256	1.99562012882618\\
0.282051282051282	1.99387856386027\\
0.307692307692308	1.99156535032599\\
0.333333333333333	1.9885359493825\\
0.358974358974359	1.9845845344225\\
0.384615384615385	1.97949280693712\\
0.41025641025641	1.97304268246579\\
0.435897435897436	1.96498078101905\\
0.461538461538462	1.95500225576446\\
0.487179487179487	1.94280016770535\\
0.512820512820513	1.92806654447899\\
0.538461538461538	1.91049520432988\\
0.564102564102564	1.88978104818002\\
0.58974358974359	1.86561772513929\\
0.615384615384615	1.83769867743298\\
0.641025641025641	1.80581419112835\\
0.666666666666667	1.76982238274976\\
0.692307692307692	1.72957177947254\\
0.717948717948718	1.68496476781748\\
0.743589743589744	1.63615660983275\\
0.769230769230769	1.58349759122027\\
0.794871794871795	1.52730376023941\\
0.82051282051282	1.46788311107591\\
0.846153846153846	1.4055444745683\\
0.871794871794872	1.34059857243085\\
0.897435897435897	1.2733557893444\\
0.923076923076923	1.20466844958916\\
0.948717948717949	1.13554264332875\\
0.974358974358974	1.06698446274007\\
1	1\\
-1.5	40\\
0	2\\
0.0256410256410256	2.00012831943794\\
0.0512820512820513	1.99999335396687\\
0.0769230769230769	1.99966570364644\\
0.102564102564103	1.99921596853627\\
0.128205128205128	1.9987147482022\\
0.153846153846154	1.9981949281243\\
0.179487179487179	1.99755713726242\\
0.205128205128205	1.99670540097248\\
0.230769230769231	1.99554534067221\\
0.256410256410256	1.99398209492606\\
0.282051282051282	1.99192025404318\\
0.307692307692308	1.98926816425093\\
0.333333333333333	1.98588586239994\\
0.358974358974359	1.9815740616863\\
0.384615384615385	1.97612162226406\\
0.41025641025641	1.96931781700393\\
0.435897435897436	1.96091771900361\\
0.461538461538462	1.95062637047895\\
0.487179487179487	1.93814650682049\\
0.512820512820513	1.92317947850951\\
0.538461538461538	1.90542948286932\\
0.564102564102564	1.88460260161587\\
0.58974358974359	1.86040205362611\\
0.615384615384615	1.83252816663549\\
0.641025641025641	1.80077190482086\\
0.666666666666667	1.76498727958011\\
0.692307692307692	1.72501555019006\\
0.717948717948718	1.68075259656617\\
0.743589743589744	1.63235265200176\\
0.769230769230769	1.58016143905701\\
0.794871794871795	1.52447578342924\\
0.82051282051282	1.46557565811269\\
0.846153846153846	1.40373987280676\\
0.871794871794872	1.33925408646466\\
0.897435897435897	1.27241258107256\\
0.923076923076923	1.20406268688841\\
0.948717948717949	1.13520813593471\\
0.974358974358974	1.0668526622818\\
1	1\\
-1.25	40\\
0	2\\
0.0256410256410256	2.00009001439114\\
0.0512820512820513	1.99986879621358\\
0.0769230769230769	1.99940362384912\\
0.102564102564103	1.9987617756796\\
0.128205128205128	1.99801052961542\\
0.153846153846154	1.99718115872339\\
0.179487179487179	1.99617869926379\\
0.205128205128205	1.99491157063931\\
0.230769230769231	1.99328989048399\\
0.256410256410256	1.99122351370576\\
0.282051282051282	1.98862199647685\\
0.307692307692308	1.98539869342563\\
0.333333333333333	1.98142125979708\\
0.358974358974359	1.97650130127578\\
0.384615384615385	1.97043959043035\\
0.41025641025641	1.9630376364647\\
0.435897435897436	1.95406458892394\\
0.461538461538462	1.94324199139023\\
0.487179487179487	1.93028876184669\\
0.512820512820513	1.9149218983562\\
0.538461538461538	1.89686298213332\\
0.564102564102564	1.8758368329842\\
0.58974358974359	1.85156289632932\\
0.615384615384615	1.82375347792297\\
0.641025641025641	1.79220139849941\\
0.666666666666667	1.75675466476446\\
0.692307692307692	1.71724329390498\\
0.717948717948718	1.67355296122143\\
0.743589743589744	1.62583640023492\\
0.769230769230769	1.5744319778175\\
0.794871794871795	1.51960501434099\\
0.82051282051282	1.46158892002312\\
0.846153846153846	1.40061181782888\\
0.871794871794872	1.33691614647898\\
0.897435897435897	1.2707675005138\\
0.923076923076923	1.20300328021116\\
0.948717948717949	1.13462176740062\\
0.974358974358974	1.06662124601823\\
1	1\\
-1	40\\
0	2\\
0.0256410256410256	2.00002797151004\\
0.0512820512820513	1.99966705174076\\
0.0769230769230769	1.99897913133003\\
0.102564102564103	1.99802610091575\\
0.128205128205128	1.99686985070056\\
0.153846153846154	1.99553903062053\\
0.179487179487179	1.99394578655742\\
0.205128205128205	1.99200560584963\\
0.230769230769231	1.98963581435355\\
0.256410256410256	1.98675380322279\\
0.282051282051282	1.98327703712502\\
0.307692307692308	1.97912681961626\\
0.333333333333333	1.97418291617312\\
0.358974358974359	1.96827425725971\\
0.384615384615385	1.96122053200515\\
0.41025641025641	1.9528426604774\\
0.435897435897436	1.94293216760126\\
0.461538461538462	1.93123691494594\\
0.487179487179487	1.9175017312062\\
0.512820512820513	1.90146880525686\\
0.538461538461538	1.88288759949756\\
0.564102564102564	1.86151300628789\\
0.58974358974359	1.83709100457129\\
0.615384615384615	1.80935427827462\\
0.641025641025641	1.77810054017527\\
0.666666666666667	1.74317051037946\\
0.692307692307692	1.7043787273755\\
0.717948717948718	1.66159643472599\\
0.743589743589744	1.61497458178859\\
0.769230769230769	1.56484053727227\\
0.794871794871795	1.51141111120235\\
0.82051282051282	1.45484628850451\\
0.846153846153846	1.39529205677576\\
0.871794871794872	1.33291829163739\\
0.897435897435897	1.2679398462931\\
0.923076923076923	1.20117366058006\\
0.948717948717949	1.13360486749455\\
0.974358974358974	1.06621860223504\\
1	1\\
-0.75	40\\
0	2\\
0.0256410256410256	1.99993352523612\\
0.0512820512820513	1.99935994696478\\
0.0769230769230769	1.99833293614074\\
0.102564102564103	1.99690616371873\\
0.128205128205128	1.99513330027324\\
0.153846153846154	1.99303897531336\\
0.179487179487179	1.99054608885499\\
0.205128205128205	1.98758077084954\\
0.230769230769231	1.98407114334851\\
0.256410256410256	1.97994582502743\\
0.282051282051282	1.97513399841598\\
0.307692307692308	1.96956867792072\\
0.333333333333333	1.96314751563811\\
0.358974358974359	1.95572505292298\\
0.384615384615385	1.94714886434981\\
0.41025641025641	1.9372684275284\\
0.435897435897436	1.9259082880382\\
0.461538461538462	1.91285534924736\\
0.487179487179487	1.89789310283758\\
0.512820512820513	1.8808016337867\\
0.538461538461538	1.86137206708307\\
0.564102564102564	1.83940423352942\\
0.58974358974359	1.81468489064631\\
0.615384615384615	1.78697986520035\\
0.641025641025641	1.75609869967059\\
0.666666666666667	1.72187688087878\\
0.692307692307692	1.68411216100229\\
0.717948717948718	1.64265916010436\\
0.743589743589744	1.5976677150272\\
0.769230769230769	1.54945076062665\\
0.794871794871795	1.49815814733078\\
0.82051282051282	1.44384428878772\\
0.846153846153846	1.3865315542822\\
0.871794871794872	1.32627414362937\\
0.897435897435897	1.26319935885284\\
0.923076923076923	1.19808128361708\\
0.948717948717949	1.13187318546821\\
0.974358974358974	1.06552833429844\\
1	1\\
-0.5	40\\
0	2\\
0.0256410256410256	1.99980251642167\\
0.0512820512820513	1.99893396544387\\
0.0769230769230769	1.99743658018393\\
0.102564102564103	1.99535259375921\\
0.128205128205128	1.99272423898282\\
0.153846153846154	1.98957051393417\\
0.179487179487179	1.98582909863577\\
0.205128205128205	1.98144065210371\\
0.230769230769231	1.97634791818464\\
0.256410256410256	1.97049459879686\\
0.282051282051282	1.96382548556173\\
0.307692307692308	1.95628905232288\\
0.333333333333333	1.94780665395889\\
0.358974358974359	1.93826681144625\\
0.384615384615385	1.92755392018214\\
0.41025641025641	1.91555502515159\\
0.435897435897436	1.90213829471383\\
0.461538461538462	1.88714259632316\\
0.487179487179487	1.87040328649883\\
0.512820512820513	1.85175184654922\\
0.538461538461538	1.83103523754229\\
0.564102564102564	1.80811336829272\\
0.58974358974359	1.78282964715407\\
0.615384615384615	1.75499942511257\\
0.641025641025641	1.72445701622871\\
0.666666666666667	1.69104243506089\\
0.692307692307692	1.65454448907309\\
0.717948717948718	1.61480647297489\\
0.743589743589744	1.57197992629545\\
0.769230769230769	1.52636254539054\\
0.794871794871795	1.47802787145879\\
0.82051282051282	1.42690133916572\\
0.846153846153846	1.37284240661947\\
0.871794871794872	1.31573841303537\\
0.897435897435897	1.2555752034064\\
0.923076923076923	1.1930397766064\\
0.948717948717949	1.12901211230664\\
0.974358974358974	1.06437219270516\\
1	1\\
-0.25	40\\
0	2\\
0.0256410256410256	1.99964242670444\\
0.0512820512820513	1.9984134438528\\
0.0769230769230769	1.99634125079027\\
0.102564102564103	1.99345404686202\\
0.128205128205128	1.98978003120155\\
0.153846153846154	1.98533123421124\\
0.179487179487179	1.98006317982952\\
0.205128205128205	1.97393391207261\\
0.230769230769231	1.96690348231884\\
0.256410256410256	1.95893324598711\\
0.282051282051282	1.94998604351274\\
0.307692307692308	1.94002798768264\\
0.333333333333333	1.92900762329619\\
0.358974358974359	1.91685250590908\\
0.384615384615385	1.90348900541364\\
0.41025641025641	1.8888467056853\\
0.435897435897436	1.87284341981541\\
0.461538461538462	1.85537772453072\\
0.487179487179487	1.83634510787765\\
0.512820512820513	1.81563739991995\\
0.538461538461538	1.79316579414065\\
0.564102564102564	1.76885869100351\\
0.58974358974359	1.74262717671713\\
0.615384615384615	1.71435091996458\\
0.641025641025641	1.68390642824046\\
0.666666666666667	1.65115755014625\\
0.692307692307692	1.61590613997064\\
0.717948717948718	1.57800214694675\\
0.743589743589744	1.53760361963604\\
0.769230769230769	1.49499642525873\\
0.794871794871795	1.45019295095284\\
0.82051282051282	1.403001467965\\
0.846153846153846	1.35311273492507\\
0.871794871794872	1.3002148225803\\
0.897435897435897	1.24409581353211\\
0.923076923076923	1.18528349524609\\
0.948717948717949	1.12450919816012\\
0.974358974358974	1.06250425537712\\
1	1\\
0	40\\
0	2\\
0.0256410256410256	1.9994745288717\\
0.0512820512820513	1.99786755596096\\
0.0769230769230769	1.99519249496276\\
0.102564102564103	1.9914627595721\\
0.128205128205128	1.98669176336885\\
0.153846153846154	1.98088412728795\\
0.179487179487179	1.97401383283746\\
0.205128205128205	1.96605671978741\\
0.230769230769231	1.95699032721646\\
0.256410256410256	1.94679360078792\\
0.282051282051282	1.93544708822324\\
0.307692307692308	1.92293390265251\\
0.333333333333333	1.90922897151169\\
0.358974358974359	1.89429777962739\\
0.384615384615385	1.87810702778901\\
0.41025641025641	1.86062673144711\\
0.435897435897436	1.84182212202274\\
0.461538461538462	1.82164942993295\\
0.487179487179487	1.80006268035308\\
0.512820512820513	1.7770131576273\\
0.538461538461538	1.75247326908212\\
0.564102564102564	1.72643525145074\\
0.58974358974359	1.6988765219464\\
0.615384615384615	1.66974545307062\\
0.641025641025641	1.63897416813054\\
0.666666666666667	1.60647163512849\\
0.692307692307692	1.57208267850718\\
0.717948717948718	1.53569035650529\\
0.743589743589744	1.49746152627776\\
0.769230769230769	1.45767308362922\\
0.794871794871795	1.41631784623734\\
0.82051282051282	1.37315023804334\\
0.846153846153846	1.3277539675172\\
0.871794871794872	1.27965114044292\\
0.897435897435897	1.22842079754326\\
0.923076923076923	1.17435294740609\\
0.948717948717949	1.11793515363979\\
0.974358974358974	1.05965498243941\\
1	1\\
0.25	40\\
0	2\\
0.0256410256410256	1.99932443827839\\
0.0512820512820513	1.99737958350829\\
0.0769230769230769	1.99416557344058\\
0.102564102564103	1.98968254582617\\
0.128205128205128	1.98393063838675\\
0.153846153846154	1.97690775999381\\
0.179487179487179	1.96860416937041\\
0.205128205128205	1.95901123489937\\
0.230769230769231	1.94812155108647\\
0.256410256410256	1.93592898255252\\
0.282051282051282	1.92242883783601\\
0.307692307692308	1.90761811069807\\
0.333333333333333	1.89149343049259\\
0.358974358974359	1.8740515168276\\
0.384615384615385	1.85529181204162\\
0.41025641025641	1.83521666190974\\
0.435897435897436	1.81382935696499\\
0.461538461538462	1.79113289477402\\
0.487179487179487	1.76712903546899\\
0.512820512820513	1.74181795873054\\
0.538461538461538	1.71522001861522\\
0.564102564102564	1.68737532521292\\
0.58974358974359	1.65831350275752\\
0.615384615384615	1.62804165262701\\
0.641025641025641	1.59654781491877\\
0.666666666666667	1.5637958745684\\
0.692307692307692	1.52969378839959\\
0.717948717948718	1.49417693028791\\
0.743589743589744	1.45741492067448\\
0.769230769230769	1.41966891813673\\
0.794871794871795	1.3809531248903\\
0.82051282051282	1.34105254903174\\
0.846153846153846	1.29955738068087\\
0.871794871794872	1.25593827575015\\
0.897435897435897	1.20964793079559\\
0.923076923076923	1.16071580766826\\
0.948717948717949	1.10933502858606\\
0.974358974358974	1.05569871790972\\
1	1\\
0.5	40\\
0	2\\
0.0256410256410256	1.99920825529712\\
0.0512820512820513	1.99700186313628\\
0.0769230769230769	1.99337064663224\\
0.102564102564103	1.98830442889978\\
0.128205128205128	1.98179303309074\\
0.153846153846154	1.9738291149869\\
0.179487179487179	1.96441538870783\\
0.205128205128205	1.95355499648733\\
0.230769230769231	1.94125181145152\\
0.256410256410256	1.92751072182204\\
0.282051282051282	1.91233776410195\\
0.307692307692308	1.8957398129719\\
0.333333333333333	1.87772905666933\\
0.358974358974359	1.85832462331552\\
0.384615384615385	1.83754908530518\\
0.41025641025641	1.81542722899222\\
0.435897435897436	1.79198881444999\\
0.461538461538462	1.76726965894764\\
0.487179487179487	1.74130507350649\\
0.512820512820513	1.71412973311654\\
0.538461538461538	1.68579576763687\\
0.564102564102564	1.65637286434042\\
0.58974358974359	1.62592431923506\\
0.615384615384615	1.59449814666267\\
0.641025641025641	1.56212666657035\\
0.666666666666667	1.52882285473459\\
0.692307692307692	1.49455780327369\\
0.717948717948718	1.45932048035833\\
0.743589743589744	1.42327153041481\\
0.769230769230769	1.38664176380712\\
0.794871794871795	1.34947912423018\\
0.82051282051282	1.31164848848135\\
0.846153846153846	1.27283856458985\\
0.871794871794872	1.23259760779697\\
0.897435897435897	1.19039618069386\\
0.923076923076923	1.14607997973995\\
0.948717948717949	1.09960473098164\\
0.974358974358974	1.05092616190597\\
1	1\\
0.75	40\\
0	2\\
0.0256410256410256	1.99912794523143\\
0.0512820512820513	1.99674077476122\\
0.0769230769230769	1.99282116248661\\
0.102564102564103	1.98735178230482\\
0.128205128205128	1.98031530819582\\
0.153846153846154	1.97170073552005\\
0.179487179487179	1.96151931422416\\
0.205128205128205	1.94978219907112\\
0.230769230769231	1.93650086732341\\
0.256410256410256	1.92168755824819\\
0.282051282051282	1.90535536549327\\
0.307692307692308	1.88751754887205\\
0.333333333333333	1.86819641064816\\
0.358974358974359	1.84742565003338\\
0.384615384615385	1.82524266989698\\
0.41025641025641	1.80168640164817\\
0.435897435897436	1.77680329838221\\
0.461538461538462	1.75065001538144\\
0.487179487179487	1.72328313853907\\
0.512820512820513	1.69475921405777\\
0.538461538461538	1.66514918940778\\
0.564102564102564	1.63453868599817\\
0.58974358974359	1.60300982362663\\
0.615384615384615	1.57063518897228\\
0.641025641025641	1.5374766007446\\
0.666666666666667	1.50358259475703\\
0.692307692307692	1.4689737565508\\
0.717948717948718	1.43368117232976\\
0.743589743589744	1.39784830616987\\
0.769230769230769	1.36166621660285\\
0.794871794871795	1.32520570467486\\
0.82051282051282	1.28841084714965\\
0.846153846153846	1.25109320563237\\
0.871794871794872	1.21294196706808\\
0.897435897435897	1.17355212414229\\
0.923076923076923	1.13271332234322\\
0.948717948717949	1.0902752449858\\
0.974358974358974	1.04608757617105\\
1	1\\
1	40\\
0	2\\
0.0256410256410256	1.99907667917084\\
0.0512820512820513	1.99657411124016\\
0.0769230769230769	1.99247039811569\\
0.102564102564103	1.98674364170516\\
0.128205128205128	1.97937194402817\\
0.153846153846154	1.97034195128422\\
0.179487179487179	1.9596703307047\\
0.205128205128205	1.94737329794247\\
0.230769230769231	1.93346710211849\\
0.256410256410256	1.9179685597756\\
0.282051282051282	1.90089511558396\\
0.307692307692308	1.88226391381452\\
0.333333333333333	1.8621034365664\\
0.358974358974359	1.84045628705109\\
0.384615384615385	1.81736883151\\
0.41025641025641	1.79288843180078\\
0.435897435897436	1.76707148711871\\
0.461538461538462	1.73998712579353\\
0.487179487179487	1.71170463690756\\
0.512820512820513	1.68229359964445\\
0.538461538461538	1.6518354894762\\
0.564102564102564	1.62042393956483\\
0.58974358974359	1.58815083673662\\
0.615384615384615	1.55510237700292\\
0.641025641025641	1.52135813159799\\
0.666666666666667	1.48698933477655\\
0.692307692307692	1.45205000399233\\
0.717948717948718	1.41659946180855\\
0.743589743589744	1.38076315337346\\
0.769230769230769	1.3446943013058\\
0.794871794871795	1.3084727713121\\
0.82051282051282	1.27209843944723\\
0.846153846153846	1.23548310782447\\
0.871794871794872	1.19845008934292\\
0.897435897435897	1.16074323963079\\
0.923076923076923	1.12218975958715\\
0.948717948717949	1.08264469531945\\
0.974358974358974	1.04196309329973\\
1	1\\
1.25	40\\
0	2\\
0.0256410256410256	1.99904559720509\\
0.0512820512820513	1.99647306626411\\
0.0769230769230769	1.99225773400458\\
0.102564102564103	1.986374927254\\
0.128205128205128	1.97879997296941\\
0.153846153846154	1.96951808832955\\
0.179487179487179	1.95854921301072\\
0.205128205128205	1.94591261054386\\
0.230769230769231	1.93162739191821\\
0.256410256410256	1.91571310499454\\
0.282051282051282	1.89818977383286\\
0.307692307692308	1.87907682480761\\
0.333333333333333	1.85840638123406\\
0.358974358974359	1.83622629266483\\
0.384615384615385	1.81258816639873\\
0.41025641025641	1.78754424449834\\
0.435897435897436	1.76115668078641\\
0.461538461538462	1.73350184137782\\
0.487179487179487	1.70465636911298\\
0.512820512820513	1.67469737049113\\
0.538461538461538	1.64371210437058\\
0.564102564102564	1.61179818044457\\
0.58974358974359	1.57905246040496\\
0.615384615384615	1.54556846802867\\
0.641025641025641	1.51143595608147\\
0.666666666666667	1.47673971144617\\
0.692307692307692	1.44155450272817\\
0.717948717948718	1.40595700524666\\
0.743589743589744	1.37005830347472\\
0.769230769230769	1.33398247210865\\
0.794871794871795	1.29781077890287\\
0.82051282051282	1.26157655640615\\
0.846153846153846	1.22525880150052\\
0.871794871794872	1.18877920384995\\
0.897435897435897	1.15200360704859\\
0.923076923076923	1.1148290284467\\
0.948717948717949	1.07716432779847\\
0.974358974358974	1.03891836501316\\
1	1\\
1.5	40\\
0	2\\
0.0256410256410256	1.99902733806709\\
0.0512820512820513	1.99641370762938\\
0.0769230769230769	1.99213280418456\\
0.102564102564103	1.98615832323032\\
0.128205128205128	1.97846396040418\\
0.153846153846154	1.96903409173123\\
0.179487179487179	1.95789057506914\\
0.205128205128205	1.94505445721802\\
0.230769230769231	1.93054651936734\\
0.256410256410256	1.91438789844901\\
0.282051282051282	1.89660011313079\\
0.307692307692308	1.87720390464675\\
0.333333333333333	1.85623349806175\\
0.358974358974359	1.83373977111455\\
0.384615384615385	1.80977734144071\\
0.41025641025641	1.78440123582726\\
0.435897435897436	1.75767689030005\\
0.461538461538462	1.72968480641614\\
0.487179487179487	1.70050582115677\\
0.512820512820513	1.67022132622513\\
0.538461538461538	1.63892175869507\\
0.564102564102564	1.60670674032086\\
0.58974358974359	1.57367570066804\\
0.615384615384615	1.53992611170004\\
0.641025641025641	1.50555346161993\\
0.666666666666667	1.47065037416782\\
0.692307692307692	1.43530392261291\\
0.717948717948718	1.39960098470008\\
0.743589743589744	1.36364280403527\\
0.769230769230769	1.32753375719631\\
0.794871794871795	1.29135393300505\\
0.82051282051282	1.2551554378981\\
0.846153846153846	1.2189582500545\\
0.871794871794872	1.18274747543911\\
0.897435897435897	1.14647266213439\\
0.923076923076923	1.11009460188734\\
0.948717948717949	1.07357920072195\\
0.974358974358974	1.03689236472919\\
1	1\\
1.75	40\\
0	2\\
0.0256410256410256	1.99901681002287\\
0.0512820512820513	1.99637948212909\\
0.0769230769230769	1.99206077083518\\
0.102564102564103	1.98603343065769\\
0.128205128205128	1.97827021625898\\
0.153846153846154	1.96875501809988\\
0.179487179487179	1.9575107986231\\
0.205128205128205	1.9445596304184\\
0.230769230769231	1.92992325399213\\
0.256410256410256	1.91362371731993\\
0.282051282051282	1.89568339388909\\
0.307692307692308	1.87612377438748\\
0.333333333333333	1.85498028315738\\
0.358974358974359	1.83230552581449\\
0.384615384615385	1.80815583257572\\
0.41025641025641	1.78258780806975\\
0.435897435897436	1.75566874744604\\
0.461538461538462	1.72748149711614\\
0.487179487179487	1.69810926935673\\
0.512820512820513	1.66763587967906\\
0.538461538461538	1.63615352226073\\
0.564102564102564	1.60376286716218\\
0.58974358974359	1.57056470085415\\
0.615384615384615	1.5366586452914\\
0.641025641025641	1.50214340435858\\
0.666666666666667	1.4671160675567\\
0.692307692307692	1.43167082920071\\
0.717948717948718	1.39590043598725\\
0.743589743589744	1.35989994159834\\
0.769230769230769	1.32376141080917\\
0.794871794871795	1.28756344772338\\
0.82051282051282	1.2513685003421\\
0.846153846153846	1.215220473397\\
0.871794871794872	1.17914276591146\\
0.897435897435897	1.14313732627936\\
0.923076923076923	1.10721096898831\\
0.948717948717949	1.07137305490882\\
0.974358974358974	1.03563294494476\\
1	1\\
2	40\\
0	2\\
0.0256410256410256	1.9990108048739\\
0.0512820512820513	1.9963599600974\\
0.0769230769230769	1.99201968333379\\
0.102564102564103	1.98596219224633\\
0.128205128205128	1.97815970464751\\
0.153846153846154	1.96859583385005\\
0.179487179487179	1.95729417188175\\
0.205128205128205	1.94427737564324\\
0.230769230769231	1.9295677316169\\
0.256410256410256	1.91318780572967\\
0.282051282051282	1.8951604567735\\
0.307692307692308	1.87550760007453\\
0.333333333333333	1.85426533991394\\
0.358974358974359	1.83148726152589\\
0.384615384615385	1.80723066437375\\
0.41025641025641	1.78155304393993\\
0.435897435897436	1.75452274482755\\
0.461538461538462	1.72622393788435\\
0.487179487179487	1.69674117602684\\
0.512820512820513	1.66615964170564\\
0.538461538461538	1.63457250631185\\
0.564102564102564	1.60208100067034\\
0.58974358974359	1.56878664384371\\
0.615384615384615	1.53479024141397\\
0.641025641025641	1.50019229986041\\
0.666666666666667	1.46509243642976\\
0.692307692307692	1.42958891340877\\
0.717948717948718	1.39377781727177\\
0.743589743589744	1.35775049329285\\
0.769230769230769	1.32159165318376\\
0.794871794871795	1.28537876308882\\
0.82051282051282	1.24917997088056\\
0.846153846153846	1.21305287802503\\
0.871794871794872	1.1770432471625\\
0.897435897435897	1.14118427143622\\
0.923076923076923	1.10551240056274\\
0.948717948717949	1.07006569282772\\
0.974358974358974	1.0348822065379\\
1	1\\
};

\end{axis}
\end{tikzpicture}%

%% file: texfigures/Velocity2/Vpared.tex
%
%
\begin{tikzpicture}[text centered]

\begin{axis}[%
width=0.3\textwidth,
height=0.3\textwidth,
scale only axis,
point meta min=-2,
point meta max=2,
colormap/\mapacolor,
xmin=0,
xmax=1,
ymin=0.25,
ymax=2,
xlabel=$d$,
ylabel=$V_0$,
axis background/.style={fill=white},
legend style={legend cell align=left, align=left, draw=white!15!black},
colorbar sampled line={thin, scatter,samples=17,scatter/use mapped color={draw=mapped color},only marks,mark=-,},colorbar style={ytick={-2,-1,0,1,2},},,
colorbar,
colorbar style={title={$\log_{10} \lambda$},xshift=-.025\textwidth},
xtick={0,.25,...,1},
minor xtick={0,.05,...,1},
ytick={0,.5,...,2},
minor ytick={0,.125,...,2},
]

\node[anchor=north west,xshift=0.0cm,yshift=0.0cm] at (rel axis cs:0,1) {(b)};
\node[anchor=north west,xshift=0.0cm,yshift=0.0cm, text width=2cm] at (rel axis cs:.35,.85) {bubble limit ($\lambda \to 0$)};
\node[anchor=north west,xshift=0.0cm,yshift=0.0cm, text width=1.5cm] at (rel axis cs:.4,.45) {bead limit ($\lambda \to \infty$)};

\addplot[thin, contour prepared, contour prepared format=matlab, contour/labels=false] table[row sep=crcr] {%
-2	40\\
0	0.38\\
0.0256410256410256	0.462024017126785\\
0.0512820512820513	0.539428283737385\\
0.0769230769230769	0.612576480706841\\
0.102564102564103	0.681832288910189\\
0.128205128205128	0.747559388363696\\
0.153846153846154	0.81005587001388\\
0.179487179487179	0.869389746870003\\
0.205128205128205	0.925634591402405\\
0.230769230769231	0.978866388319546\\
0.256410256410256	1.0291594927134\\
0.282051282051282	1.0765878562527\\
0.307692307692308	1.12122835278876\\
0.333333333333333	1.16311331088016\\
0.358974358974359	1.20222033149718\\
0.384615384615385	1.23851526889528\\
0.41025641025641	1.27196482873678\\
0.435897435897436	1.30251694357487\\
0.461538461538462	1.33009248249761\\
0.487179487179487	1.35461163153017\\
0.512820512820513	1.37599235824777\\
0.538461538461538	1.3941267992428\\
0.564102564102564	1.40888112811192\\
0.58974358974359	1.42011917938627\\
0.615384615384615	1.42770424744489\\
0.641025641025641	1.43148863871921\\
0.666666666666667	1.43131751139605\\
0.692307692307692	1.42703481647778\\
0.717948717948718	1.41848542312867\\
0.743589743589744	1.40551716957631\\
0.769230769230769	1.38797797162588\\
0.794871794871795	1.36571157276874\\
0.82051282051282	1.3385585347868\\
0.846153846153846	1.30635856833491\\
0.871794871794872	1.26895077509813\\
0.897435897435897	1.22617397856754\\
0.923076923076923	1.17794786607462\\
0.948717948717949	1.124215074217\\
0.974358974358974	1.06491823989276\\
1	1\\
-1.75	40\\
0	0.38\\
0.0256410256410256	0.462014406875633\\
0.0512820512820513	0.539367594803539\\
0.0769230769230769	0.612431596260002\\
0.102564102564103	0.681578443721305\\
0.128205128205128	0.747180168788457\\
0.153846153846154	0.809541953582174\\
0.179487179487179	0.868734482828951\\
0.205128205128205	0.924834116628785\\
0.230769230769231	0.977919675454557\\
0.256410256410256	1.02806830438343\\
0.282051282051282	1.07535672253043\\
0.307692307692308	1.11986454703057\\
0.333333333333333	1.16162636742563\\
0.358974358974359	1.2006214753998\\
0.384615384615385	1.23681724653805\\
0.41025641025641	1.27018183599256\\
0.435897435897436	1.30066440772476\\
0.461538461538462	1.32818678510676\\
0.487179487179487	1.35266993189278\\
0.512820512820513	1.37403242684946\\
0.538461538461538	1.39216712559556\\
0.564102564102564	1.40694095735\\
0.58974358974359	1.41821813225257\\
0.615384615384615	1.42586180409206\\
0.641025641025641	1.42972312279478\\
0.666666666666667	1.42964539459685\\
0.692307692307692	1.42547022390125\\
0.717948717948718	1.41704067872071\\
0.743589743589744	1.40420493326319\\
0.769230769230769	1.38681121585739\\
0.794871794871795	1.36470035053764\\
0.82051282051282	1.33770777240828\\
0.846153846153846	1.30566690170482\\
0.871794871794872	1.26841067957406\\
0.897435897435897	1.22577293139885\\
0.923076923076923	1.17767144484087\\
0.948717948717949	1.12404776428869\\
0.974358974358974	1.06484343444187\\
1	1\\
-1.5	40\\
0	0.38\\
0.0256410256410256	0.461997759193346\\
0.0512820512820513	0.539262410793544\\
0.0769230769230769	0.612180451150617\\
0.102564102564103	0.681138376614592\\
0.128205128205128	0.746522682631616\\
0.153846153846154	0.808650830876961\\
0.179487179487179	0.867598126705949\\
0.205128205128205	0.923445751739714\\
0.230769230769231	0.976277431862617\\
0.256410256410256	1.02617513937432\\
0.282051282051282	1.07322038338178\\
0.307692307692308	1.1174975322112\\
0.333333333333333	1.15904509151131\\
0.358974358974359	1.1978452807551\\
0.384615384615385	1.23386811268484\\
0.41025641025641	1.26708426014987\\
0.435897435897436	1.29744503291769\\
0.461538461538462	1.32487392643782\\
0.487179487179487	1.34929327771971\\
0.512820512820513	1.37062275565379\\
0.538461538461538	1.38875648059869\\
0.564102564102564	1.40356270307032\\
0.58974358974359	1.41490631171618\\
0.615384615384615	1.4226502627166\\
0.641025641025641	1.42664376798245\\
0.666666666666667	1.42672699808452\\
0.692307692307692	1.4227375638255\\
0.717948717948718	1.41451546111764\\
0.743589743589744	1.40190945797865\\
0.769230769230769	1.3847683495186\\
0.794871794871795	1.3629279722522\\
0.82051282051282	1.3362149380164\\
0.846153846153846	1.30445177318784\\
0.871794871794872	1.26746067215697\\
0.897435897435897	1.22506665921009\\
0.923076923076923	1.17718408134064\\
0.948717948717949	1.12375244259717\\
0.974358974358974	1.06471124735764\\
1	1\\
-1.25	40\\
0	0.38\\
0.0256410256410256	0.46196947764055\\
0.0512820512820513	0.539083565627309\\
0.0769230769230769	0.611753321672983\\
0.102564102564103	0.680389803490274\\
0.128205128205128	0.74540406783939\\
0.153846153846154	0.807134424340757\\
0.179487179487179	0.865664005818028\\
0.205128205128205	0.921082162869279\\
0.230769230769231	0.973480934276453\\
0.256410256410256	1.02295047557144\\
0.282051282051282	1.06958042104231\\
0.307692307692308	1.11346320710815\\
0.333333333333333	1.15464398989749\\
0.358974358974359	1.19310995893572\\
0.384615384615385	1.22883561130541\\
0.41025641025641	1.26179591572376\\
0.435897435897436	1.29194586322081\\
0.461538461538462	1.31921184427126\\
0.487179487179487	1.34351860394961\\
0.512820512820513	1.36478775411014\\
0.538461538461538	1.38291561867214\\
0.564102564102564	1.3977727272213\\
0.58974358974359	1.40922520116878\\
0.615384615384615	1.41713579488785\\
0.641025641025641	1.42135062066502\\
0.666666666666667	1.42170473749474\\
0.692307692307692	1.41802918619363\\
0.717948717948718	1.410158889611\\
0.743589743589744	1.39794366019603\\
0.769230769230769	1.38123333194176\\
0.794871794871795	1.35985548309716\\
0.82051282051282	1.33362192141094\\
0.846153846153846	1.30233669315241\\
0.871794871794872	1.2658035296252\\
0.897435897435897	1.22383208851172\\
0.923076923076923	1.17633042355904\\
0.948717948717949	1.12323412605635\\
0.974358974358974	1.06447878765342\\
1	1\\
-1	40\\
0	0.38\\
0.0256410256410256	0.461922987353961\\
0.0512820512820513	0.53878914784854\\
0.0769230769230769	0.611049885113659\\
0.102564102564103	0.679156602779239\\
0.128205128205128	0.743560703442684\\
0.153846153846154	0.804634731118655\\
0.179487179487179	0.862474623828426\\
0.205128205128205	0.917183099089495\\
0.230769230769231	0.968865802706204\\
0.256410256410256	1.01762629239738\\
0.282051282051282	1.06356752308294\\
0.307692307692308	1.10679518185297\\
0.333333333333333	1.14736534756772\\
0.358974358974359	1.18527336552874\\
0.384615384615385	1.22050111166124\\
0.41025641025641	1.25303066314112\\
0.435897435897436	1.28282315635851\\
0.461538461538462	1.30980988608887\\
0.487179487179487	1.33391975907115\\
0.512820512820513	1.35507782916516\\
0.538461538461538	1.37318418905771\\
0.564102564102564	1.38811320155403\\
0.58974358974359	1.39973322907212\\
0.615384615384615	1.40790706160753\\
0.641025641025641	1.41247625536338\\
0.666666666666667	1.41326807807648\\
0.692307692307692	1.41010338996317\\
0.717948717948718	1.40280920380138\\
0.743589743589744	1.39123715226385\\
0.769230769230769	1.37523899449539\\
0.794871794871795	1.35462937696928\\
0.82051282051282	1.32919636518551\\
0.846153846153846	1.29871379168703\\
0.871794871794872	1.26295453584986\\
0.897435897435897	1.22170186855054\\
0.923076923076923	1.17485221573924\\
0.948717948717949	1.12233344282796\\
0.974358974358974	1.06407341564034\\
1	1\\
-0.75	40\\
0	0.38\\
0.0256410256410256	0.461850547763995\\
0.0512820512820513	0.538329343432191\\
0.0769230769230769	0.609950580924925\\
0.102564102564103	0.677228454162531\\
0.128205128205128	0.740677155907907\\
0.153846153846154	0.800722479446081\\
0.179487179487179	0.857480157194275\\
0.205128205128205	0.911073575991637\\
0.230769230769231	0.961629432475059\\
0.256410256410256	1.00927203520479\\
0.282051282051282	1.0541249957859\\
0.307692307692308	1.09631460701926\\
0.333333333333333	1.1359139299915\\
0.358974358974359	1.17293096665363\\
0.384615384615385	1.20735909028414\\
0.41025641025641	1.23919155099936\\
0.435897435897436	1.26839927514737\\
0.461538461538462	1.29492154340498\\
0.487179487179487	1.31869423946616\\
0.512820512820513	1.33964838959491\\
0.538461538461538	1.35769030328793\\
0.564102564102564	1.37270049775072\\
0.58974358974359	1.38455133592667\\
0.615384615384615	1.39310657321812\\
0.641025641025641	1.39820204774499\\
0.666666666666667	1.3996544904598\\
0.692307692307692	1.39727056842131\\
0.717948717948718	1.39086615837619\\
0.743589743589744	1.38029600476162\\
0.769230769230769	1.36541543763263\\
0.794871794871795	1.34602078735408\\
0.82051282051282	1.32186498055654\\
0.846153846153846	1.29267560471651\\
0.871794871794872	1.25817653958897\\
0.897435897435897	1.21810723707716\\
0.923076923076923	1.17234260576935\\
0.948717948717949	1.12079497156257\\
0.974358974358974	1.0633766608438\\
1	1\\
-0.5	40\\
0	0.38\\
0.0256410256410256	0.461746580072173\\
0.0512820512820513	0.537667181653895\\
0.0769230769230769	0.608365945345997\\
0.102564102564103	0.674447011749312\\
0.128205128205128	0.736514520127424\\
0.153846153846154	0.795070477204483\\
0.179487179487179	0.85025867217557\\
0.205128205128205	0.902231799763313\\
0.230769230769231	0.951146423255709\\
0.256410256410256	0.99715632323732\\
0.282051282051282	1.0404145131949\\
0.307692307692308	1.08107673400804\\
0.333333333333333	1.11924029506751\\
0.358974358974359	1.15493119827436\\
0.384615384615385	1.18815925466405\\
0.41025641025641	1.21893388875268\\
0.435897435897436	1.24724046773566\\
0.461538461538462	1.27303040633184\\
0.487179487179487	1.2962506042843\\
0.512820512820513	1.3168419277503\\
0.538461538461538	1.33472026669219\\
0.564102564102564	1.34977522651446\\
0.58974358974359	1.36188589303466\\
0.615384615384615	1.37091931408414\\
0.641025641025641	1.37670618556595\\
0.666666666666667	1.3790518072869\\
0.692307692307692	1.37774648350032\\
0.717948717948718	1.37259306821478\\
0.743589743589744	1.36345139197399\\
0.769230769230769	1.35018318873101\\
0.794871794871795	1.332563243276\\
0.82051282051282	1.31029945122458\\
0.846153846153846	1.28305651782989\\
0.871794871794872	1.25048745538549\\
0.897435897435897	1.21226371281316\\
0.923076923076923	1.16822145891959\\
0.948717948717949	1.11824233053551\\
0.974358974358974	1.06220796508696\\
1	1\\
-0.25	40\\
0	0.38\\
0.0256410256410256	0.461613694038101\\
0.0512820512820513	0.536817020890997\\
0.0769230769230769	0.606328762640587\\
0.102564102564103	0.670867701368769\\
0.128205128205128	0.731152617589454\\
0.153846153846154	0.78778253638785\\
0.179487179487179	0.840936463086887\\
0.205128205128205	0.890803889843481\\
0.230769230769231	0.937578899266424\\
0.256410256410256	0.981452335834071\\
0.282051282051282	1.02261428530824\\
0.307692307692308	1.0612578130196\\
0.333333333333333	1.09751088801941\\
0.358974358974359	1.13142208963494\\
0.384615384615385	1.16302196191187\\
0.41025641025641	1.19234065831775\\
0.435897435897436	1.21938248718052\\
0.461538461538462	1.24411535003927\\
0.487179487179487	1.26650180101371\\
0.512820512820513	1.28649734908077\\
0.538461538461538	1.30403101717338\\
0.564102564102564	1.31900405802085\\
0.58974358974359	1.33130547874676\\
0.615384615384615	1.34080961227111\\
0.641025641025641	1.34734603997568\\
0.666666666666667	1.35071239797676\\
0.692307692307692	1.35068606906563\\
0.717948717948718	1.34705942140385\\
0.743589743589744	1.33969975889633\\
0.769230769230769	1.32847909116495\\
0.794871794871795	1.31315497166846\\
0.82051282051282	1.29339124947477\\
0.846153846153846	1.26878379675534\\
0.871794871794872	1.23889945479443\\
0.897435897435897	1.20331740846313\\
0.923076923076923	1.1618101469941\\
0.948717948717949	1.1142039629295\\
0.974358974358974	1.06032514951593\\
1	1\\
0	40\\
0	0.38\\
0.0256410256410256	0.461466899523204\\
0.0512820512820513	0.535872925256606\\
0.0769230769230769	0.604063029853532\\
0.102564102564103	0.666882165967306\\
0.128205128205128	0.725175284427134\\
0.153846153846154	0.779648018620054\\
0.179487179487179	0.830517369302517\\
0.205128205128205	0.878012530515991\\
0.230769230769231	0.922368073885732\\
0.256410256410256	0.963814870834345\\
0.282051282051282	1.00258314625814\\
0.307692307692308	1.03890661228623\\
0.333333333333333	1.07294645998978\\
0.358974358974359	1.10477538578162\\
0.384615384615385	1.13444616608041\\
0.41025641025641	1.16201158040286\\
0.435897435897436	1.1874971272485\\
0.461538461538462	1.21088983225662\\
0.487179487179487	1.2321712672377\\
0.512820512820513	1.25131552675632\\
0.538461538461538	1.2682667502569\\
0.564102564102564	1.28293819656043\\
0.58974358974359	1.29523068551658\\
0.615384615384615	1.30502994235827\\
0.641025641025641	1.31217144138154\\
0.666666666666667	1.31645417274003\\
0.692307692307692	1.31765344174674\\
0.717948717948718	1.3155598185352\\
0.743589743589744	1.31004931562548\\
0.769230769230769	1.30100681906873\\
0.794871794871795	1.28818777188976\\
0.82051282051282	1.27123352376133\\
0.846153846153846	1.24969185591674\\
0.871794871794872	1.22305479852875\\
0.897435897435897	1.19080508102796\\
0.923076923076923	1.15262970643117\\
0.948717948717949	1.10827371042798\\
0.974358974358974	1.05748212946778\\
1	1\\
0.25	40\\
0	0.38\\
0.0256410256410256	0.461328710526329\\
0.0512820512820513	0.534979462284261\\
0.0769230769230769	0.601915479496737\\
0.102564102564103	0.663099986386698\\
0.128205128205128	0.719496205110613\\
0.153846153846154	0.771909530767843\\
0.179487179487179	0.820591790762645\\
0.205128205128205	0.8658085540618\\
0.230769230769231	0.907831474908751\\
0.256410256410256	0.946928084870825\\
0.282051282051282	0.983365443230681\\
0.307692307692308	1.01741476513481\\
0.333333333333333	1.04926764805444\\
0.358974358974359	1.0790183892936\\
0.384615384615385	1.10673967651554\\
0.41025641025641	1.13250489839702\\
0.435897435897436	1.15635928772088\\
0.461538461538462	1.1783082557888\\
0.487179487179487	1.1983524532856\\
0.512820512820513	1.21648532394118\\
0.538461538461538	1.23266500557608\\
0.564102564102564	1.24681391565287\\
0.58974358974359	1.25884340021635\\
0.615384615384615	1.26865181390941\\
0.641025641025641	1.27608640640055\\
0.666666666666667	1.28095710311568\\
0.692307692307692	1.28305048855112\\
0.717948717948718	1.28216634024819\\
0.743589743589744	1.27818688741647\\
0.769230769230769	1.27100681086737\\
0.794871794871795	1.26039809587124\\
0.82051282051282	1.24601624654924\\
0.846153846153846	1.22740976106636\\
0.871794871794872	1.20404612463145\\
0.897435897435897	1.1753509520965\\
0.923076923076923	1.14093437517716\\
0.948717948717949	1.10046077177804\\
0.974358974358974	1.05359452051403\\
1	1\\
0.5	40\\
0	0.38\\
0.0256410256410256	0.461216883137486\\
0.0512820512820513	0.5342531250778\\
0.0769230769230769	0.600167266431194\\
0.102564102564103	0.660017847807918\\
0.128205128205128	0.714863407554726\\
0.153846153846154	0.7655896087765\\
0.179487179487179	0.812475743004455\\
0.205128205128205	0.855816040808332\\
0.230769230769231	0.895911349746414\\
0.256410256410256	0.933058041413809\\
0.282051282051282	0.967552181513085\\
0.307692307692308	0.999694644569493\\
0.333333333333333	1.02970083092386\\
0.358974358974359	1.05768149654592\\
0.384615384615385	1.08372442750971\\
0.41025641025641	1.10791885685781\\
0.435897435897436	1.13032544770234\\
0.461538461538462	1.15096433774399\\
0.487179487179487	1.16985201485231\\
0.512820512820513	1.18699844585266\\
0.538461538461538	1.2023721669617\\
0.564102564102564	1.21590025750722\\
0.58974358974359	1.22750074603948\\
0.615384615384615	1.23708208217065\\
0.641025641025641	1.24450439210511\\
0.666666666666667	1.24959277907327\\
0.692307692307692	1.2521528172293\\
0.717948717948718	1.25199982818665\\
0.743589743589744	1.24901596147165\\
0.769230769230769	1.24309608774829\\
0.794871794871795	1.23403715311108\\
0.82051282051282	1.22153696281798\\
0.846153846153846	1.20519360870899\\
0.871794871794872	1.18451575837986\\
0.897435897435897	1.15894531782156\\
0.923076923076923	1.12807265092138\\
0.948717948717949	1.09153362822405\\
0.974358974358974	1.04896412087008\\
1	1\\
0.75	40\\
0	0.38\\
0.0256410256410256	0.461136952851726\\
0.0512820512820513	0.53373216295194\\
0.0769230769230769	0.598912067043662\\
0.102564102564103	0.657803101869908\\
0.128205128205128	0.711531701768854\\
0.153846153846154	0.761040630584283\\
0.179487179487179	0.806628392280585\\
0.205128205128205	0.848609248454599\\
0.230769230769231	0.887304426691061\\
0.256410256410256	0.923030409624049\\
0.282051282051282	0.956103494850969\\
0.307692307692308	0.986845304310114\\
0.333333333333333	1.01548771758173\\
0.358974358974359	1.04215257302887\\
0.384615384615385	1.06693774123192\\
0.41025641025641	1.08994314186293\\
0.435897435897436	1.1112399607232\\
0.461538461538462	1.13085854012379\\
0.487179487179487	1.14882664928174\\
0.512820512820513	1.16516626103072\\
0.538461538461538	1.17985237339816\\
0.564102564102564	1.19281315060075\\
0.58974358974359	1.20396949199699\\
0.615384615384615	1.21323597381361\\
0.641025641025641	1.22048285554018\\
0.666666666666667	1.22554883850568\\
0.692307692307692	1.22825821637046\\
0.717948717948718	1.22844142146846\\
0.743589743589744	1.22597449788183\\
0.769230769230769	1.22074265503791\\
0.794871794871795	1.21256311344319\\
0.82051282051282	1.20118061963537\\
0.846153846153846	1.18626251619561\\
0.871794871794872	1.16739786933534\\
0.897435897435897	1.14410484469836\\
0.923076923076923	1.11603079321793\\
0.948717948717949	1.08286346691087\\
0.974358974358974	1.04429061832297\\
1	1\\
1	40\\
0	0.38\\
0.0256410256410256	0.461084754213693\\
0.0512820512820513	0.533391142589104\\
0.0769230769230769	0.59808983101763\\
0.102564102564103	0.656351485390666\\
0.128205128205128	0.709346769102223\\
0.153846153846154	0.758055607073127\\
0.179487179487179	0.802788853726295\\
0.205128205128205	0.843873636475304\\
0.230769230769231	0.881644259617329\\
0.256410256410256	0.916430095750543\\
0.282051282051282	0.948560407846933\\
0.307692307692308	0.97837013714809\\
0.333333333333333	1.00610165085183\\
0.358974358974359	1.03188367377323\\
0.384615384615385	1.05582028830669\\
0.41025641025641	1.07801803788672\\
0.435897435897436	1.09855467190589\\
0.461538461538462	1.11746695913876\\
0.487179487179487	1.13478991520338\\
0.512820512820513	1.15055332531285\\
0.538461538461538	1.16473578037173\\
0.564102564102564	1.17726483562918\\
0.58974358974359	1.1880620209578\\
0.615384615384615	1.19704496014625\\
0.641025641025641	1.20409048956475\\
0.666666666666667	1.20904699374238\\
0.692307692307692	1.21175300518846\\
0.717948717948718	1.21205019133954\\
0.743589743589744	1.20980611371992\\
0.769230769230769	1.20489212050096\\
0.794871794871795	1.19713672955667\\
0.82051282051282	1.18632071571465\\
0.846153846153846	1.17217249871338\\
0.871794871794872	1.15436241788514\\
0.897435897435897	1.13250157860627\\
0.923076923076923	1.10634100247349\\
0.948717948717949	1.07567460396699\\
0.974358974358974	1.04029629812856\\
1	1\\
1.25	40\\
0	0.38\\
0.0256410256410256	0.461052647755348\\
0.0512820512820513	0.533181072730185\\
0.0769230769230769	0.597583100412595\\
0.102564102564103	0.65545655629066\\
0.128205128205128	0.707999263298063\\
0.153846153846154	0.756213951284451\\
0.179487179487179	0.800418991988939\\
0.205128205128205	0.840949338915293\\
0.230769230769231	0.878147244646829\\
0.256410256410256	0.9123499098259\\
0.282051282051282	0.943894469236711\\
0.307692307692308	0.97312395905804\\
0.333333333333333	1.0002870801894\\
0.358974358974359	1.02551664302807\\
0.384615384615385	1.04892038821687\\
0.41025641025641	1.07060877445666\\
0.435897435897436	1.09066344387999\\
0.461538461538462	1.10912499720766\\
0.487179487179487	1.12603282840682\\
0.512820512820513	1.14142148089177\\
0.538461538461538	1.15527145084716\\
0.564102564102564	1.16750930734698\\
0.58974358974359	1.17805635718155\\
0.615384615384615	1.18683157227897\\
0.641025641025641	1.19371565763603\\
0.666666666666667	1.19856310283758\\
0.692307692307692	1.20122184830673\\
0.717948717948718	1.20154080916411\\
0.743589743589744	1.19937989926827\\
0.769230769230769	1.19459776316711\\
0.794871794871795	1.18702773444156\\
0.82051282051282	1.17647348822022\\
0.846153846153846	1.16270652359315\\
0.871794871794872	1.14545932739138\\
0.897435897435897	1.12442114150749\\
0.923076923076923	1.09944905102734\\
0.948717948717949	1.07045012048823\\
0.974358974358974	1.03733141508189\\
1	1\\
1.5	40\\
0	0.38\\
0.0256410256410256	0.461033622206611\\
0.0512820512820513	0.533056477602149\\
0.0769230769230769	0.59728246930306\\
0.102564102564103	0.654925500425796\\
0.128205128205128	0.707199471498579\\
0.153846153846154	0.755120606901941\\
0.179487179487179	0.799011703980834\\
0.205128205128205	0.839212321710154\\
0.230769230769231	0.876069387654831\\
0.256410256410256	0.909924704013413\\
0.282051282051282	0.941120031830298\\
0.307692307692308	0.970003167393027\\
0.333333333333333	0.996826520525055\\
0.358974358974359	1.02172526575074\\
0.384615384615385	1.0448092493484\\
0.41025641025641	1.06619118834777\\
0.435897435897436	1.08595497407964\\
0.461538461538462	1.10414342782841\\
0.487179487179487	1.1207985020144\\
0.512820512820513	1.13595753371844\\
0.538461538461538	1.14960202402487\\
0.564102564102564	1.16165771409974\\
0.58974358974359	1.17204552773382\\
0.615384615384615	1.18068500688847\\
0.641025641025641	1.18745904198801\\
0.666666666666667	1.19222574671693\\
0.692307692307692	1.19483882618303\\
0.717948717948718	1.19515154916006\\
0.743589743589744	1.19301823418295\\
0.769230769230769	1.18828814265876\\
0.794871794871795	1.18079626025874\\
0.82051282051282	1.17035959224411\\
0.846153846153846	1.15677673645756\\
0.871794871794872	1.13982129479684\\
0.897435897435897	1.11923711787205\\
0.923076923076923	1.0949648227408\\
0.948717948717949	1.06700248629131\\
0.974358974358974	1.03534818616419\\
1	1\\
1.75	40\\
0	0.38\\
0.0256410256410256	0.461022596285887\\
0.0512820512820513	0.532984232439231\\
0.0769230769230769	0.597108123379221\\
0.102564102564103	0.65461748402505\\
0.128205128205128	0.706735526688073\\
0.153846153846154	0.754486289664832\\
0.179487179487179	0.798195125105648\\
0.205128205128205	0.838204249937984\\
0.230769230769231	0.874863288934126\\
0.256410256410256	0.908516698157118\\
0.282051282051282	0.939508906351592\\
0.307692307692308	0.968190455561909\\
0.333333333333333	0.994815889505887\\
0.358974358974359	1.0195217328014\\
0.384615384615385	1.04241903004888\\
0.41025641025641	1.06362178502592\\
0.435897435897436	1.08321517174016\\
0.461538461538462	1.10124327983671\\
0.487179487179487	1.11774953048638\\
0.512820512820513	1.13277286898714\\
0.538461538461538	1.14629533332546\\
0.564102564102564	1.15824209674263\\
0.58974358974359	1.16853376874573\\
0.615384615384615	1.17709013482822\\
0.641025641025641	1.18379530552428\\
0.666666666666667	1.18850948587111\\
0.692307692307692	1.19108977562883\\
0.717948717948718	1.19139197122762\\
0.743589743589744	1.18926674124427\\
0.769230769230769	1.18455719229475\\
0.794871794871795	1.17709874849877\\
0.82051282051282	1.16671592017576\\
0.846153846153846	1.15322341293007\\
0.871794871794872	1.1364200368599\\
0.897435897435897	1.11608422103385\\
0.923076923076923	1.09221349416502\\
0.948717948717949	1.06486867119177\\
0.974358974358974	1.0341105678811\\
1	1\\
2	40\\
0	0.38\\
0.0256410256410256	0.461016288697083\\
0.0512820512820513	0.53294289060467\\
0.0769230769230769	0.597008345740577\\
0.102564102564103	0.654441194122616\\
0.128205128205128	0.706469973149545\\
0.153846153846154	0.754123189024369\\
0.179487179487179	0.797727652110528\\
0.205128205128205	0.837627096189799\\
0.230769230769231	0.874172684997762\\
0.256410256410256	0.907710388494964\\
0.282051282051282	0.938586157067773\\
0.307692307692308	0.967152099203329\\
0.333333333333333	0.993663974133688\\
0.358974358974359	1.01825907210528\\
0.384615384615385	1.04104911583974\\
0.41025641025641	1.06214883776058\\
0.435897435897436	1.08164413855181\\
0.461538461538462	1.09957982704865\\
0.487179487179487	1.11600015970969\\
0.512820512820513	1.13094499788491\\
0.538461538461538	1.1443966710629\\
0.564102564102564	1.15627999747531\\
0.58974358974359	1.16651537498169\\
0.615384615384615	1.17502269793924\\
0.641025641025641	1.18168675750066\\
0.666666666666667	1.1863689498295\\
0.692307692307692	1.18892833210173\\
0.717948717948718	1.18922214491238\\
0.743589743589744	1.18709882492567\\
0.769230769230769	1.18239770313945\\
0.794871794871795	1.1749542592744\\
0.82051282051282	1.16459719276496\\
0.846153846153846	1.15115052442579\\
0.871794871794872	1.13442791510668\\
0.897435897435897	1.11422855075504\\
0.923076923076923	1.09058565369235\\
0.948717948717949	1.06359964694285\\
0.974358974358974	1.03337095441069\\
1	1\\
};

\end{axis}
\end{tikzpicture}%

%% file: texfigures/Nonlinear/mu_03_def_paperWe.tex
%
\begin{tikzpicture}[%
baseline
]

\begin{axis}[%
name=refcorner,
width=0.3\textwidth,
height=0.3\textwidth,
scale only axis,
point meta min=-1,
point meta max=5,
colormap/\mapacolor,
xmode=log,
xmin=1e-3,
xmax=1e3,
xminorticks=true,
ymin=0,
ymax=1,
axis background/.style={fill=white},
legend style={legend cell align=left, align=left, draw=white!15!black},
colorbar sampled line={scatter,samples=7,scatter/use mapped color={draw=mapped color, fill=mapped color},only marks,mark=-,},colorbar style={ytick={-1,1,3,5},},,
colorbar,
colorbar style={title={$\log_{10} \La$},xshift=-.025\textwidth},
xlabel=$\We$,
ylabel=$d_c$,
ytick={0,.2,...,1},
minor ytick={0,.05,...,1},
xtick={1e-3,1e-2,1e-1,1e0,1e1,1e2,1e3},
]

\node[anchor=north west,xshift=0.0cm,yshift=0.0cm] at (rel axis cs:0,1) {(a)};

\node[anchor=south west,xshift=0.0cm,yshift=0.0cm] at (rel axis cs:0.1,.05) {Unstable};
\node[anchor=north east,xshift=0.0cm,yshift=0.0cm] at (rel axis cs:0.9,.9) {Stable};

\addplot[contour prepared, contour prepared format=matlab, contour/labels=false, line width=1.0pt] table[row sep=crcr] {%
\\
};

\addplot[contour prepared, contour/labels=false] table[row sep=crcr, x expr=10*\thisrow{x}^2, meta expr=1] {%
x y\\
0.0001	0.14604625958442\\
0.01	0.14604625958442\\
0.0102920052719443	0.146045776921477\\
0.0105925372517729	0.146045262971285\\
0.0109018449238513	0.146044721354776\\
0.0112201845430196	0.146044145051454\\
0.0115478198468946	0.146043536552425\\
0.0118850222743702	0.146042890830474\\
0.0122320711904993	0.146042206062479\\
0.0125892541179417	0.146041482185605\\
0.0129568669751702	0.146040714410411\\
0.0133352143216332	0.146039902655828\\
0.0137246096100756	0.146039042164896\\
0.0141253754462275	0.146038129989029\\
0.0145378438560766	0.146037163992686\\
0.0149623565609443	0.146036141025655\\
0.0153992652605949	0.146035057368454\\
0.0158489319246111	0.146033910067657\\
0.0163117290922784	0.146032694130716\\
0.0167880401812256	0.146031406569531\\
0.0172782598050786	0.146030043012633\\
0.0177827941003892	0.146028597446216\\
0.0183020610631106	0.146027067519554\\
0.018836490894898	0.146025446458531\\
0.0193865263595221	0.14602372921484\\
0.0199526231496888	0.14602191074397\\
0.0205352502645715	0.146019984578429\\
0.0211348903983665	0.146017943758766\\
0.0217520403401952	0.146015782739019\\
0.0223872113856834	0.146013493720113\\
0.0230409297605585	0.146011069017276\\
0.0237137370566166	0.146008500514949\\
0.0244061906804198	0.146005780575014\\
0.0251188643150958	0.14600289912443\\
0.0258523483956219	0.145999847311485\\
0.0266072505979881	0.145996615149753\\
0.0273841963426436	0.145993191179633\\
0.0281838293126445	0.145989564834741\\
0.0290068119869315	0.145985723784715\\
0.0298538261891796	0.145981655722346\\
0.0307255736526745	0.14597734686153\\
0.0316227766016838	0.14597278303914\\
0.0325461783498046	0.145967949570349\\
0.0334965439157828	0.145962829664708\\
0.0344746606573149	0.145957407550477\\
0.0354813389233576	0.145951664413133\\
0.0365174127254838	0.145945581867671\\
0.0375837404288444	0.145939139634166\\
0.0386812054633052	0.14593231683898\\
0.0398107170553497	0.145925090562468\\
0.0409732109813542	0.145917437377135\\
0.0421696503428582	0.145909332094267\\
0.0434010263644744	0.145900747845348\\
0.0446683592150963	0.145891656734436\\
0.0459726988530872	0.145882028716031\\
0.0473151258961481	0.145871832342773\\
0.0486967525165863	0.145861034101805\\
0.0501187233627272	0.145849598699752\\
0.0515822165072306	0.145837488591749\\
0.0530884444230988	0.14582466425433\\
0.0546386549881854	0.145811083640094\\
0.0562341325190349	0.145796702483651\\
0.0578761988349121	0.145781473817077\\
0.059566214352901	0.145765348077944\\
0.0613055792149821	0.14574827258359\\
0.0630957344480193	0.145730191893696\\
0.0649381631576211	0.145711047190687\\
0.0668343917568615	0.145690776343342\\
0.0687859912308807	0.145669313598534\\
0.0707945784384138	0.145646589434009\\
0.0728618174513228	0.145622530401026\\
0.0749894209332456	0.145597058826287\\
0.0771791515585013	0.145570092548469\\
0.0794328234724281	0.145541544744209\\
0.081752303794365	0.145511323749606\\
0.0841395141645195	0.145479332567406\\
0.0865964323360065	0.145445468854006\\
0.0891250938133746	0.145409624426324\\
0.0917275935389779	0.14537168502566\\
0.0944060876285924	0.145331530017671\\
0.0971627951577106	0.145289032034705\\
0.1	0.145244056629703\\
0.102920052719443	0.145196461937477\\
0.105925372517729	0.145146098199087\\
0.109018449238513	0.145092807538984\\
0.112201845430196	0.145036423380049\\
0.115478198468946	0.14497677013643\\
0.118850222743702	0.144913662743039\\
0.122320711904993	0.14484690615638\\
0.125892541179417	0.144776294976596\\
0.129568669751702	0.144701612857481\\
0.133352143216332	0.144622632071862\\
0.137246096100756	0.144539113002457\\
0.141253754462275	0.144450803658659\\
0.145378438560766	0.144357439003629\\
0.149623565609443	0.144258740600198\\
0.153992652605949	0.144154415911666\\
0.158489319246111	0.144044157860303\\
0.163117290922784	0.143927644214251\\
0.167880401812256	0.143804537064754\\
0.172782598050786	0.143674482252277\\
0.177827941003892	0.143537108866276\\
0.183020610631106	0.143392028706939\\
0.18836490894898	0.143238835781026\\
0.193865263595221	0.143077105810001\\
0.199526231496888	0.142906395821967\\
0.205352502645715	0.142726243681302\\
0.211348903983665	0.142536167793954\\
0.217520403401952	0.142335666753772\\
0.223872113856834	0.142124219100955\\
0.230409297605585	0.141901283206485\\
0.237137370566166	0.141666297139609\\
0.244061906804198	0.141418678722011\\
0.251188643150958	0.141157825663269\\
0.258523483956219	0.140883115824654\\
0.266072505979881	0.140593907647567\\
0.273841963426436	0.140289540720819\\
0.281838293126445	0.139969336545478\\
0.290068119869315	0.139632599493997\\
0.298538261891796	0.139278617982185\\
0.307255736526745	0.138906665898502\\
0.316227766016838	0.138516004255504\\
0.325461783498046	0.138105883168373\\
0.334965439157828	0.137675544096395\\
0.344746606573149	0.137224222412115\\
0.354813389233575	0.136751150326171\\
0.365174127254838	0.136255560144693\\
0.375837404288444	0.135736687903205\\
0.386812054633052	0.135193777403568\\
0.398107170553497	0.134626084618354\\
0.409732109813542	0.134032882520553\\
0.421696503428582	0.133413466289129\\
0.434010263644744	0.132767158937685\\
0.446683592150963	0.132093317319262\\
0.459726988530872	0.131391338489567\\
0.473151258961481	0.13066066645421\\
0.486967525165863	0.12990079917179\\
0.501187233627272	0.129111295884454\\
0.515822165072306	0.128291784618608\\
0.530884444230988	0.127441969866162\\
0.546386549881854	0.126561640329491\\
0.562341325190349	0.125650676675473\\
0.578761988349121	0.124709059179347\\
0.595662143529011	0.123736875163453\\
0.613055792149821	0.122734326124143\\
0.630957344480193	0.121701734419671\\
0.649381631576211	0.120639549391988\\
0.668343917568615	0.119548352783497\\
0.687859912308808	0.118428863342194\\
0.707945784384138	0.11728194044574\\
0.728618174513228	0.116108586656362\\
0.749894209332456	0.114909949059785\\
0.771791515585013	0.113687319289992\\
0.794328234724282	0.112442132150305\\
0.81752303794365	0.111175962746983\\
0.841395141645195	0.109890522091252\\
0.865964323360065	0.10858765112711\\
0.891250938133746	0.107269313197294\\
0.91727593538978	0.105937584957559\\
0.944060876285924	0.104594645812424\\
0.971627951577106	0.103242765935425\\
1	0.101884293019408\\
};

\addplot[contour prepared, contour/labels=false, line width=1.0pt] table[row sep=crcr, x expr=100*\thisrow{x}^2, meta expr=2] {%
x y\\
0.0001	0.606118995325642\\
0.00102920052719443	0.606118942810036\\
0.00105925372517729	0.606118887170572\\
0.00109018449238513	0.606118828234493\\
0.00112201845430196	0.606118765834836\\
0.00115478198468946	0.606118699723069\\
0.00118850222743702	0.606118629683778\\
0.00122320711904993	0.606118555515426\\
0.00125892541179417	0.606118476935435\\
0.00129568669751702	0.606118393704421\\
0.00133352143216332	0.606118305544792\\
0.00137246096100756	0.606118212147831\\
0.00141253754462275	0.60611811321784\\
0.00145378438560766	0.606118008440776\\
0.00149623565609443	0.606117897443082\\
0.00153992652605949	0.606117779869717\\
0.00158489319246111	0.606117655325299\\
0.00163117290922784	0.606117523403593\\
0.00167880401812256	0.606117383662557\\
0.00172782598050786	0.606117235641459\\
0.00177827941003892	0.606117078856346\\
0.00183020610631106	0.606116912774948\\
0.0018836490894898	0.606116736850424\\
0.00193865263595221	0.606116550504124\\
0.00199526231496888	0.606116353118342\\
0.00205352502645715	0.606116144028973\\
0.00211348903983665	0.606115922553827\\
0.00217520403401952	0.606115687954199\\
0.00223872113856834	0.606115439457344\\
0.00230409297605584	0.606115176234048\\
0.00237137370566166	0.606114897413871\\
0.00244061906804198	0.60611460207015\\
0.00251188643150958	0.60611428922759\\
0.00258523483956219	0.606113957848882\\
0.00266072505979881	0.606113606833035\\
0.00273841963426436	0.606113235022332\\
0.00281838293126445	0.606112841176239\\
0.00290068119869315	0.606112423997484\\
0.00298538261891796	0.606111982097339\\
0.00307255736526745	0.606111514016051\\
0.00316227766016838	0.606111018200021\\
0.00325461783498046	0.606110493008673\\
0.00334965439157828	0.606109936697721\\
0.00344746606573149	0.606109347427234\\
0.00354813389233575	0.60610872324142\\
0.00365174127254838	0.606108062074948\\
0.00375837404288444	0.606107361735355\\
0.00386812054633052	0.606106619903852\\
0.00398107170553497	0.606105834118925\\
0.00409732109813541	0.606105001781082\\
0.00421696503428582	0.606104120130784\\
0.00434010263644744	0.606103186247626\\
0.00446683592150963	0.606102197036971\\
0.00459726988530872	0.606101149223244\\
0.0047315125896148	0.60610003933426\\
0.00486967525165863	0.606098863694069\\
0.00501187233627272	0.606097618408524\\
0.00515822165072306	0.606096299352152\\
0.00530884444230988	0.606094902156562\\
0.00546386549881854	0.606093422193863\\
0.00562341325190349	0.606091854564917\\
0.00578761988349121	0.606090194076342\\
0.0059566214352901	0.606088435230695\\
0.00613055792149821	0.606086572202753\\
0.00630957344480193	0.606084598825433\\
0.00649381631576211	0.606082508564951\\
0.00668343917568615	0.606080294502149\\
0.00687859912308807	0.606077949307187\\
0.00707945784384138	0.606075465218039\\
0.00728618174513227	0.606072834012749\\
0.00749894209332456	0.606070046983662\\
0.00771791515585013	0.606067094908735\\
0.00794328234724281	0.606063968020357\\
0.0081752303794365	0.606060655973261\\
0.00841395141645195	0.60605714781158\\
0.00865964323360065	0.60605343193178\\
0.00891250938133746	0.606049496044099\\
0.00917275935389779	0.606045327134393\\
0.00944060876285924	0.606040911419153\\
0.00971627951577106	0.60603623430113\\
0.01	0.606031280321015\\
0.0102920052719443	0.606026033108018\\
0.0105925372517729	0.606020475325161\\
0.0109018449238513	0.606014588612197\\
0.0112201845430196	0.606008353527044\\
0.0115478198468946	0.606001749481014\\
0.0118850222743702	0.605994754672453\\
0.0122320711904993	0.605987346014202\\
0.0125892541179417	0.60597949906048\\
0.0129568669751702	0.605971187925841\\
0.0133352143216332	0.605962385201058\\
0.0137246096100756	0.605953061864987\\
0.0141253754462275	0.605943187189663\\
0.0145378438560766	0.605932728641233\\
0.0149623565609443	0.605921651773877\\
0.0153992652605949	0.605909920119939\\
0.0158489319246111	0.605897495071311\\
0.0163117290922784	0.605884335755026\\
0.0167880401812256	0.605870398902558\\
0.0172782598050786	0.605855638710517\\
0.0177827941003892	0.605840006694258\\
0.0183020610631106	0.605823451532738\\
0.018836490894898	0.605805918904859\\
0.0193865263595221	0.605787351317167\\
0.0199526231496888	0.60576768792062\\
0.0205352502645715	0.605746864318626\\
0.0211348903983665	0.605724812363675\\
0.0217520403401952	0.605701459942799\\
0.0223872113856834	0.605676730751441\\
0.0230409297605585	0.605650544054937\\
0.0237137370566166	0.60562281443692\\
0.0244061906804198	0.605593451534277\\
0.0251188643150958	0.605562359757839\\
0.0258523483956219	0.605529437997948\\
0.0266072505979881	0.605494579314934\\
0.0273841963426436	0.605457670612866\\
0.0281838293126445	0.605418592296198\\
0.0290068119869315	0.605377217909059\\
0.0298538261891796	0.605333413755755\\
0.0307255736526745	0.605287038501862\\
0.0316227766016838	0.605237942755703\\
0.0325461783498046	0.605185968628519\\
0.0334965439157828	0.605130949273423\\
0.0344746606573149	0.605072708401471\\
0.0354813389233576	0.605011059775227\\
0.0365174127254838	0.6049458066781\\
0.0375837404288444	0.604876741358842\\
0.0386812054633052	0.604803644451273\\
0.0398107170553497	0.604726284367858\\
0.0409732109813542	0.604644416666467\\
0.0421696503428582	0.604557783390441\\
0.0434010263644744	0.604466112381159\\
0.0446683592150963	0.604369116562112\\
0.0459726988530872	0.604266493195616\\
0.0473151258961481	0.604157923110472\\
0.0486967525165863	0.604043069901753\\
0.0501187233627272	0.603921579102182\\
0.0515822165072306	0.603793077325247\\
0.0530884444230988	0.603657171381464\\
0.0546386549881854	0.603513447367301\\
0.0562341325190349	0.603361469728357\\
0.0578761988349121	0.603200780298048\\
0.0595662143529011	0.603030897312299\\
0.0613055792149821	0.602851314403017\\
0.0630957344480193	0.602661499571252\\
0.0649381631576211	0.602460894143086\\
0.0668343917568615	0.602248911709297\\
0.0687859912308808	0.602024937053053\\
0.0707945784384138	0.60178832506676\\
0.0728618174513228	0.601538399662018\\
0.0749894209332456	0.601274452674566\\
0.0771791515585013	0.600995742767494\\
0.0794328234724282	0.600701494334544\\
0.081752303794365	0.600390896405099\\
0.0841395141645195	0.600063101551925\\
0.0845864114996922	0.6\\
0.0865964323360065	0.599474422530128\\
0.0891250938133746	0.598797601598215\\
0.091727593538978	0.598085441464665\\
0.0944060876285924	0.59733640744021\\
0.0971627951577106	0.596548937400236\\
0.1	0.59572144687745\\
0.102920052719443	0.594852335052652\\
0.105925372517729	0.593939991719907\\
0.109018449238513	0.592982805302637\\
0.112201845430196	0.591979171990558\\
0.115478198468946	0.590927506058519\\
0.118850222743702	0.589826251415792\\
0.122320711904993	0.588673894414527\\
0.125892541179417	0.587468977920686\\
0.129568669751702	0.586210116616926\\
0.133352143216332	0.584896013462734\\
0.137246096100756	0.583525477182556\\
0.141253754462275	0.582097440584924\\
0.145378438560766	0.580610979432818\\
0.149623565609443	0.579065331487827\\
0.153992652605949	0.577459915234339\\
0.158489319246111	0.575794347657193\\
0.163117290922784	0.574068460293839\\
0.167880401812256	0.572282312612841\\
0.172782598050786	0.570436201585217\\
0.177827941003892	0.568530666116085\\
0.183020610631106	0.566566484797142\\
0.18836490894898	0.564544665228696\\
0.193865263595221	0.562466422952171\\
0.199526231496888	0.560333147834503\\
0.205352502645715	0.558146355562978\\
0.211348903983665	0.555907621744311\\
0.217520403401952	0.553618495953904\\
0.223872113856834	0.551280392941526\\
0.227346165568289	0.55\\
0.230409297605585	0.548043209044909\\
0.237137370566166	0.543792690823481\\
0.244061906804198	0.539545287114381\\
0.251188643150958	0.535312725474446\\
0.258523483956219	0.531106305479702\\
0.266072505979881	0.526936623271838\\
0.273841963426436	0.522813284600322\\
0.281838293126445	0.51874461199089\\
0.290068119869316	0.514737351372959\\
0.298538261891796	0.510796382568282\\
0.307255736526745	0.506924436381386\\
0.316227766016838	0.503121818540995\\
0.323902954159249	0.5\\
0.325461783498046	0.498909916960769\\
0.334965439157828	0.492504331115194\\
0.344746606573149	0.486406414457418\\
0.354813389233575	0.480610392745612\\
0.365174127254838	0.475107180387783\\
0.375837404288444	0.469884579924408\\
0.386812054633052	0.464927484494096\\
0.398107170553497	0.460218066557926\\
0.409732109813542	0.455735935191213\\
0.421696503428582	0.451458242772237\\
0.42597673901691	0.45\\
0.434010263644744	0.445272280692118\\
0.446683592150963	0.438449024887557\\
0.459726988530872	0.432143472826554\\
0.473151258961481	0.426307665246798\\
0.486967525165863	0.420895597833167\\
0.501187233627272	0.415863091354465\\
0.515822165072306	0.41116760714552\\
0.530884444230988	0.406768006498326\\
0.546386549881854	0.402624248750706\\
0.556922501264742	0.4\\
0.562341325190349	0.397580131474438\\
0.578761988349121	0.390836235307476\\
0.595662143529011	0.384676659520251\\
0.613055792149821	0.379033339962393\\
0.630957344480193	0.373845305387939\\
0.649381631576211	0.369057542214457\\
0.668343917568615	0.364620003643914\\
0.687859912308808	0.360486728358346\\
0.707945784384138	0.356615037679093\\
0.728618174513228	0.352964781785078\\
0.746728493616698	0.35\\
0.749894209332456	0.349025012555592\\
0.771791515585013	0.342772690933986\\
0.794328234724282	0.337067422682764\\
0.81752303794365	0.331843312279281\\
0.841395141645195	0.327042546013378\\
0.865964323360065	0.322613942177636\\
0.891250938133746	0.318511747706864\\
0.91727593538978	0.314694622490258\\
0.944060876285924	0.311124764232618\\
0.971627951577106	0.307767134651512\\
1	0.304588752741082\\
};

\addplot[contour prepared, contour/labels=false, line width=1.0pt] table[row sep=crcr, x expr=1000*\thisrow{x}^2, meta expr=3] {%
x y\\
0.0001	0.780227485252681\\
0.00102920052719443	0.780225976518986\\
0.00105925372517729	0.780224378353632\\
0.00109018449238513	0.780222685449835\\
0.00112201845430196	0.78022089219231\\
0.00115478198468946	0.780218992627729\\
0.00118850222743702	0.780216980452671\\
0.00122320711904993	0.780214848989209\\
0.00125892541179417	0.780212591161294\\
0.00129568669751702	0.780210199473785\\
0.00133352143216332	0.780207665987047\\
0.00137246096100756	0.780204982289533\\
0.00141253754462275	0.780202139469884\\
0.00145378438560766	0.780199128088202\\
0.00149623565609443	0.780195938144133\\
0.00153992652605949	0.780192559040817\\
0.00158489319246111	0.780188979555716\\
0.00163117290922784	0.780185187796071\\
0.00167880401812256	0.780181171164441\\
0.00172782598050786	0.780176916313921\\
0.00177827941003892	0.780172409104306\\
0.00183020610631106	0.78016763455435\\
0.0018836490894898	0.780162576793251\\
0.00193865263595221	0.780157219004525\\
0.00199526231496888	0.780151543373383\\
0.00205352502645715	0.780145531023829\\
0.00211348903983665	0.78013916195739\\
0.00217520403401952	0.780132414985808\\
0.00223872113856834	0.780125267659364\\
0.00230409297605584	0.780117696191946\\
0.00237137370566166	0.780109675381834\\
0.00244061906804198	0.780101178525883\\
0.00251188643150958	0.780092177331821\\
0.00258523483956219	0.780082641820942\\
0.00266072505979881	0.780072540229537\\
0.00273841963426436	0.780061838901604\\
0.00281838293126445	0.780050502174895\\
0.00290068119869315	0.780038492263108\\
0.00298538261891796	0.780025769127787\\
0.00307255736526745	0.780012290343785\\
0.00316227766016838	0.779998010958242\\
0.00325461783498046	0.779982883339023\\
0.00334965439157828	0.779966857015402\\
0.00344746606573149	0.779949878509199\\
0.00354813389233575	0.779931891156123\\
0.00365174127254838	0.779912834916611\\
0.00375837404288444	0.77989264617573\\
0.00386812054633052	0.779871257531727\\
0.00398107170553497	0.779848597572309\\
0.00409732109813541	0.779824590638065\\
0.00421696503428582	0.779799156573057\\
0.00434010263644744	0.779772210461116\\
0.00446683592150963	0.779743662347606\\
0.00459726988530872	0.779713416946566\\
0.0047315125896148	0.779681373330767\\
0.00486967525165863	0.77964742460724\\
0.00501187233627272	0.779611457573856\\
0.00515822165072306	0.779573352359986\\
0.00530884444230988	0.779532982048232\\
0.00546386549881854	0.779490212277901\\
0.00562341325190349	0.779444900829893\\
0.00578761988349121	0.779396897191439\\
0.0059566214352901	0.779346042103213\\
0.00613055792149821	0.779292167084943\\
0.00630957344480193	0.779235093943269\\
0.00649381631576211	0.779174634259685\\
0.00668343917568615	0.779110588859809\\
0.00687859912308807	0.779042747264709\\
0.00707945784384138	0.778970887125258\\
0.00728618174513227	0.778894773639948\\
0.00749894209332456	0.778814158958855\\
0.00771791515585013	0.778728781574368\\
0.00794328234724281	0.77863836570178\\
0.0081752303794365	0.77854262065196\\
0.00841395141645195	0.778441240198912\\
0.00865964323360065	0.778333901946113\\
0.00891250938133746	0.778220266695227\\
0.00917275935389779	0.778099977821753\\
0.00944060876285924	0.777972660662194\\
0.00971627951577106	0.777837921918494\\
0.01	0.777695349084961\\
0.0102920052719443	0.777544509904331\\
0.0105925372517729	0.777384951858958\\
0.0109018449238513	0.777216201704502\\
0.0112201845430196	0.777037765052599\\
0.0115478198468946	0.776849126009885\\
0.0118850222743702	0.77664974688035\\
0.0122320711904993	0.776439067937983\\
0.0125892541179417	0.776216507275882\\
0.0129568669751702	0.77598146073834\\
0.0133352143216332	0.775733301940339\\
0.0137246096100756	0.775471382379283\\
0.0141253754462275	0.775195031641023\\
0.0145378438560766	0.774903557701878\\
0.0149623565609443	0.774596247325471\\
0.0153992652605949	0.774272366551356\\
0.0158489319246111	0.773931161269671\\
0.0163117290922784	0.773571857872663\\
0.0167880401812256	0.773193663970813\\
0.0172782598050786	0.772795769157127\\
0.0177827941003892	0.772377345798416\\
0.0183020610631106	0.77193754982806\\
0.018836490894898	0.771475521508541\\
0.0193865263595221	0.770990386126438\\
0.0199526231496888	0.770481254575489\\
0.0205352502645715	0.769947223776241\\
0.0211348903983665	0.769387376872327\\
0.0217520403401952	0.768800783134608\\
0.0223872113856834	0.768186497494359\\
0.0230409297605585	0.767543559615876\\
0.0237137370566166	0.766870992406233\\
0.0244061906804198	0.766167799846756\\
0.0251188643150958	0.7654329640152\\
0.0258523483956219	0.764665441150359\\
0.0266072505979881	0.763864156591563\\
0.0273841963426436	0.763027998403284\\
0.0281838293126445	0.762155809470236\\
0.0290068119869315	0.761246377819588\\
0.0298538261891796	0.760298424895002\\
0.0307255736526745	0.759310591470568\\
0.0316227766016838	0.758281420852342\\
0.0325461783498046	0.757209338970055\\
0.0334965439157828	0.756092630912785\\
0.0344746606573149	0.75492941341126\\
0.0354813389233576	0.753717602716643\\
0.0365174127254838	0.752454877276142\\
0.0375837404288444	0.751138634562735\\
0.0384910557980165	0.75\\
0.0386812054633052	0.749534055888309\\
0.0398107170553497	0.746711985419956\\
0.0409732109813542	0.743820329533635\\
0.0421696503428582	0.740860262788527\\
0.0434010263644744	0.737833204003969\\
0.0446683592150963	0.734740808935857\\
0.0459726988530872	0.731584950161619\\
0.0473151258961481	0.728367680683084\\
0.0486967525165863	0.72509117754876\\
0.0501187233627272	0.721757661743573\\
0.0515822165072306	0.718369290718634\\
0.0530884444230988	0.714928020235524\\
0.0546386549881854	0.711435432635052\\
0.0562341325190349	0.707892529108579\\
0.0578761988349121	0.704299483878354\\
0.0595662143529011	0.700655358128875\\
0.0598673385948669	0.7\\
0.0613055792149821	0.694366591545531\\
0.0630957344480193	0.68770276819736\\
0.0649381631576211	0.681261955396305\\
0.0668343917568615	0.675051035697845\\
0.0687859912308808	0.669070381018294\\
0.0707945784384138	0.663313686796442\\
0.0728618174513228	0.657767966043388\\
0.0749894209332456	0.652413655000149\\
0.0759888439158324	0.65\\
0.0771791515585013	0.644973474437407\\
0.0794328234724282	0.636305588579092\\
0.081752303794365	0.628413123019663\\
0.0841395141645195	0.621191526380682\\
0.0865964323360065	0.614541233812451\\
0.0891250938133746	0.608367690144587\\
0.091727593538978	0.602581066007149\\
0.0929658073753854	0.6\\
0.0944060876285924	0.594970586774827\\
0.0971627951577106	0.586411271757349\\
0.1	0.578771200468908\\
0.102920052719443	0.571879287539726\\
0.105925372517729	0.565589984244546\\
0.109018449238513	0.559777774550048\\
0.112201845430196	0.55433284631119\\
0.114933357774538	0.55\\
0.115478198468946	0.548582980819769\\
0.118850222743702	0.540607593861279\\
0.122320711904993	0.533489419178576\\
0.125892541179417	0.52706466640314\\
0.129568669751702	0.521196356512843\\
0.133352143216332	0.515767959031104\\
0.137246096100756	0.510678620764972\\
0.141253754462275	0.505839525058203\\
0.145378438560766	0.501171056964179\\
0.146459110803652	0.5\\
0.149623565609443	0.494670098357966\\
0.153992652605949	0.488156391984445\\
0.158489319246111	0.482251470206029\\
0.163117290922784	0.476840149530278\\
0.167880401812256	0.471823112877726\\
0.172782598050786	0.467113198975799\\
0.177827941003892	0.462632548842197\\
0.183020610631106	0.458310371963307\\
0.18836490894898	0.454081158397931\\
0.193712314230796	0.45\\
0.193865263595221	0.44981950117495\\
0.199526231496888	0.443665077302419\\
0.205352502645715	0.438091634065387\\
0.211348903983665	0.432994957523704\\
0.217520403401952	0.428284161053192\\
0.223872113856834	0.423878549920919\\
0.230409297605585	0.419705161218591\\
0.237137370566166	0.415696793781955\\
0.244061906804198	0.411790391010739\\
0.251188643150958	0.40792567031914\\
0.258523483956219	0.404043910992273\\
0.266072505979881	0.400086820104932\\
0.266241259643164	0.4\\
0.273841963426436	0.394615479728565\\
0.281838293126445	0.389624529311118\\
0.290068119869315	0.385059104658059\\
0.298538261891796	0.380836587184402\\
0.307255736526745	0.376882466805565\\
0.316227766016838	0.373128203538245\\
0.325461783498046	0.369509451103891\\
0.334965439157828	0.36596452985552\\
0.344746606573149	0.362433057595693\\
0.354813389233575	0.358854657899679\\
0.365174127254838	0.355167667640382\\
0.375837404288444	0.351307758432126\\
0.379434611812254	0.35\\
0.386812054633052	0.34655364641782\\
0.398107170553497	0.341880119239042\\
0.409732109813542	0.337648541657387\\
0.421696503428582	0.333773940466904\\
0.434010263644744	0.33018008553532\\
0.446683592150963	0.326797114630943\\
0.459726988530872	0.323559546354269\\
0.473151258961481	0.320404548435254\\
0.486967525165863	0.317270353384996\\
0.501187233627272	0.314094725920988\\
0.515822165072306	0.310813387979626\\
0.530884444230988	0.307358296766648\\
0.546386549881854	0.303655646299493\\
0.560926306089177	0.3\\
0.562341325190349	0.299555663730344\\
0.578761988349121	0.294927790294561\\
0.595662143529011	0.290840628270865\\
0.613055792149821	0.287193993354716\\
0.630957344480193	0.283901226857182\\
0.649381631576211	0.28088574751064\\
0.668343917568615	0.278078299809102\\
0.687859912308808	0.275414671902867\\
0.707945784384138	0.272833711637176\\
0.728618174513228	0.270275502015899\\
0.749894209332456	0.267679572440006\\
0.771791515585013	0.264983021188661\\
0.794328234724282	0.26211840618925\\
0.81752303794365	0.259011219720007\\
0.841395141645195	0.255576687214141\\
0.865964323360065	0.251715500182422\\
0.876340634392209	0.25\\
0.891250938133746	0.247334089821876\\
0.91727593538978	0.243365649250853\\
0.944060876285924	0.239942260568725\\
0.971627951577106	0.236961045609688\\
};

\addplot[contour prepared, contour/labels=false, line width=1.0pt] table[row sep=crcr, x expr=10000*\thisrow{x}^2, meta expr=4] {%
x y\\
0.0001	0.852191500994557\\
0.00102920052719443	0.852173822679992\\
0.00105925372517729	0.852154724611017\\
0.00109018449238513	0.852134098625891\\
0.00112201845430196	0.8521118297166\\
0.00115478198468946	0.85208779578864\\
0.00118850222743702	0.852061867432096\\
0.00122320711904993	0.852033907708788\\
0.00125892541179417	0.852003771954843\\
0.00129568669751702	0.85197130760184\\
0.00133352143216332	0.851936354017566\\
0.00137246096100756	0.851898742368105\\
0.00141253754462275	0.851858295501124\\
0.00145378438560766	0.851814827852944\\
0.00149623565609443	0.85176814537719\\
0.00153992652605949	0.851718045497019\\
0.00158489319246111	0.851664317079156\\
0.00163117290922784	0.85160674042918\\
0.00167880401812256	0.85154508730648\\
0.00172782598050786	0.851479120957261\\
0.00177827941003892	0.851408596162633\\
0.00183020610631106	0.851333259299885\\
0.0018836490894898	0.851252848412624\\
0.00193865263595221	0.851167093286862\\
0.00199526231496888	0.851075715528037\\
0.00205352502645715	0.850978428634436\\
0.00211348903983665	0.850874938061307\\
0.00217520403401952	0.850764941269183\\
0.00223872113856834	0.850648127749754\\
0.00230409297605584	0.850524179020205\\
0.00237137370566166	0.850392768577839\\
0.00244061906804198	0.850253561803216\\
0.00251188643150958	0.850106215800269\\
0.00256182068205423	0.85\\
0.00258523483956219	0.849872554783729\\
0.00266072505979881	0.849451801932484\\
0.00273841963426436	0.849011136999794\\
0.00281838293126445	0.848550104032578\\
0.00290068119869315	0.848068259073872\\
0.00298538261891796	0.847565168470529\\
0.00307255736526745	0.847040406401676\\
0.00316227766016838	0.846493551515756\\
0.00325461783498046	0.845924182550793\\
0.00334965439157828	0.845331872798899\\
0.00344746606573149	0.844716183259647\\
0.00354813389233575	0.844076654307909\\
0.00365174127254838	0.84341279567872\\
0.00375837404288444	0.84272407454469\\
0.00386812054633052	0.84200990142879\\
0.00398107170553497	0.841269613655674\\
0.00409732109813541	0.84050245599733\\
0.00421696503428582	0.839707558111039\\
0.00434010263644744	0.838883908297279\\
0.00446683592150963	0.838030323021542\\
0.00459726988530872	0.837145411541263\\
0.0047315125896148	0.836227534855431\\
0.00486967525165863	0.835274758048782\\
0.00501187233627272	0.834284794924947\\
0.00515822165072306	0.833254943620381\\
0.00530884444230988	0.832182011655104\\
0.00546386549881854	0.831062228620343\\
0.00562341325190349	0.829891144440329\\
0.00578761988349121	0.828663510919028\\
0.0059566214352901	0.827373144168003\\
0.00613055792149821	0.826012765664799\\
0.00630957344480193	0.824573820391434\\
0.00649381631576211	0.823046272269431\\
0.00668343917568615	0.821418380879843\\
0.00687859912308807	0.819676470936253\\
0.00707945784384138	0.817804720236266\\
0.00728618174513227	0.815785018404472\\
0.00749894209332456	0.813596997657031\\
0.00771791515585013	0.811218426006523\\
0.00794328234724281	0.80862631455953\\
0.0081752303794365	0.805799378623361\\
0.00841395141645195	0.802722996466964\\
0.00861375974080287	0.8\\
0.00865964323360065	0.798924517915483\\
0.00891250938133746	0.792505884137458\\
0.00917275935389779	0.785592300046845\\
0.00944060876285924	0.778238693717961\\
0.00971627951577106	0.770554398319035\\
0.01	0.76271424268053\\
0.0102920052719443	0.754962153143739\\
0.0104878370659272	0.75\\
0.0105925372517729	0.745050400866087\\
0.0109018449238513	0.73130803247297\\
0.0112201845430196	0.719205417620154\\
0.0115478198468946	0.7087517347524\\
0.0118799908782812	0.7\\
0.0118850222743702	0.699765279425935\\
0.0122320711904993	0.684976694669153\\
0.0125892541179417	0.672846188866822\\
0.0129568669751702	0.662831852154229\\
0.0133352143216332	0.654509846755819\\
0.0135770568808354	0.65\\
0.0137246096100756	0.64513551284277\\
0.0141253754462275	0.63363147144511\\
0.0145378438560766	0.623980160868174\\
0.0149623565609443	0.615782010687139\\
0.0153992652605949	0.608732879864179\\
0.0158489319246111	0.60259553818786\\
0.0160615733106696	0.6\\
0.0163117290922784	0.594580042196331\\
0.0167880401812256	0.585564969790237\\
0.0172782598050786	0.577744989874593\\
0.0177827941003892	0.570883843799907\\
0.0183020610631106	0.564792898662146\\
0.018836490894898	0.559317627156828\\
0.0193865263595221	0.554327955937843\\
0.0199154298499851	0.55\\
0.0199526231496888	0.549463029747663\\
0.0205352502645715	0.54170439997861\\
0.0211348903983665	0.534860973393594\\
0.0217520403401952	0.528765834493014\\
0.0223872113856834	0.523279836466709\\
0.0230409297605585	0.518283815065412\\
0.0237137370566166	0.513672692539921\\
0.0244061906804198	0.509350859990322\\
0.0251188643150958	0.505228424503917\\
0.0258523483956219	0.501218035280728\\
0.0260807591697471	0.5\\
0.0266072505979881	0.495491540185072\\
0.0273841963426436	0.489621724554548\\
0.0281838293126445	0.484375994708932\\
0.0290068119869315	0.479641463643971\\
0.0298538261891796	0.475316919886624\\
0.0307255736526745	0.471308455001778\\
0.0316227766016838	0.467525698868293\\
0.0325461783498046	0.463878398969939\\
0.0334965439157828	0.460273138748474\\
0.0344746606573149	0.45661003096811\\
0.0354813389233576	0.452779256704451\\
0.0361868008066619	0.45\\
0.0365174127254838	0.448252563927719\\
0.0375837404288444	0.443202241988562\\
0.0386812054633052	0.438781359553091\\
0.0398107170553497	0.434875414097906\\
0.0409732109813542	0.431377370860536\\
0.0421696503428582	0.428182842666405\\
0.0434010263644744	0.425185505709431\\
0.0446683592150963	0.422272403748818\\
0.0459726988530872	0.419318802472203\\
0.0473151258961481	0.416182244007107\\
0.0486967525165863	0.412695436205181\\
0.0501187233627272	0.408657624014525\\
0.0515822165072306	0.403824194344739\\
0.0525928154751457	0.4\\
};

\addplot[contour prepared, contour/labels=false, line width=1.0pt] table[row sep=crcr, x expr=100000*\thisrow{x}^2, meta expr=5] {%
x y\\
0.0001	0.871345955992159\\
0.000102920052719443	0.871338202711464\\
0.000105925372517729	0.871329009768897\\
0.000109018449238513	0.871318225528574\\
0.000112201845430196	0.871305687596684\\
0.000115478198468946	0.871291222533175\\
0.000118850222743702	0.87127464562434\\
0.000122320711904993	0.871255760698351\\
0.000125892541179417	0.871234360002315\\
0.000129568669751702	0.871210224158085\\
0.000133352143216332	0.871183122169306\\
0.000137246096100756	0.871152811522251\\
0.000141253754462275	0.871119038360212\\
0.000145378438560766	0.871081537743962\\
0.000149623565609443	0.871040033999786\\
0.000153992652605949	0.870994241155571\\
0.000158489319246111	0.870943863472137\\
0.000163117290922784	0.87088859606258\\
0.000167880401812256	0.870828125607833\\
0.000172782598050786	0.870762131159869\\
0.000177827941003892	0.87069028503771\\
0.000183020610631106	0.870612253808884\\
0.00018836490894898	0.87052769935817\\
0.000193865263595221	0.870436280040208\\
0.000199526231496888	0.870337651908383\\
0.000205352502645715	0.870231470025062\\
0.000211348903983665	0.870117389838595\\
0.000217520403401952	0.869995068627963\\
0.000223872113856834	0.869864167016563\\
0.000230409297605584	0.869724350541158\\
0.000237137370566166	0.86957529126598\\
0.000244061906804198	0.869416669450061\\
0.000251188643150958	0.869248175244792\\
0.000258523483956219	0.869069510414495\\
0.000266072505979881	0.868880390071551\\
0.000273841963426436	0.868680544400168\\
0.000281838293126446	0.868469720354954\\
0.000290068119869315	0.868247683300983\\
0.000298538261891796	0.868014218570243\\
0.000307255736526745	0.867769132884013\\
0.000316227766016838	0.867512255594678\\
0.000325461783498046	0.867243439685498\\
0.000334965439157828	0.866962562448257\\
0.000344746606573149	0.866669525745898\\
0.000354813389233575	0.866364255751316\\
0.000365174127254838	0.866046702021855\\
0.000375837404288444	0.865716835748185\\
0.000386812054633052	0.865374646988562\\
0.000398107170553497	0.865020140650013\\
0.000409732109813541	0.864653330948449\\
0.000421696503428582	0.864274234020011\\
0.000434010263644744	0.863882858297069\\
0.000446683592150963	0.863479192186919\\
0.000459726988530872	0.863063188502391\\
0.00047315125896148	0.862634744979197\\
0.000486967525165863	0.862193680068759\\
0.000501187233627273	0.861739703015266\\
0.000515822165072306	0.861272376980452\\
0.000530884444230988	0.860791073658513\\
0.000546386549881854	0.860294917395783\\
0.000562341325190349	0.859782716242336\\
0.000578761988349121	0.859252876569666\\
0.00059566214352901	0.858703296788485\\
0.000613055792149821	0.858131234194479\\
0.000630957344480193	0.857533136881893\\
0.000649381631576211	0.856904429788039\\
0.000668343917568615	0.856239239987873\\
0.000687859912308807	0.855530041016944\\
0.000707945784384138	0.854767188983122\\
0.000728618174513227	0.853938314521062\\
0.000749894209332456	0.853027525291181\\
0.000771791515585013	0.852014367729891\\
0.000794328234724281	0.850872506263021\\
0.00080982468565847	0.85\\
0.00081752303794365	0.849516672929687\\
0.000841395141645195	0.847849373091562\\
0.000865964323360065	0.845935655916466\\
0.000891250938133746	0.843717620891519\\
0.000917275935389779	0.841129095444761\\
0.000944060876285924	0.838101097960148\\
0.000971627951577106	0.834575559108033\\
0.001	0.828519025963909\\
0.00102920052719443	0.823292794517433\\
0.00105925372517729	0.81747987873947\\
0.00109018449238513	0.811317535594275\\
0.00112201845430196	0.805220657510924\\
0.00115282417580686	0.8\\
0.00115478198468946	0.799307320261096\\
0.00118850222743702	0.787154355771372\\
0.00122320711904993	0.775523994649653\\
0.00125892541179417	0.76463178511116\\
0.00129568669751702	0.754469508296951\\
0.00131279526361624	0.75\\
0.00133352143216332	0.739977288171541\\
0.00137246096100756	0.724038764981904\\
0.00141253754462275	0.710603030744857\\
0.00144934677959703	0.7\\
0.00145378438560766	0.698148316339116\\
0.00149623565609443	0.682705283803711\\
0.00153992652605949	0.670083185590005\\
0.00158489319246111	0.659426123755714\\
0.00163117290922784	0.650059154933105\\
0.00163148628790815	0.65\\
0.00167880401812256	0.638086916996821\\
0.00172782598050786	0.627941816851349\\
0.00177827941003892	0.619187509132794\\
0.00183020610631106	0.611454399596768\\
0.0018836490894898	0.604429357204192\\
0.00192017847222543	0.6\\
0.00193865263595221	0.596887787294756\\
0.00199526231496888	0.588379688975074\\
0.00205352502645715	0.580902718656499\\
0.00211348903983665	0.574241143081731\\
0.00217520403401952	0.568210251541473\\
0.00223872113856834	0.562643869208499\\
0.00230409297605584	0.557383672818128\\
0.00237137370566166	0.552268831961022\\
0.00240194514390025	0.55\\
0.00244061906804198	0.546170631959318\\
0.00251188643150958	0.539888232768901\\
0.00258523483956219	0.534266755460167\\
0.00266072505979881	0.529185198849454\\
0.00273841963426436	0.524532968798408\\
0.00281838293126445	0.520204149056031\\
0.00290068119869315	0.516091958607777\\
0.00298538261891796	0.51208279885729\\
0.00307255736526745	0.508049160783106\\
0.00316227766016838	0.503840353753342\\
0.00324035372714955	0.5\\
0.00325461783498046	0.499195938881098\\
0.00334965439157828	0.494301435510071\\
0.00344746606573149	0.489981640664362\\
0.00354813389233575	0.486150984035554\\
0.00365174127254838	0.482726958011443\\
0.00375837404288444	0.479625648122998\\
0.00386812054633052	0.476756899573578\\
0.00398107170553497	0.474018609228014\\
0.00409732109813541	0.471289420173894\\
0.00421696503428582	0.468418700403639\\
0.00434010263644744	0.465211964955418\\
0.00446683592150963	0.461408559544666\\
0.00459726988530872	0.45664584121449\\
0.0047315125896148	0.45039886836219\\
0.00473914245864724	0.45\\
};

\addplot [scatter, scatter src=1, mark=*,scatter/use mapped color={draw=mapped color, fill=mapped color},mark size=1pt] table[row sep=crcr] {1.365286837599484   0.136046226569237\\};
\addplot [scatter, scatter src=2, mark=*,scatter/use mapped color={draw=mapped color, fill=mapped color},mark size=1pt] table[row sep=crcr] {0.973072482165624   0.596118922420575\\};
\addplot [scatter, scatter src=3, mark=*,scatter/use mapped color={draw=mapped color, fill=mapped color},mark size=1pt] table[row sep=crcr] {0.409351533863862   0.770226026147784\\};
\addplot [scatter, scatter src=4, mark=*,scatter/use mapped color={draw=mapped color, fill=mapped color},mark size=1pt] table[row sep=crcr] {0.147478309651488   0.842191485062032\\};
\addplot [scatter, scatter src=5, mark=*,scatter/use mapped color={draw=mapped color, fill=mapped color},mark size=1pt] table[row sep=crcr] {0.026406903353171   0.861334521010319\\};

\end{axis}

%% file: texfigures/Nonlinear/mu_3_def_paperWe.tex
%


\begin{axis}[%
at=(refcorner.south west), anchor=south west, xshift=.45\textwidth, 
width=0.3\textwidth,
height=0.3\textwidth,
scale only axis,
point meta min=-1,
point meta max=5,
colormap/\mapacolor,
xmode=log,
xmin=1e-3,
xmax=1e3,
xminorticks=true,
ymin=0,
ymax=1,
axis background/.style={fill=white},
legend style={legend cell align=left, align=left, draw=white!15!black},
colorbar sampled line={scatter,samples=7,scatter/use mapped color={draw=mapped color},only marks,mark=-,},colorbar style={ytick={-1,1,3,5},},,
colorbar,
colorbar style={title={$\log_{10} \La$},xshift=-.025\textwidth},
xlabel=$\We$,
ylabel=$d_c$,
ytick={0,.2,...,1},
minor ytick={0,.05,...,1},
xtick={1e-3,1e-2,1e-1,1e0,1e1,1e2,1e3},
]

\node[anchor=north west,xshift=0.0cm,yshift=0.0cm] at (rel axis cs:0,1) {(b)};

\node[anchor=north west,xshift=0.0cm,yshift=0.0cm] at (rel axis cs:0.05,.93) {Stable};
\node[anchor=south west,xshift=0.0cm,yshift=0.0cm] at (rel axis cs:0.05,.33) {Unstable};

\addplot[contour prepared, contour/labels=false, line width=1.0pt] table[row sep=crcr, x expr=0.1*\thisrow{x}^2, meta expr=-1] {%
x y\\
0.0001	0.45279255390037\\
0.01	0.45279255390037\\
0.0125892541179417	0.452794801766521\\
0.0158489319246111	0.452798364755142\\
0.0199526231496888	0.45280401272144\\
0.0251188643150958	0.452812966773282\\
0.0316227766016838	0.452827164753373\\
0.0398107170553497	0.452849684193961\\
0.0501187233627272	0.452885418404819\\
0.0630957344480193	0.45294216224261\\
0.0794328234724281	0.453032368646688\\
0.1	0.453176021555242\\
0.125892541179417	0.453405410221339\\
0.158489319246111	0.453773241249381\\
0.199526231496888	0.454366825407016\\
0.251188643150958	0.45533374690562\\
0.316227766016838	0.456929916287516\\
0.398107170553497	0.459611568534791\\
0.501187233627272	0.464208127135499\\
0.630957344480193	0.472191899548417\\
0.794328234724282	0.485607747399464\\
0.97465355085712	0.5\\
0.996150447839835	0.500935848475605\\
};

\addplot[contour prepared, contour/labels=false, line width=1.0pt] table[row sep=crcr, x expr=1*\thisrow{x}^2, meta expr=0] {%
x y\\
0.0001	0.461933937550413\\
0.01	0.461933937550413\\
0.0125892541179417	0.461936883524672\\
0.0158489319246111	0.461941552751529\\
0.0199526231496888	0.461948953742152\\
0.0251188643150958	0.461960685838311\\
0.0316227766016838	0.461979286285287\\
0.0398107170553497	0.462008782542442\\
0.0501187233627272	0.46205557327037\\
0.0630957344480193	0.462129838748556\\
0.0794328234724281	0.462247811546003\\
0.1	0.462435463340627\\
0.125892541179417	0.462734566526394\\
0.158489319246111	0.463212839043414\\
0.199526231496888	0.463981331568773\\
0.251188643150958	0.465225105346652\\
0.316227766016838	0.467259021076686\\
0.398107170553497	0.470631270474014\\
0.501187233627272	0.476311376821828\\
0.630957344480193	0.485967830549624\\
0.778739885163426	0.5\\
0.794328234724282	0.500997683300803\\
0.965606746410849	0.508361199589804\\
};

\addplot[contour prepared, contour/labels=false, line width=1.0pt] table[row sep=crcr, x expr=10*\thisrow{x}^2, meta expr=1] {%
x y\\
0.0001		0.518616220962632\\
0.01	0.518616220962632\\
0.0125892541179417	0.518620615066281\\
0.0158489319246111	0.518627579380131\\
0.0199526231496888	0.51863861805786\\
0.0251188643150958	0.518656116476523\\
0.0316227766016838	0.51868385890972\\
0.0398107170553497	0.518727852378788\\
0.0501187233627272	0.518797640883308\\
0.0630957344480193	0.518908409601068\\
0.0794328234724281	0.519084372831045\\
0.1	0.519364274942866\\
0.125892541179417	0.519810432076881\\
0.158489319246111	0.520523847984168\\
0.199526231496888	0.521670043295024\\
0.251188643150958	0.523524298671292\\
0.316227766016838	0.52655253711199\\
0.398107170553497	0.531554938761569\\
0.501187233627272	0.539890024019789\\
0.601562465776148	0.55\\
0.630957344480193	0.552139808691726\\
0.763365803909798	0.559476116206573\\
};

\addplot[contour prepared, contour/labels=false, line width=1.0pt] table[row sep=crcr, x expr=100*\thisrow{x}^2, meta expr=2] {%
x y\\
0.0001	0.635660041510456\\
0.01	0.635660041510456\\
0.0105925372517729	0.635662427551866\\
0.0112201845430196	0.635665104540722\\
0.0118850222743702	0.6356681079385\\
0.0125892541179417	0.635671477521708\\
0.0133352143216332	0.635675257909769\\
0.0141253754462275	0.635679499149257\\
0.0149623565609443	0.635684257374953\\
0.0158489319246111	0.635689595545758\\
0.0167880401812256	0.635695584271536\\
0.0177827941003892	0.635702302739727\\
0.018836490894898	0.635709839750531\\
0.0199526231496888	0.635718294877834\\
0.0211348903983665	0.635727779768461\\
0.0223872113856834	0.635738419594248\\
0.0237137370566166	0.635750354678167\\
0.0251188643150958	0.635763742311483\\
0.0266072505979881	0.635778758783815\\
0.0281838293126445	0.635795601652946\\
0.0298538261891796	0.635814492275202\\
0.0316227766016838	0.635835678632547\\
0.0334965439157828	0.635859438481274\\
0.0354813389233576	0.635886082861009\\
0.0375837404288444	0.635915959999447\\
0.0398107170553497	0.635949459653907\\
0.0421696503428582	0.635987017931721\\
0.0446683592150963	0.636029122636216\\
0.0473151258961481	0.63607631918527\\
0.0501187233627272	0.636129217153346\\
0.0530884444230988	0.636188497485936\\
0.0562341325190349	0.636254920438566\\
0.059566214352901	0.636329334287697\\
0.0630957344480193	0.636412684857527\\
0.0668343917568615	0.636506025899222\\
0.0707945784384138	0.636610530346997\\
0.0749894209332456	0.63672750245603\\
0.0794328234724281	0.636858390803237\\
0.0841395141645195	0.637004802090661\\
0.0891250938133746	0.637168515641354\\
0.0944060876285924	0.63735149839927\\
0.1	0.637555920142204\\
0.105925372517729	0.637784168466495\\
0.112201845430196	0.638038862894026\\
0.118850222743702	0.638322867153398\\
0.125892541179417	0.638639298258238\\
0.133352143216332	0.638991530384709\\
0.141253754462275	0.639383190638339\\
0.149623565609443	0.639818142452181\\
0.158489319246111	0.640300450342355\\
0.167880401812256	0.6408343167075\\
0.177827941003892	0.641423976734963\\
0.18836490894898	0.642073530397289\\
0.199526231496888	0.642786679607986\\
0.211348903983665	0.643566321698685\\
0.223872113856834	0.644413924099099\\
0.237137370566166	0.645328564195819\\
0.251188643150958	0.646305454842809\\
0.266072505979881	0.647333678233193\\
0.281838293126445	0.648392703185047\\
0.298538261891796	0.649447046408999\\
0.308519088280046	0.65\\
0.316227766016838	0.650315683248838\\
0.334965439157828	0.650918475115666\\
0.354813389233575	0.651282930024656\\
0.375837404288444	0.651238955028087\\
0.398107170553497	0.650623943361753\\
0.408678200821033	0.65\\
0.421696503428582	0.648819812862446\\
0.446683592150963	0.644944886770307\\
0.473151258961481	0.639482194456121\\
0.479549134418539	0.638589882225738\\
};

\addplot[contour prepared, contour/labels=false, line width=1.0pt] table[row sep=crcr, x expr=1000*\thisrow{x}^2, meta expr=3] {%
x y\\
0.0001	0.742789783216931\\
0.001	0.742789783216931\\
0.00105925372517729	0.742790971440468\\
0.00112201845430196	0.742792304547625\\
0.00118850222743702	0.742793800188632\\
0.00125892541179417	0.742795478160637\\
0.00133352143216332	0.742797360676391\\
0.00141253754462275	0.74279947263723\\
0.00149623565609443	0.742801841974844\\
0.00158489319246111	0.742804500014119\\
0.00167880401812256	0.742807481881896\\
0.00177827941003892	0.742810826962709\\
0.0018836490894898	0.742814579412248\\
0.00199526231496888	0.742818788737123\\
0.00211348903983665	0.742823510427333\\
0.00223872113856834	0.742828806681133\\
0.00237137370566166	0.742834747204214\\
0.00251188643150958	0.742841410101096\\
0.00266072505979881	0.742848882878567\\
0.00281838293126445	0.742857263553965\\
0.00298538261891796	0.742866661900645\\
0.00316227766016838	0.742877200825195\\
0.00334965439157828	0.742889017905073\\
0.00354813389233575	0.742902267096837\\
0.00375837404288444	0.74291712062406\\
0.00398107170553497	0.74293377108145\\
0.00421696503428582	0.742952433750026\\
0.00446683592150963	0.742973349160451\\
0.0047315125896148	0.742996785913304\\
0.00501187233627272	0.743023043778706\\
0.00530884444230988	0.743052457093607\\
0.00562341325190349	0.743085398469244\\
0.0059566214352901	0.743122282824721\\
0.00630957344480193	0.743163571755363\\
0.00668343917568615	0.743209778237064\\
0.00707945784384138	0.74326147165949\\
0.00749894209332456	0.743319283173784\\
0.00794328234724281	0.743383911317486\\
0.00841395141645195	0.743456127867535\\
0.00891250938133746	0.743536783841269\\
0.00944060876285924	0.743626815536303\\
0.01	0.743727250460923\\
0.0105925372517729	0.74383921295503\\
0.0112201845430196	0.74396392924801\\
0.0118850222743702	0.744102731623401\\
0.0125892541179417	0.744257061279034\\
0.0133352143216332	0.74442846937419\\
0.0141253754462275	0.744618615637899\\
0.0149623565609443	0.744829263789226\\
0.0158489319246111	0.745062272874144\\
0.0167880401812256	0.745319583468149\\
0.0177827941003892	0.745603197529841\\
0.018836490894898	0.745915150514193\\
0.0199526231496888	0.746257474182378\\
0.0211348903983665	0.74663214836786\\
0.0223872113856834	0.747041039791836\\
0.0237137370566166	0.74748582585402\\
0.0251188643150958	0.747967901159766\\
0.0266072505979881	0.7484882643579\\
0.0281838293126445	0.749047382633237\\
0.0298538261891796	0.749645030874959\\
0.0308390825772287	0.75\\
0.0316227766016838	0.750138036136017\\
0.0334965439157828	0.750470532480721\\
0.0354813389233576	0.750822097272902\\
0.0375837404288444	0.751190694994651\\
0.0398107170553497	0.751573279778795\\
0.0421696503428582	0.751965568300121\\
0.0446683592150963	0.752361773820947\\
0.0473151258961481	0.752754293075174\\
0.0501187233627272	0.753133334032077\\
0.0530884444230988	0.753486467227055\\
0.0562341325190349	0.753798075748564\\
0.0595662143529011	0.75404866859427\\
0.0630957344480193	0.754214008607163\\
0.0668343917568615	0.754263989846328\\
0.0707945784384138	0.754161181805802\\
0.0749894209332456	0.753858944150609\\
0.0794328234724282	0.753299015206781\\
0.0841395141645195	0.752408504178169\\
0.0891250938133746	0.751096275142282\\
0.0922950363617174	0.75\\
0.0944060876285924	0.748847031155683\\
0.1	0.745243987016763\\
0.105925372517729	0.740954334066126\\
0.112201845430196	0.736014807628641\\
0.118850222743702	0.730486784127518\\
0.125892541179417	0.72443208175793\\
0.133352143216332	0.717875756991995\\
0.141253754462275	0.710753446617101\\
0.149623565609443	0.702816415553824\\
0.152519740300876	0.7\\
0.158489319246111	0.692181683634321\\
0.167880401812256	0.681639262796306\\
0.177827941003892	0.671907160870526\\
0.18836490894898	0.662399494064518\\
0.199526231496888	0.65195561554003\\
0.201553493492258	0.65\\
0.211348903983665	0.639353021966358\\
0.223872113856834	0.628264911370446\\
0.237137370566166	0.618189287597086\\
0.251188643150958	0.6082564381537\\
0.258626430923231	0.6\\
0.266072505979881	0.593633626662123\\
0.281838293126445	0.58318731301039\\
0.298538261891796	0.577882137143515\\
0.310035486772539	0.567502693123922\\
};

\addplot[contour prepared, contour/labels=false, line width=1.0pt] table[row sep=crcr, x expr=10000*\thisrow{x}^2, meta expr=4] {%
x y\\
0.0001		0.802269422433585\\
0.001	0.802269422433585\\
0.00105925372517729	0.802354843374189\\
0.00112201845430196	0.802450369181509\\
0.00118850222743702	0.802557152673504\\
0.00125892541179417	0.802676467464439\\
0.00133352143216332	0.802809717913379\\
0.00141253754462275	0.802958449185525\\
0.00149623565609443	0.803124357210464\\
0.00158489319246111	0.803309298253598\\
0.00167880401812256	0.803515297736485\\
0.00177827941003892	0.803744557859276\\
0.0018836490894898	0.803999463472866\\
0.00199526231496888	0.804282585542776\\
0.00211348903983665	0.80459668143547\\
0.00223872113856834	0.804944691142711\\
0.00237137370566166	0.805329728467001\\
0.00251188643150958	0.805755066104467\\
0.00266072505979881	0.806224113530419\\
0.00281838293126445	0.806740386603469\\
0.00298538261891796	0.80730746789882\\
0.00316227766016838	0.807928956964139\\
0.00334965439157828	0.808608409988407\\
0.00354813389233575	0.80934926878071\\
0.00375837404288444	0.810154779476091\\
0.00398107170553497	0.81102790198234\\
0.00421696503428582	0.811971211817011\\
0.00446683592150963	0.812986796575667\\
0.0047315125896148	0.814076149737477\\
0.00501187233627272	0.815240064746886\\
0.00530884444230988	0.816478532214305\\
0.00562341325190349	0.817790642584686\\
0.0059566214352901	0.81917449568201\\
0.00630957344480193	0.820627117152958\\
0.00668343917568615	0.822144380018172\\
0.00707945784384138	0.823720927304292\\
0.00740718573330193	0.825\\
0.00749894209332456	0.825249601318489\\
0.00794328234724281	0.826442719300992\\
0.00841395141645195	0.827658523160108\\
0.00891250938133746	0.828885903886982\\
0.00944060876285924	0.830111932836866\\
0.01	0.831321791435061\\
0.0105925372517729	0.832498697932383\\
0.0112201845430196	0.833623825901543\\
0.0118850222743702	0.834676213047207\\
0.0125892541179417	0.835632665929072\\
0.0133352143216332	0.836467676049641\\
0.0141253754462275	0.837153374072769\\
0.0149623565609443	0.83765955885124\\
0.0158489319246111	0.837953841312137\\
0.0167880401812256	0.838001931440581\\
0.0177827941003892	0.837768055751909\\
0.018836490894898	0.837215402048807\\
0.0199526231496888	0.836306320812428\\
0.0211348903983665	0.835001741702387\\
0.0223872113856834	0.833258883990914\\
0.0237137370566166	0.831025907286843\\
0.0251188643150958	0.828231832476109\\
0.0265093175417507	0.825\\
0.0266072505979881	0.824702871418843\\
0.0281838293126445	0.819564516141309\\
0.0298538261891796	0.814074644694881\\
0.0316227766016838	0.808163975968364\\
0.0334965439157828	0.801585011440993\\
0.033939945399342	0.8\\
0.0354813389233576	0.793041348008963\\
0.0375837404288444	0.784584881775935\\
0.0398107170553497	0.775955729743555\\
0.0400676527332844	0.775\\
0.0421696503428582	0.764432223348427\\
0.0446683592150963	0.75335744395849\\
0.0454726738629093	0.75\\
0.0473151258961481	0.739415986545967\\
0.0501187233627272	0.725242058939214\\
0.0501703918176429	0.725\\
0.0530884444230988	0.709382757996925\\
0.0550612034416894	0.7\\
0.0562341325190349	0.694219019126626\\
0.0595662143529011	0.679529064087103\\
0.0606616130125771	0.675\\
0.0630957344480193	0.665743860034161\\
0.0668343917568615	0.652318412978899\\
0.0674899011122136	0.65\\
0.0707945784384138	0.640599161042394\\
0.0749894209332456	0.628358151393731\\
0.0760742617660571	0.625\\
0.0794328234724282	0.618388623921333\\
0.0841395141645195	0.607857481303545\\
0.0868911281221207	0.6\\
};

\addplot[contour prepared, contour/labels=false, line width=1.0pt] table[row sep=crcr, x expr=100000*\thisrow{x}^2, meta expr=5] {%
x y\\
0.0001	0.817737549374314\\
0.000112201845430196	0.817973753185207\\
0.000125892541179417	0.818268817334385\\
0.000141253754462275	0.818636698035321\\
0.000158489319246111	0.81909427129992\\
0.000177827941003892	0.819661736423699\\
0.000199526231496888	0.820362964988992\\
0.000223872113856834	0.821225727962307\\
0.000251188643150958	0.822281713545013\\
0.000281838293126446	0.82356623907168\\
0.000313575544942886	0.825\\
0.000316227766016838	0.825093331248465\\
0.000312693962693371	0.825\\
0.000316227766016838	0.825127480467855\\
0.000334965439157828	0.825832911151058\\
0.000354813389233575	0.826605892066505\\
0.000375837404288444	0.827450875765658\\
0.000398107170553497	0.828372209588441\\
0.000421696503428582	0.829374052140176\\
0.000446683592150963	0.83046028400876\\
0.00047315125896148	0.831634416231611\\
0.000501187233627273	0.832899501150187\\
0.000530884444230988	0.834258051431866\\
0.000562341325190349	0.835711974131681\\
0.00059566214352901	0.837262527636824\\
0.000630957344480193	0.838910310099642\\
0.000668343917568615	0.840655288390769\\
0.000707945784384138	0.842496876562516\\
0.000749894209332456	0.844434072084316\\
0.000794328234724281	0.846465656586454\\
0.000841395141645195	0.848590465456886\\
0.000872765908269114	0.85\\
0.000891250938133746	0.850638825173102\\
0.000944060876285924	0.852463149939196\\
0.001	0.854353673124838\\
0.00105925372517729	0.856305998566061\\
0.00112201845430196	0.858315905409423\\
0.00118850222743702	0.860379559800142\\
0.00125892541179417	0.862493694453524\\
0.00133352143216332	0.864655745232953\\
0.00141253754462276	0.866863938519036\\
0.00149623565609443	0.869117327073651\\
0.00158489319246111	0.871415774373529\\
0.00167880401812256	0.873759887523986\\
0.00172942864657933	0.875\\
0.00177827941003892	0.875869115376232\\
0.0018836490894898	0.87770234864145\\
0.00199526231496888	0.879554532781305\\
0.00211348903983665	0.881417218845863\\
0.00223872113856834	0.883279702760722\\
0.00237137370566166	0.885128393812694\\
0.00251188643150958	0.886946148117212\\
0.00266072505979881	0.888711618814577\\
0.00281838293126445	0.890398689832534\\
0.00298538261891796	0.891976072193766\\
0.00316227766016838	0.893407149957748\\
0.00334965439157828	0.894650163254407\\
0.00354813389233575	0.895658794537522\\
0.00375837404288444	0.896383143167549\\
0.00398107170553497	0.896770851376735\\
0.00421696503428582	0.89676763799459\\
0.00446683592150963	0.896315495931165\\
0.0047315125896148	0.895345071340142\\
0.00501187233627272	0.893756024305472\\
0.00530884444230988	0.89137495306005\\
0.00562341325190349	0.887870686727258\\
0.00595662143529011	0.88253959586005\\
0.00623905475454748	0.875\\
0.00630957344480193	0.872853694053947\\
0.00668343917568615	0.858843515116428\\
0.00684217703546802	0.85\\
0.00707945784384138	0.837148788087714\\
0.00736208263368695	0.825\\
};

\addplot [scatter, scatter src=-1, mark=*,scatter/use mapped color={draw=mapped color, fill=mapped color},mark size=1pt] table[row sep=crcr] {0.022316541117218   0.462792572759965\\};
\addplot [scatter, scatter src=0, mark=*,scatter/use mapped color={draw=mapped color, fill=mapped color},mark size=1pt] table[row sep=crcr] 
{0.180435348721748   0.471936472963784\\};
\addplot [scatter, scatter src=1, mark=*,scatter/use mapped color={draw=mapped color, fill=mapped color},mark size=1pt] table[row sep=crcr] 
{1.244291914298851   0.528617698456151\\};
\addplot [scatter, scatter src=2, mark=*,scatter/use mapped color={draw=mapped color, fill=mapped color},mark size=1pt] table[row sep=crcr] {5.851561000000000   0.645660040878889\\};
\addplot [scatter, scatter src=3, mark=*,scatter/use mapped color={draw=mapped color, fill=mapped color},mark size=1pt] table[row sep=crcr] {2.263856399999999   0.752787488433247\\};
\addplot [scatter, scatter src=4, mark=*,scatter/use mapped color={draw=mapped color, fill=mapped color},mark size=1pt] table[row sep=crcr] 
{0.183869440000000   0.812313411573168\\};
\addplot [scatter, scatter src=5, mark=*,scatter/use mapped color={draw=mapped color, fill=mapped color},mark size=1pt] table[row sep=crcr] {0.014639806440000   0.827700000000000\\};

\end{axis}

%% file: texfigures/Nonlinear/WeLa_nonlin_paper3.tex
%
%

\begin{axis}[%
at=(refcorner.south west), anchor=south west, xshift=.0\textwidth, yshift=\ydesp, 
width=0.3\textwidth,
height=0.3\textwidth,
scale only axis,
xmode=log,
xmin=0.1,
xmax=100000,
xminorticks=true,
ymode=log,
ymin=0.01,
ymax=10,
yminorticks=true,
axis background/.style={fill=white},
legend style={legend cell align=left, align=left, draw=white!15!black},
xlabel={$\La$},
ylabel={$\We$},,
xtick={1e-4,1e-3,1e-2,1e-1,1e0,1e1,1e2,1e3,1e4,1e5},
ytick={1e-4,1e-3,1e-2,1e-1,1e0,1e1,1e2,1e3,1e4,1e5},
legend style={at={(0.02,0.02)},anchor=south west,draw=none,fill=none}
]

\node[anchor=north west,xshift=0.0cm,yshift=0.0cm] at (rel axis cs:0,1) {(c)};

\node[anchor=north west,xshift=0.0cm,yshift=0.0cm] at (rel axis cs:.7,.8) {Nonlinear};
\node[anchor=north west,xshift=0.0cm,yshift=0.0cm] at (rel axis cs:.45,.5) {Linear};


\addplot [mark=*, mark size=1pt, only marks, blue,forget plot]
  table[row sep=crcr]{%
0.1	0.0223165411172176\\
1	0.180435348721748\\
10	1.24429191429885\\
100	5.851561\\
1000	2.2638564\\
10000	0.18386944\\
100000	0.01463980644\\
};
\addplot [mark=*, mark size=1pt, only marks, red,forget plot]
  table[row sep=crcr]{%
10	1.36528683759948\\
100	0.973072482165624\\
1000	0.409351533863862\\
10000	0.147478309651488\\
100000	0.0264069033531711\\
};

\addplot [color=red,[-,forget plot]
  table[row sep=crcr]{%
9.7656	1.3634427749541\\
11.7907603009749	1.37366660612919\\
};

\addplot [color=red,-]
  table[row sep=crcr]{%
9.7656	1.3634427749541\\
11.7907603009749	1.37366660612919\\
14.2358921597286	1.37394073156729\\
17.1880879951962	1.36470898031208\\
20.7525011861455	1.34660633405761\\
25.0560914978637	1.32042266292836\\
30.252147223991	1.28706405025802\\
36.5257451163145	1.24751387330573\\
44.1003425781281	1.20279552097369\\
53.2457369265162	1.15393826454974\\
64.2876752221389	1.10194739379827\\
77.6194569561688	1.0477793245828\\
93.7159428670051	0.992322005108778\\
113.150468862611	0.93638061731874\\
136.615267500407	0.880668301260489\\
164.946124411287	0.825801429129732\\
199.152147897545	0.772298821561494\\
240.45171205909	0.720584226160771\\
290.315853695397	0.670991358516605\\
350.520668724433	0.623770828553496\\
423.210574410913	0.579098329050263\\
510.974690722226	0.537083537984592\\
616.939061416661	0.497779272687534\\
744.877999659241	0.461190523827412\\
899.348524151283	0.427283084929534\\
1085.8526742139	0.395987421883486\\
1311.03348527778	0.367089868589498\\
1582.91160517142	0.340262709499677\\
1911.17097917257	0.315226824965036\\
2307.50377955305	0.29175260177086\\
2786.02686555907	0.269652822079694\\
3363.78460759025	0.248776737497632\\
4061.35598552117	0.229005037906707\\
4903.5875852185	0.210245485446668\\
5920.47860163714	0.192429033985241\\
7148.24937115533	0.175506296648662\\
8630.63149288186	0.15944426061762\\
10420.4254913794	0.144223180788938\\
12581.3815027262	0.129833612795102\\
15190.4699715086	0.116273571486918\\
18340.6232380207	0.103545823146953\\
22144.0456674441	0.0916553379620995\\
26736.2102235065	0.080606943041595\\
32280.683839379	0.0704032249078944\\
38974.953459252	0.0610427334241741\\
47057.4602666187	0.0525185363234907\\
56816.0926493357	0.044817165006163\\
68598.4404098365	0.037917978646426\\
82824.1754621416	0.0317929559272591\\
100000	0.0264069033531711\\
};

\addlegendentry{$\lambda=0.3$}

\addplot [color=blue]
  table[row sep=crcr]{%
0.1	0.0223165411172176\\
0.132571136559011	0.0295557578878893\\
0.175751062485479	0.0387990332662955\\
0.232995181051537	0.0505305533987523\\
0.308884359647748	0.0653480011576391\\
0.409491506238043	0.0839936440154265\\
0.542867543932386	0.107395765586457\\
0.719685673001152	0.136724162299901\\
0.954095476349994	0.173464864016704\\
1.2648552168553	0.219521298465897\\
1.67683293681101	0.277352099594689\\
2.22299648252619	0.350160104436451\\
2.94705170255181	0.442153468972103\\
3.90693993705462	0.558909300607673\\
5.17947467923121	0.707884366352657\\
6.866488450043	0.899138818815203\\
9.10298177991522	1.14637148715484\\
12.0679264063933	1.46784544418966\\
15.9985871960606	1.8799806452718\\
21.2095088792019	2.39189188803944\\
28.1176869797423	3.00206974058102\\
37.2759372031494	3.69119523004165\\
49.4171336132383	4.41525923042756\\
65.5128556859551	5.10228008376705\\
86.8511373751353	5.6567531074423\\
115.139539932645	5.97636352771496\\
152.641796717523	6.00696658649881\\
202.358964772516	5.76385420754209\\
268.269579527972	5.29904395276913\\
355.648030622313	4.68485358511147\\
471.486636345739	3.9975664507927\\
625.055192527398	3.3043324214841\\
828.642772854684	2.65550635247195\\
1098.54114198756	2.08243319593525\\
1456.34847750124	1.59874861443239\\
1930.69772888325	1.20490550742302\\
2559.54792269953	0.893812861560835\\
3393.22177189533	0.654362055626203\\
4498.43266896944	0.474050502118332\\
5963.62331659465	0.34074006715571\\
7906.0432109077	0.243652563832607\\
10481.1313415469	0.173790106791917\\
13894.9549437314	0.123977391488305\\
18420.6996932672	0.0886911763272322\\
24420.5309454865	0.0637963611979013\\
32374.5754281765	0.0462643079739761\\
42919.3426012878	0.0339147013991515\\
56898.6602901829	0.0251987377446402\\
75431.2006335462	0.0190272284933888\\
100000	0.01463980644\\
};
\addlegendentry{$\lambda=3$}


\end{axis}

%% file: texfigures/Nonlinear/ReCa_nonlin_paper3.tex
%
%

\begin{axis}[%
at=(refcorner.south west), anchor=south west, xshift=.45\textwidth, yshift=\ydesp, 
width=.3\textwidth,
height=.3\textwidth,
scale only axis,
xmode=log,
xmin=0.01,
xmax=1000,
xminorticks=true,
xlabel style={font=\color{white!15!black}},
xlabel={$\Re$},
ymode=log,
ymin=0.0001,
ymax=10,
yminorticks=true,
ylabel style={font=\color{white!15!black}},
ylabel={$\Ca$},
axis background/.style={fill=white},
legend style={legend cell align=left, align=left, draw=white!15!black},
xtick={1e-4,1e-3,1e-2,1e-1,1e0,1e1,1e2,1e3,1e4,1e5},
ytick={1e-4,1e-3,1e-2,1e-1,1e0,1e1,1e2,1e3,1e4,1e5},
legend style={at={(0.02,0.02)},anchor=south west,draw=none,fill=none},
clip=false
]

\node[anchor=north west,xshift=0.0cm,yshift=0.0cm] at (rel axis cs:0,1) {(d)};

\node[anchor=north west,xshift=0.0cm,yshift=0.0cm] at (rel axis cs:.7,.8) {Nonlinear};
\node[anchor=north west,xshift=0.0cm,yshift=0.0cm] at (rel axis cs:.4,.45) {Linear};

\node[anchor=north west,xshift=0.0cm,yshift=0.0cm] at (rel axis cs:.64,0) {$\Re_*$};
\node[anchor=north east,xshift=0.0cm,yshift=0.0cm] at (rel axis cs:0,.77) {$\Ca_*$};

\addplot [mark=*, mark size=1pt, only marks, blue,forget plot]
  table[row sep=crcr]{%
0.0472403864476336	0.472403864476336\\
0.424776822251106	0.424776822251106\\
3.52745221696744	0.352745221696744\\
24.19	0.2419\\
47.58	0.04758\\
42.88	0.004288\\
38.262	0.00038262\\
};

\addplot [color=red,-,smooth]
  table[row sep=crcr]{%
3.64895009051807	0.373653445821871\\
4.02449669975303	0.341326309501879\\
4.42258658348829	0.310664518518845\\
4.84321567158003	0.281777453835101\\
5.28634557561285	0.254732936921457\\
5.75192411790922	0.22956190587026\\
6.23990794283952	0.20626330741545\\
6.75028693948867	0.184809013970631\\
7.28311022341266	0.165148608778036\\
7.8385134600673	0.147213916315688\\
8.41674736251394	0.130923187585036\\
9.01820725998348	0.116184879586004\\
9.64346371053711	0.102900994382807\\
10.2932942192224	0.0909699652391245\\
10.9687162218662	0.0802890952274681\\
11.6710216034523	0.070756567607197\\
12.401813138921	0.062273057407853\\
13.1630433738985	0.0547429804561495\\
13.957056604814	0.0480754200198034\\
14.7866347745308	0.0421847728076646\\
15.6550482745267	0.036991155753416\\
16.5661128425976	0.0324206132776998\\
17.524253972267	0.0284051619815197\\
18.5345805145506	0.0248827063264449\\
19.6029694645016	0.0217968550990852\\
20.7360555797686	0.0190965644531661\\
21.9378009341651	0.0167332117604279\\
23.2078821022969	0.0146615149111776\\
24.5449049647761	0.0128428619062658\\
25.9464878409525	0.0112443966813256\\
27.4091226909554	0.00983806833659347\\
28.9280376161442	0.00859984838234565\\
30.4970651279251	0.0075090844625903\\
32.108521490167	0.00654796532786644\\
33.7531032357519	0.00570107680592215\\
35.4198076597347	0.00495503245908865\\
37.0958846375941	0.00429816574467222\\
38.7668274371385	0.00372027298397837\\
40.4164102123434	0.00321239843204703\\
42.0267795120217	0.002766654329382\\
43.5786063362401	0.00237607009154958\\
45.0513039711145	0.00203446581747925\\
46.423314989715	0.00173634612391324\\
47.6724684123282	0.00147681098236751\\
48.7764050974722	0.00125148077850734\\
49.713066052132	0.00105643325777607\\
50.4612346189699	0.000888150386025532\\
51.0011195822043	0.000743473456211275\\
51.3149623421809	0.000619565025016598\\
51.387647692	0.00051387647692\\
51.387647692	0.0001\\
};
\addlegendentry{$\lambda=0.3$}

\addplot [color=red,[-,smooth,forget plot]
  table[row sep=crcr]{%
3.64895009051807	0.373653445821871\\
4.02449669975303	0.341326309501879\\
};


\addplot [color=blue,smooth]
  table[row sep=crcr]{%
0.01	0.472403864476335\\
0.0472403864476336	0.472403864476335\\
0.0625958498229749	0.472168010682415\\
0.0825770629167742	0.46985242506626\\
0.108505186225252	0.465697126161813\\
0.142073838168178	0.459958019014621\\
0.185458037847646	0.452898375234765\\
0.241457398918853	0.44478142342016\\
0.313685225600957	0.435864207623951\\
0.406819421935585	0.426392779359904\\
0.526937054661596	0.416598712358301\\
0.681962708433578	0.406696870906252\\
0.882272452524296	0.396884322336709\\
1.1415117754642	0.387340260938001\\
1.47770939894669	0.378226802242757\\
1.91480264031237	0.369690510891074\\
2.48473868130633	0.361864539550892\\
3.23038678188616	0.354871278443473\\
4.20878257889861	0.348757726652103\\
5.48425330198078	0.342796100353861\\
7.12255938813174	0.3358191568083\\
9.18755991855363	0.32675375912581\\
11.7299941005781	0.314680058522769\\
14.7712374338484	0.298909231552258\\
18.2829138486499	0.279073681908961\\
22.1651853416951	0.255208924276457\\
26.2319604118971	0.227827559735279\\
30.2805906908255	0.198376796801352\\
34.1521239535395	0.168770007258794\\
37.7037437545613	0.140544238451866\\
40.8186103548219	0.114772490890495\\
43.4142736833501	0.0920795423171116\\
45.4465635431906	0.0727080809607043\\
46.9091264813745	0.0566095886165422\\
47.829264484994	0.0435388923153647\\
48.2528248969512	0.0331327464836863\\
48.231818612826	0.0249815483238405\\
47.8305012841195	0.0186870895676269\\
47.1210735643089	0.0138868240073766\\
46.1788291912049	0.0102655374859226\\
45.078214354473	0.0075588634562207\\
43.889949853116	0.00555144320392311\\
42.679233066534	0.00407200632028672\\
41.504942702904	0.00298705126220137\\
40.4197170272944	0.00219425525090493\\
39.4707614930696	0.00161629415761596\\
38.7012574490271	0.00119542131247054\\
38.1522828749546	0.000888929805597952\\
37.8652138337247	0.000665485156251698\\
37.8846497936188	0.000502241108128014\\
38.262	0.00038262\\
38.262	0.0001\\
};
\addlegendentry{$\lambda=3$}

\addplot [mark=*, mark size=1pt, only marks, red,forget plot]
  table[row sep=crcr]{%
3.6949788058925	0.36949788058925\\
9.8644436344156	0.098644436344156\\
20.232437664895	0.020232437664895\\
38.40290479267	0.003840290479267\\
51.387647692	0.00051387647692\\
};

\end{axis}
\end{tikzpicture}%

%% file: texfigures/stability_linear/limited01Lin_paper.tex
%
%

\begin{axis}[%
at=(referenceplot.south west), anchor=south west, xshift=.475\textwidth, yshift=.0\textwidth,  
width=.3\textwidth,
height=.09\textwidth,
scale only axis,
xmode=log,
xmin=0.01,
xmax=100,
xminorticks=true,
xtick={1e-3,1e-2,1e-1,1,1e1,1e2,1e3},
xlabel={$\lambda$},
ymode=log,
ymin=0.01,
ymax=3.16227766016838,
yminorticks=true,
ylabel={$\Oh = \La^{-\frac12}$},
axis background/.style={fill=white},
legend style={legend cell align=left, align=left, draw=white!15!black},
]

\node[anchor=north west,xshift=0.0cm,yshift=0.0cm] at (rel axis cs:0,1) {(d)};

\node[anchor=center,xshift=0.0cm,yshift=0.0cm] at (rel axis cs:0.5,.4) {Unstable};
\node[anchor=east,xshift=0.0cm,yshift=0.0cm] at (rel axis cs:0.99,.925) {Stable};
\node[anchor=center,xshift=0.0cm,yshift=0.0cm] at (rel axis cs:0.25,.85) {Stable};

\addplot [smooth, color=blue]
  table[row sep=crcr]{%
9.92209456607453	3.16227766016838\\
10	2.96243123136991\\
10.3513405426594	2.37137370566166\\
11.2124894238406	1.77827941003892\\
12.9448339207282	1.33352143216332\\
16.7971166653964	1\\
17.7827941003892	0.96571495161023\\
31.6227766016838	0.766957391618201\\
34.7572231416566	0.749894209332456\\
56.2341325190349	0.703560764862999\\
100	0.675298560480116\\
177.827941003892	0.664045898083553\\
316.227766016838	0.66574867179773\\
562.341325190349	0.68122351418567\\
1000	0.716912060105831\\
};

\addplot [smooth,color=blue]
  table[row sep=crcr]{%
0.001	0.170382055357612\\
0.00177827941003892	0.170594885969781\\
0.00316227766016838	0.170972490571752\\
0.00562341325190349	0.171640982370673\\
0.01	0.172820011083756\\
0.0177827941003892	0.174885787175858\\
0.0285638472736824	0.177827941003892\\
0.0316227766016838	0.178958272996281\\
0.0562341325190349	0.189908032412099\\
0.1	0.208327308388645\\
0.176529324846788	0.237137370566166\\
0.177827941003892	0.237823118357529\\
0.316227766016838	0.315498799181285\\
0.31769019761343	0.316227766016838\\
0.445283656680355	0.421696503428582\\
0.545443603131533	0.562341325190349\\
0.562341325190349	0.606047119931129\\
0.631162934668174	0.749894209332456\\
0.691111129137416	1\\
0.728152198589045	1.33352143216332\\
0.750134286795651	1.77827941003892\\
0.762877215562035	2.37137370566166\\
0.770169777347182	3.16227766016838\\
};
\end{axis}

%% file: texfigures/La0Lainf_linear/La0_linear_paper.tex
%

\begin{axis}[%
at=(referenceplot.south west), anchor=south west, xshift=.475\textwidth, yshift=.105\textwidth,  
width=0.3\textwidth,
height=0.09\textwidth,
scale only axis,
xmode=log,
xmin=0.01,
xmax=100,
xminorticks=true,
xlabel style={font=\color{white!15!black}},
ymin=0.1,
ymax=.5,
ylabel style={font=\color{white!15!black}},
ylabel={$d_c$},
axis background/.style={fill=white},
legend style={legend cell align=left, align=left, draw=white!15!black},
legend pos=north east,
xticklabels=\empty,
ytick={0.2,.4},
minor ytick={0,.025,...,2},
]

\node[anchor=north west,xshift=0.0cm,yshift=0.0cm] at (rel axis cs:0,1) {(c)};

\node[anchor=center,xshift=0.0cm,yshift=0.0cm] at (3,.2) {Unstable};
\node[anchor=center,xshift=0.0cm,yshift=0.0cm] at (30,.3) {Stable};

\addplot[blue] table[row sep=crcr] {%
0	228\\
0.724055857222019	0.1\\
0.724024832897343	0.104020100502513\\
0.724208449784016	0.108040201005025\\
0.72457951410995	0.112060301507538\\
0.725114421503913	0.11608040201005\\
0.725792616532628	0.120100502512563\\
0.726596145104347	0.124120603015075\\
0.727509281780306	0.128140703517588\\
0.728518217886098	0.1321608040201\\
0.729610799262077	0.136180904522613\\
0.730776304764986	0.140201005025126\\
0.73168071434272	0.143131158633298\\
0.732046705296443	0.144221105527638\\
0.733494488436783	0.148241206030151\\
0.734971677628417	0.152261306532663\\
0.736379308747477	0.156281407035176\\
0.73774360141303	0.160301507537688\\
0.739101630981156	0.164321608040201\\
0.740485743058068	0.168341708542714\\
0.741924131704325	0.172361809045226\\
0.743441321290044	0.176381909547739\\
0.745058571528293	0.180402010050251\\
0.746794220956946	0.184422110552764\\
0.748663980879288	0.188442211055276\\
0.750681189256462	0.192462311557789\\
0.75285703208605	0.196482412060302\\
0.755198527417729	0.200502512562814\\
0.757549510573443	0.204522613065327\\
0.759842552144322	0.208542713567839\\
0.762119006956982	0.212562814070352\\
0.764415618105329	0.216582914572864\\
0.766764979558591	0.220603015075377\\
0.769195928892487	0.224623115577889\\
0.771733883562684	0.228643216080402\\
0.774401131493908	0.232663316582915\\
0.777217084662851	0.236683417085427\\
0.780198502681	0.24070351758794\\
0.783359692039778	0.244723618090452\\
0.784282206133768	0.245809352104992\\
0.78705469834612	0.248743718592965\\
0.791025896560499	0.252763819095477\\
0.794924222184132	0.25678391959799\\
0.798784451905063	0.260804020100503\\
0.802653805844156	0.264824120603015\\
0.806574482363536	0.268844221105528\\
0.810584077885813	0.27286432160804\\
0.814715947450187	0.276884422110553\\
0.818999517632827	0.280904522613065\\
0.823460561296237	0.284924623115578\\
0.828121441876483	0.28894472361809\\
0.833001333479601	0.292964824120603\\
0.83811642187285	0.296984924623116\\
0.840665288561832	0.298876097512447\\
0.843913146828297	0.301005025125628\\
0.850087666264024	0.305025125628141\\
0.856167732077582	0.309045226130653\\
0.862221235776327	0.313065326633166\\
0.868309084635343	0.317085427135678\\
0.874485752835008	0.321105527638191\\
0.880799751856328	0.325125628140704\\
0.887294036655167	0.329145728643216\\
0.894006361136564	0.333165829145729\\
0.900969593994845	0.337185929648241\\
0.901101825166502	0.337257694451823\\
0.909518612865543	0.341206030150754\\
0.918449485899808	0.345226130653266\\
0.927761251562034	0.349246231155779\\
0.937252506763983	0.353266331658291\\
0.946555682588116	0.357286432160804\\
0.95576114802018	0.361306532663317\\
0.964961120264695	0.365326633165829\\
0.96588322411587	0.36571818018514\\
0.975990657911231	0.369346733668342\\
0.987393969084934	0.373366834170854\\
0.999056372046701	0.377386934673367\\
1.01104715590127	0.381407035175879\\
1.02342661945459	0.385427135678392\\
1.03532184329566	0.389151076834371\\
1.0364217971193	0.389447236180904\\
1.05223951186243	0.393467336683417\\
1.06866123877766	0.39748743718593\\
1.085633905805	0.401507537688442\\
1.10234401142458	0.405527638190955\\
1.10975249641207	0.407320958393379\\
1.12047117186077	0.409547738693467\\
1.13985036695374	0.41356783919598\\
1.15924158715355	0.417587939698493\\
1.17881391434949	0.421608040201005\\
1.18953406737032	0.423749876761683\\
1.20062123158765	0.425628140703518\\
1.22519233232122	0.42964824120603\\
1.25043109418015	0.433668341708543\\
1.27505124071301	0.437467222265799\\
1.27677040564269	0.437688442211055\\
1.30980025509504	0.441708542713568\\
1.3439696675226	0.44572864321608\\
1.36671635646201	0.448289944236063\\
1.38261747505684	0.449748743718593\\
1.42721910798589	0.453768844221106\\
1.46497139830729	0.457273592115478\\
1.47203441043258	0.457788944723618\\
1.52723786191194	0.461809045226131\\
1.57029012472938	0.464953541848648\\
1.58643481223677	0.465829145728643\\
1.66224738209234	0.469849246231156\\
1.68318035333096	0.470925877812579\\
1.76869307645787	0.473869346733668\\
1.80418640939207	0.475036425226474\\
1.93389175045523	0.477042575844673\\
2.07292177959537	0.477055272716883\\
2.22194686093952	0.475310514031613\\
2.2938866381665	0.473869346733668\\
2.38168555197616	0.472021887850385\\
2.46320441408005	0.469849246231156\\
2.55290806823952	0.467379949215994\\
2.60332729321266	0.465829145728643\\
2.73192050338893	0.461809045226131\\
2.73643999707467	0.461665639727851\\
2.85418555865872	0.457788944723618\\
2.93316627839004	0.45522232482724\\
2.97856183796817	0.453768844221106\\
3.10673413915381	0.449748743718593\\
3.1440354715915	0.448560114721912\\
3.23070252121237	0.44572864321608\\
3.34933334966733	0.441708542713568\\
3.37006432927193	0.44096948114963\\
3.45537795822305	0.437688442211055\\
3.55667665482829	0.433668341708543\\
3.61234269970943	0.431378298935945\\
3.65340434548459	0.42964824120603\\
3.7454315107311	0.425628140703518\\
3.836016032798	0.421608040201005\\
3.87203878181256	0.419967315263732\\
3.92502330393417	0.417587939698493\\
4.01336442172351	0.41356783919598\\
4.1022854831389	0.409547738693467\\
4.15040475785048	0.407380529638878\\
4.1934416333339	0.405527638190955\\
4.28804306009529	0.401507537688442\\
4.38500604106199	0.39748743718593\\
4.44878283112758	0.394752712921315\\
4.48099526870894	0.393467336683417\\
4.57699665869992	0.389447236180904\\
4.6703132389158	0.385427135678392\\
4.76118230463044	0.381407035175879\\
4.76861169771447	0.38106202320976\\
4.85572227769865	0.377386934673367\\
4.94894946064964	0.373366834170854\\
5.04084320326991	0.369346733668342\\
5.11143348344017	0.366215284470917\\
5.13403821820143	0.365326633165829\\
5.2341971507008	0.361306532663317\\
5.3353007834681	0.357286432160804\\
5.43841051532823	0.353266331658291\\
5.47890117959394	0.351707047490736\\
5.55150062468182	0.349246231155779\\
5.6676893918319	0.345226130653266\\
5.78009133622048	0.341206030150754\\
5.87278661318948	0.337766943223353\\
5.88956196235189	0.337185929648241\\
5.99946108014829	0.333165829145729\\
6.10625570142965	0.329145728643216\\
6.2103892785624	0.325125628140704\\
6.29498899022189	0.321776530192268\\
6.31325455925933	0.321105527638191\\
6.41872179204893	0.317085427135678\\
6.52349278355038	0.313065326633166\\
6.62851232700562	0.309045226130653\\
6.73487432383374	0.305025125628141\\
6.74754405311069	0.304545944036368\\
6.84982557579608	0.301005025125628\\
6.96712861138827	0.296984924623116\\
7.08013068685349	0.292964824120603\\
7.18871740741134	0.28894472361809\\
7.23263389648354	0.287226542087562\\
7.29769168027978	0.284924623115578\\
7.4058731613055	0.280904522613065\\
7.51042583300554	0.276884422110553\\
7.61191736435983	0.27286432160804\\
7.71103035360329	0.268844221105528\\
7.75259748862946	0.267103359528676\\
7.81351655051865	0.264824120603015\\
7.91900164991646	0.260804020100503\\
8.0250044248735	0.25678391959799\\
8.13289203514827	0.252763819095477\\
8.24371444145999	0.248743718592965\\
8.30994194935339	0.246238214969845\\
8.35538455359942	0.244723618090452\\
8.46772199125612	0.24070351758794\\
8.57427605311702	0.236683417085427\\
8.67545574090363	0.232663316582915\\
8.77177900321295	0.228643216080402\\
8.86388174140686	0.224623115577889\\
8.90735463861044	0.222621758040542\\
8.95782102520019	0.220603015075377\\
9.05400807039251	0.216582914572864\\
9.14852997432574	0.212562814070352\\
9.24275377558065	0.208542713567839\\
9.33825580899446	0.204522613065327\\
9.43684764048214	0.200502512562814\\
9.53565166094207	0.196482412060302\\
9.54771611420806	0.195941467842281\\
9.63838639320454	0.192462311557789\\
9.73545819276207	0.188442211055276\\
9.82572709890461	0.184422110552764\\
9.90976435594244	0.180402010050251\\
9.98827749192552	0.176381909547739\\
10.0621242175238	0.172361809045226\\
10.1323285090873	0.168341708542714\\
10.2000995639998	0.164321608040201\\
10.2341140210545	0.16223824026737\\
10.2705120718096	0.160301507537688\\
10.3454378454843	0.156281407035176\\
10.4232163487066	0.152261306532663\\
10.5053956759989	0.148241206030151\\
10.5863134909339	0.144221105527638\\
10.6640792843235	0.140201005025126\\
10.7381060880928	0.136180904522613\\
10.80772801898	0.1321608040201\\
10.8721893337994	0.128140703517588\\
10.9306318501072	0.124120603015075\\
10.9698579789238	0.121025108045097\\
10.9833708600277	0.120100502512563\\
11.0312930625238	0.11608040201005\\
11.0688611141386	0.112060301507538\\
11.0944855899068	0.108040201005025\\
11.1063561554998	0.104020100502513\\
11.1024093009652	0.1\\
};

\end{axis}


%% file: texfigures/La0Lainf_linear/Linf_linear_paper.tex
%

\begin{axis}[%
at=(referenceplot.south west), anchor=south west, xshift=.475\textwidth, yshift=.21\textwidth, 
width=0.3\textwidth,
height=0.09\textwidth,
scale only axis,
xmode=log,
xmin=0.01,
xmax=100,
xminorticks=true,
xlabel style={font=\color{white!15!black}},
ymin=0.7,
ymax=0.9,
ylabel style={font=\color{white!15!black}},
ylabel={$d_c$},
axis background/.style={fill=white},
legend style={legend cell align=left, align=left, draw=white!15!black},
legend pos=north east,
ytick={0,.2,...,2},
minor ytick={0,.1,...,2},
xticklabels=\empty,
ytick={0,.1,...,2},
minor ytick={0,.025,...,2},
]

\node[anchor=north west,xshift=0.0cm,yshift=0.0cm] at (rel axis cs:0,1.05) {(b)};

\node[anchor=east,xshift=0.0cm,yshift=0.0cm] at (1,.76) {Unstable};
\node[anchor=west,xshift=0.0cm,yshift=0.0cm] at (10,.8) {Stable};

\addplot[blue] table[row sep=crcr] {%
0	230\\
0.001	0.846779886996074\\
0.00107189131920513	0.846781063473459\\
0.00114895100018731	0.846782404129315\\
0.00123155060329283	0.84678390901672\\
0.00132008840083142	0.846785578195317\\
0.00141499129743458	0.84678741173132\\
0.00151671688847092	0.846789409697513\\
0.00162575566644379	0.846791572173267\\
0.00174263338600965	0.84679389924454\\
0.00186791359902078	0.846796362913818\\
0.00200220037181558	0.846798905677072\\
0.00214614119785841	0.846801566504072\\
0.00230043011977292	0.84680438979552\\
0.0024658110758226	0.846807419998541\\
0.00264308148697411	0.84681070161346\\
0.00283309610183932	0.846814279202245\\
0.00303677111803546	0.846818197398607\\
0.00325508859983506	0.84682247269243\\
0.00348910121340677	0.846826918318695\\
0.0037399373024788	0.846831552210054\\
0.00400880632889846	0.846836449451149\\
0.00429700470432084	0.846841685263859\\
0.0046059220411451	0.846847335026508\\
0.004937047852839	0.846853474298046\\
0.00529197873595844	0.846860178847233\\
0.00567242606849198	0.846867520474095\\
0.00608022426164943	0.846875256279855\\
0.00651733960488242	0.846883295341908\\
0.00698587974678525	0.846891759208433\\
0.00748810385759002	0.846900769823964\\
0.00802643352225717	0.84691044958221\\
0.00860346441668451	0.84692092139304\\
0.00922197882333432	0.846932308763818\\
0.00988495904662559	0.84694473589529\\
0.0105956017927762	0.846958045788562\\
0.0113573335834311	0.84697187516115\\
0.0121738272773966	0.846986392392098\\
0.013049019780144	0.84700177847931\\
0.0139871310264724	0.847018215652281\\
0.0149926843278605	0.847035887546993\\
0.0160705281826164	0.84705497941828\\
0.0172258596539879	0.847075678390589\\
0.0184642494289554	0.847098016922539\\
0.0197916686785356	0.847121394783576\\
0.0212145178491063	0.847145913330414\\
0.0227396575235793	0.847171789650721\\
0.0243744415012222	0.847199243906824\\
0.0261267522556333	0.847228499743049\\
0.0280050389418363	0.847259784772162\\
0.0300183581357559	0.847293331144589\\
0.0321764175025074	0.847329350712126\\
0.0344896226040576	0.847367485545526\\
0.0369691270719503	0.847407614911381\\
0.0396268863870148	0.847449838565362\\
0.042475715525369	0.847494262422067\\
0.0455293507486695	0.847540999278306\\
0.0488025158365443	0.847590169613705\\
0.0523109930805626	0.847641902478569\\
0.0560716993820546	0.847696336480725\\
0.0590281234571978	0.847738693467337\\
0.0601027678207038	0.847753095512333\\
0.0644236350872137	0.847810436472462\\
0.0690551352016233	0.84787072720076\\
0.0740195999691564	0.847933490409632\\
0.0793409666579749	0.84799825370031\\
0.0850448934180268	0.848064549460345\\
0.0911588829975082	0.848131914451261\\
0.097712415353465	0.848199889081904\\
0.104737089795945	0.848269611155893\\
0.112266777351081	0.848345064536993\\
0.120337784077759	0.848424601790874\\
0.128989026125331	0.848506120286961\\
0.138262217376466	0.848587485244697\\
0.148202070579886	0.848666522769709\\
0.158856512942805	0.848741010947618\\
0.17027691722259	0.848808668794779\\
0.182518349431904	0.848868557683328\\
0.195639834351706	0.848932266616634\\
0.209704640132323	0.848998946813237\\
0.224780583354873	0.849063503413572\\
0.240940356023953	0.849120651207543\\
0.258261876068268	0.849164884209948\\
0.276828663039207	0.849190442099046\\
0.296730240818887	0.849191273005027\\
0.318062569279412	0.849161112517586\\
0.340928506974681	0.849113145526953\\
0.365438307095726	0.849054569488167\\
0.391710149080926	0.848977557747615\\
0.419870708444391	0.848873819881265\\
0.45005576757005	0.848734552318806\\
0.482410870416537	0.848550398721188\\
0.517092024289676	0.848311422749884\\
0.554266452066311	0.848007096504076\\
0.583172303948993	0.847738693467337\\
0.594113398496503	0.847626604663786\\
0.636824994471859	0.847161818449688\\
0.682607183427239	0.846629819342032\\
0.73168071434272	0.846024127634264\\
0.784282206133768	0.845338220539912\\
0.840665288561832	0.844565707070855\\
0.899865180033446	0.843718592964824\\
0.901101825166502	0.843698934418944\\
0.96588322411587	0.84264905461689\\
1.03532184329566	0.84148354649417\\
1.10975249641207	0.840179681157534\\
1.13674601167369	0.839698492462312\\
1.18953406737032	0.838656343438362\\
1.27505124071301	0.836953671468306\\
1.33935152514486	0.835678391959799\\
1.36671635646201	0.835080416463742\\
1.46497139830729	0.832983706726909\\
1.52898396038437	0.831658291457286\\
1.57029012472938	0.830723804980439\\
1.68318035333096	0.828285682732199\\
1.71494888276599	0.827638190954774\\
1.80418640939207	0.825673652567862\\
1.90103014803876	0.823618090452261\\
1.93389175045523	0.82287356364186\\
2.07292177959537	0.819842490251656\\
2.08472146952159	0.819597989949749\\
2.22194686093952	0.816610848610701\\
2.27246484684943	0.815577889447236\\
2.38168555197616	0.813266175463235\\
2.46865729220377	0.811557788944724\\
2.55290806823952	0.809870809500898\\
2.67986011438404	0.807537688442211\\
2.73643999707467	0.806493761471492\\
2.91405041275985	0.803517587939699\\
2.93316627839004	0.803200464268962\\
3.1440354715915	0.800055700016186\\
3.18556209651232	0.799497487437186\\
3.37006432927193	0.796967590290067\\
3.48956376614648	0.795477386934673\\
3.61234269970943	0.793882892768482\\
3.8190585710609	0.791457286432161\\
3.87203878181256	0.790812779903774\\
4.15040475785048	0.787768975780609\\
4.1842697283367	0.787437185929648\\
4.44878283112758	0.784765017311661\\
4.59868787237469	0.783417085427136\\
4.76861169771447	0.781847740929349\\
5.06814257181045	0.779396984924623\\
5.11143348344017	0.779034544842204\\
5.47890117959394	0.776312414537755\\
5.62250614850656	0.775376884422111\\
5.87278661318948	0.773714506485566\\
6.27943026264805	0.771356783919598\\
6.29498899022189	0.771265167165348\\
6.74754405311069	0.768928973655419\\
7.10130737416559	0.767336683417085\\
7.23263389648354	0.766738709478193\\
7.75259748862946	0.764663762843822\\
8.13850273175486	0.763316582914573\\
8.30994194935339	0.762713257291275\\
8.90735463861044	0.76086322041245\\
9.48316227349538	0.75929648241206\\
9.54771611420806	0.759119985328333\\
10.2341140210545	0.757456729993485\\
10.9698579789238	0.755919368959483\\
11.3272931135244	0.755276381909548\\
11.7584955405216	0.754499175171242\\
12.6038292967973	0.753184409073578\\
13.5099352119803	0.75196722042179\\
14.1182970845976	0.751256281407035\\
14.4811822767453	0.750830367343022\\
15.5222535742705	0.749759195749501\\
16.6381688607613	0.748748762073522\\
17.8343087693191	0.747787256662161\\
18.6204256152688	0.747236180904523\\
19.116440753857	0.746881721425277\\
20.4907468981585	0.746041938090157\\
21.9638537241655	0.745271806289422\\
23.5428641432242	0.744562180267475\\
25.2353917043477	0.743904218192764\\
27.0495973046314	0.743289343186715\\
27.2944626722211	0.74321608040201\\
28.9942285388288	0.742696920529609\\
31.0786618778201	0.742129984087883\\
33.3129478793467	0.741592213439436\\
35.7078596490046	0.741101267257848\\
38.2749447851631	0.740652511288791\\
41.0265810582719	0.740240370319516\\
43.9760360930272	0.739859402403709\\
47.1375313411672	0.739504280662996\\
50.2603751897855	0.739195979899497\\
50.5263106533568	0.739169172051591\\
54.1587137807947	0.738842772136829\\
58.052255160949	0.738529410863023\\
62.2257083673023	0.738241525637873\\
66.6991966303012	0.737979014777766\\
71.4942898659758	0.737738598011171\\
76.6341086800746	0.737517046570569\\
82.1434358491943	0.737311175965356\\
88.0488358164346	0.73711783990547\\
94.3787827777538	0.736933925329772\\
101.163797976621	0.736756533931747\\
108.436596868961	0.73659140111361\\
116.232246867985	0.73644115413532\\
124.588336429501	0.736303939696114\\
133.54515629299	0.736177922598179\\
143.145893752348	0.736061283169679\\
153.436840893001	0.735952215078937\\
164.467617799466	0.735848923529838\\
176.291411809595	0.735749623830267\\
188.965233969121	0.735655547784882\\
202.550193923067	0.735570145287659\\
217.111794569451	0.735492449247995\\
232.720247896041	0.735421439150584\\
249.450813523032	0.735356099648679\\
267.384161583995	0.73529541981261\\
286.606761694825	0.735238392497132\\
307.211299886176	0.735184013826039\\
310.665562674577	0.735175879396985\\
329.297125509715	0.735131056109527\\
352.970730273065	0.73508280573708\\
378.346261713193	0.735039256745462\\
405.546073584083	0.734999857425386\\
434.701315812503	0.734964057707583\\
465.952566866468	0.734931308913955\\
499.450511585514	0.734901063540734\\
535.356667741073	0.734872775073384\\
573.844164830239	0.734845963984944\\
615.09857885805	0.734821335034412\\
659.318827133354	0.734799230843078\\
706.718127392749	0.734779649505922\\
757.525025877192	0.734762589390173\\
811.984499318401	0.734748049134914\\
870.359136148517	0.734736027650756\\
932.930402628468	0.734726524119594\\
1000	0.73471953799442\\
};

\end{axis}

\end{tikzpicture}%

%% file: texfigures/rhorat_def/rhoratNL_mu01blue_1red_paper.tex
%
%
%
\begin{tikzpicture}[baseline]

\begin{axis}[%
width=0.3\textwidth,
height=0.3\textwidth,
scale only axis,
xmode=log,
xmin=1e-1,
xmax=1e2,
xminorticks=true,
ymin=0,
ymax=1,
axis background/.style={fill=white},
legend style={legend cell align=left, align=left, draw=white!15!black},
legend pos=south west,
xlabel=$\Re$,
ytick={0,.2,...,2},
minor ytick={0,.05,...,2},
legend style={draw=none,fill=none},
]

\node[anchor=north west,xshift=0.0cm,yshift=0.0cm] at (rel axis cs:0,1) {(b)};

\node[anchor=center,xshift=0.0cm,yshift=0.0cm] at (1,.7) {Unstable};
\node[anchor=center,xshift=0.0cm,yshift=0.0cm] at (1,.94) {Stable};

\addplot[color=blue,forget plot,mark=*, mark size=1pt, only marks,] table[row sep=crcr, x expr=3162*\thisrowno{0}] {%
0.0025	0.834603777799144\\
0.0025	0.85\\
0.0025	0.862209569274668\\
};

\addplot[smooth,color=blue] table[row sep=crcr, x expr=3162*\thisrowno{0}] {%
0	24\\
1e-06	0.820315306724812\\
1e-05	0.820315306724812\\
1.58489319246111e-05	0.82031592357313\\
2.51188643150958e-05	0.820317471509973\\
3.98107170553497e-05	0.820321358669819\\
6.30957344480193e-05	0.820331118158251\\
0.0001	0.820355606733141\\
0.000158489319246111	0.820416957850793\\
0.000251188643150958	0.820570058568429\\
0.000398107170553497	0.820948392647325\\
0.000630957344480193	0.821860985194145\\
0.001	0.823938787080266\\
0.00158489319246111	0.828102119233403\\
0.00251188643150958	0.834603777799144\\
0.00398107170553497	0.841031752603073\\
0.00630957344480194	0.842542976356697\\
0.01	0.837643629107535\\
0.0158489319246111	0.828526413395378\\
0.0193366297457975	0.8\\
0.0251188643150958	0.732295860525613\\
};
\addlegendentry{$\varphi=0$}

\addplot[smooth, dashed, color=blue] table[row sep=crcr, x expr=3162*\thisrowno{0}] {%
0	24\\
1e-06	0.840502165670685\\
1e-05	0.840502165670685\\
1.58489319246111e-05	0.840502744192476\\
2.51188643150958e-05	0.840504195361629\\
3.98107170553497e-05	0.840507839490738\\
6.30957344480193e-05	0.840516988008642\\
0.0001	0.8405399376332\\
0.000158489319246111	0.840597398568074\\
0.000251188643150958	0.840740577392237\\
0.000398107170553497	0.84109307733939\\
0.000630957344480193	0.841935510388697\\
0.001	0.843810358972058\\
0.00158489319246111	0.847365984411756\\
0.00251188643150958	0.852206576839938\\
0.00398107170553497	0.854921191516813\\
0.00630957344480194	0.843779246194714\\
0.00689048494748536	0.8\\
0.00933101770540307	0.7\\
};
\addlegendentry{$\varphi=1$}

\addplot[smooth,dashdotted,color=blue] table[row sep=crcr, x expr=3162*\thisrowno{0}] {%
0	22\\
1e-06	0.871096961013767\\
1e-05	0.871096961013767\\
1.58489319246111e-05	0.87109672480084\\
2.51188643150958e-05	0.871096131509196\\
3.98107170553497e-05	0.871094639714928\\
6.30957344480193e-05	0.871090881776739\\
0.0001	0.871081383329555\\
0.000158489319246111	0.871057167442683\\
0.000251188643150958	0.87099423858803\\
0.000398107170553497	0.870824967185594\\
0.000630957344480193	0.870354179780422\\
0.001	0.869103356304832\\
0.00158489319246111	0.866453812031566\\
0.00251188643150958	0.862209569274668\\
0.00398107170553497	0.850251076880692\\
0.00437904157546808	0.8\\
0.00630957344480194	0.717906725530321\\
0.00670453103533424	0.7\\
};
\addlegendentry{$\varphi=2$}

\end{axis}
\end{tikzpicture}%

%% file: aaConclusion.tex
\section{Conclusion}
\label{sec:Conclusion}

In this article, we propose a general study  
on the transport of dispersed objects by an external flow in a cylindrical microchannel. By performing both experiments and numerical simulations, we provide a systematic exploration of the influence of all parameters of the system, namely the Reynolds and capillary numbers ($\Re$ and $\Ca$), the dimensionless dispersed object diameter ($d$), as well as the viscosity and density ratios ($\lambda$ and $\varphi$), on the object equilibrium velocity and lateral position. As a result, we highlight all possible object behaviours depending on the problem parameters and the forces involved, for beads, bubbles or drops, dispersed in flows from low to intermediate values of $\Re$, within the object size range $d<1$.

In particular, by varying the nature of the dispersed and continuous phases, the object and microchannel sizes, as well as the flow rates for the fluid phases, we have been able to experimentally characterize the 
equilibrium velocity and lateral position of various dispersed objects such as beads, bubbles and drops, over a very wide range of parameters, i.e. for $\Re=[10^{-2};10^2]$, $\Ca=[10^{-3}; 0.3]$, $d=[0.1; 1]$, $\lambda=[10^{-2}; \infty[$ and  $\varphi=[10^{-3}; 2]$. The experiments were supplemented with numerical simulations solving by finite element methods the steady 3D Navier-Stokes equations for incompressible two-phase fluids including both the effects of inertia and possible interfacial deformations. In addition, two reduced versions of the model were considered in order to easily and specifically compute: (i) the object equilibrium velocity (based on the inertialess and nondeformable limit) and (ii) the stability of its centered position (based on a linear stability analysis of the axisymmetric solution).

Excellently agreeing with the experimental results, the numerical simulations helped us to rationalize them. Although many parameters have been varied, our experiments can be categorized into two types of behaviors depending on the value of their Laplace number ($\La=\Re/\Ca$). For experiments characterized by small values of $\La$ (typically $\La<1$), i.e. when dominated by the capillary effects, the dispersed object remains centered whatever $d$ for these specific values of $\lambda$, meaning that capillary effects promote the stability of the centered position in these conditions. For experiments characterized by large values of $\La$ (typically $\La>10^3$), i.e. when dominated by the inertial effects, only large objects are centered. In these cases, it exists a critical diameter $d_c$ below which the centered position becomes unstable because of the inertially-induced lateral migration leading to off-centered positions. Moreover, it appears that whatever the value of $\La$, most of the experiments took place in the linear regimes ($\We<\We_*$). Interestingly note that in the small nondeformable object limit (i.e. $\La\to\infty$ and $d\to0$), our results recover both the seminal observation of \cite{Segre1962} and the analytical results of \cite{Ho1974} concerning an eccentricity $\varepsilon\approx 0.3$. 

Once validated through such a comparison, the reduced versions of the model enabled us to provide an exhaustive parametric analysis on two aspects: 

\begin{enumerate}

  \item{\textit{The velocity of the dispersed object}, which was analyzed as functions of $d$, $\varepsilon$ and $\lambda$. This velocity $V_0$ decreases with an increase of $\lambda$, indicating that bubbles ($\lambda \to 0$) are faster than drops (intermediate $\lambda$), which are themselves faster than beads ($\lambda \to \infty$). Moreover, $V_0$ is also markedly affected by the object eccentricity $\varepsilon$ depending on the potential lateral migration of the dispersed object. Basically, the velocity of the object is maximal when it is centered inside the Poiseuille flow and slow down with an increase of $\varepsilon$ (for a given size $d$). This could explain why the theoretical predictions for centered objects generally overestimate the equilibrium object velocity when lateral migration occurs. Finally, for a given $\varepsilon$, $V_0$ naturally decreases with an increase of $d$. Usefully, we propose an expression for the equilibrium velocity of a dispersed object $V_0(d,\varepsilon,\lambda)$ [see eq. \eqref{Vdropecc}], derived in nondeformable and inertialess limit (i.e. $\Re=\Ca=0$) but actually valid for a larger range of values of $\Re$ and $\Ca$ in the linear regimes, as confirmed by the comparison with the experimental results.}
  
  \item{\textit{The stability of the centered position}, which was analyzed as functions of $d$, $\lambda$, $\varphi$, $\Re$ and $\Ca$ (or $\La$ and $\We=\Re \cdot \Ca$). In the linear regime (i.e. when the Weber number $\We < \We_*$), the critical diameter $d_c$ below which the centered position is unstable is strongly affected by the values of $\La$ and $\lambda$. This threshold decreases with $\La$, from the nondeformable limit ($\La \to \infty$) for which $d_c$ always exists whatever the value of $\lambda$ and decreases from $\lim{d_c}_{\lambda \to 0}=0.83$ to $\lim{d_c}_{\lambda \to \infty}=0.73$, to the inertialess limit ($\La \to 0$) for which $d_c$ only exits when $0.7\lesssim \lambda \lesssim10$ with $\max(d_c) \lesssim 0.475$. Thus, while only large dispersed objects remain centered when the problem is dominated by inertial effects, small objects can be centered when the problem is dominated by the capillary effects for values of $\lambda$ remaining however outside the aforementioned range. This evidences that, while inertial effects tend to destabilize the centered position of dispersed objects resulting in off-centered positions because of inertia-induced migration forces pushing the dispersed objects outwards when $d<d_c$, capillary effects tend to stabilize the centered position of the dispersed objects because of capillary-induced migration forces pushing the dispersed objects inwards, except when $0.7\lesssim \lambda \lesssim10$ for which it is the opposite. These results are in agreement with the analytical analysis performed by \cite{Ho1974} and \cite{Chan1979}. {Concerning $\varphi$, its effect is negligible in the pure capillary regime and maximum in the pure inertial regime (i.e. $\Ca \to 0$). In this latter regime, the influence of $\varphi$ is highly dependent on the value of $\lambda$: while an increase of $\varphi$, which corresponds to an increase of the inertia within the dispersed object, does not affect the stability of its centered position for large values of $\lambda$ (i.e. when $\lambda \gtrsim 20$), it leads to an important stabilization of its centered position through an strong decrease of $d_c$ for large values of $\lambda$ (i.e. when $\lambda \lesssim 10^{-3}$) and a moderate destabilization of its centered position through a relative increase of $d_c$ for intermediate values of $\lambda$ (i.e. when $10^{-3}\lesssim \lambda \lesssim 20$).} 
  In the non-linear regime, the effects of the various parameters become relatively complex and would need a proper detailed investigation. 
However, it could be observed that, while just above the nonlinear transition ($\Re\gtrsim \Re_*$) the effect of an increase in $\Re$ can have opposite effects on $d_c$ (either decreasing or increasing depending on the conditions), for larger values of $\Re$ stronger nonlinearities result in a stabilization of the centered position since $d_c$ decreases in pursuing increasing $\Re$.}

\end{enumerate}

Besides its fundamental interest, we hope that this study, elucidating the role of all the parameters on the stability of the centered position of objects dispersed in microchannels, will be of primary interest for the design and optimization of multiphase microfluidic devices, in particular for those related to the manipulation and sorting of dispersed micro-objects, which may be beads, bubbles and drops as here, but also more complex micro-bio-objects such as cells, bacteria, capsules or vesicles. Finally, while in the present study we focused exclusively on situations where external body forces were absent ($\bf{f}=0$), it would be interesting as a perspective to extend the investigation to situations in which the gravitational force is to be considered in order to analyse its impact on the equilibrium lateral position of the objects. Indeed, while the equilibrium velocity of a dispersed object derived in the $\Re=\Ca=0$ limit, but valid also up to moderate values of these quantities, was fully characterized as a function of the lateral position, the relation between $\bf{f}$ and $\bf{\varepsilon}$ is missing and would require a three-dimensional study with non-zero Reynolds and capillary numbers.

%% file: aaAcknowledgements.tex
\section* {Acknowledgements}

The authors would like to thank Anna\"{\i}g Carr\'{e} for carrying out preliminary experiments. Fundings from F\'ed\'eration Wallonie-Bruxelles (ARC ESCAPE project), F.R.S.-FNRS and CNRS are also gratefully acknowledged.


%% file: aaApendices_v2.tex
\appendix

%

\section{Intermediate steps}\label{intsteps}

In this appendix, we provide some intermediate results used in \secref{SecEqs3}, such as the components of the stress tensor, perturbation of the interfacial tension, perturbation of the flux terms, and the equations for the volume and position of the drop.

The viscous stress tensors $\T_0$ and $\T_1$ are given by their components in cylindrical coordinates as
\begin{align} \label{Trz}
T_{0rr} = 2 \mu \partial_r v_{0} - p_0 \,, \quad
T_{0zz} = 2 \mu \partial_z u_{0} - p_0 \,, \quad
T_{0rz} = T_{0 z r} = \mu (\partial_r u_{0}+ \partial_z v_{0}) \,,
 \\ \nonumber
 T_{1rr} = 2 \mu \partial_r v_{1} - p_1 \,, \quad
T_{1zz} = 2 \mu \partial_z u_{1} - p_1 \,, \quad
T_{1rz} = T_{1 z r} = \mu (\partial_r u_{1}+ \partial_z v_{1}) \,,
\end{align}
for the components in the $r$-$z$ plane and
\begin{align} \label{Ttheta}
&T_{0z\theta} = T_{0 \theta z} = 0 \,, \quad
T_{0r\theta} = T_{0 \theta r} = 0 \,, \quad
T_{0\theta\theta} = 2 \mu \frac{v_{0}}{r} - p_0 \,,
\\ \nonumber
&T_{1z\theta} = T_{1 \theta z} = \mu \left(\partial_z w_{1} + \frac{i}{r} u_1 \right) \,, \quad
T_{1r\theta} = T_{1 \theta r} = \mu \left(\partial_r w_{1} - \frac{w_1}{r} + \frac{i}{r} v_1 \right) \,, \quad
T_{1\theta\theta} = 2 \mu \frac{v_{1}+ i w_1}{r} - p_1 \,, 
\end{align}
for the out-of-plane components.

Concerning the boundary conditions \eqref{velcontYL}, it is convenient to write them at the unperturbed geometry since it is the known one. To do so, 
the method first proposed by \cite{Rivero2018} is used, mainly arising two kinds of terms to perturb: (i) the interfacial tension terms and (ii) the mass and momentum flux terms. 

We first consider the perturbation of the (i) interfacial tension term in the RHS. For this purpose, we need to perturb $\mathcal{D}_{s} \cdot \id$ as
\begin{subequations}
\label{pertPsi}
\begin{align}
\lim_{\dd \theta \rightarrow 0} \frac{1}{\dd \theta}
\int_{\Sigma_d} \mathcal{D}_{s} \cdot \id \dd \Sigma = 
\int_{\Gamma} \left( \mathcal{D}_{s} \cdot \id  + \mathcal{D}_{s} \cdot \Psi  \right) r \dd \Gamma \,, 
\end{align}
\end{subequations}
where the mean value theorem has been used and $\Psi =  (\grad_s \cdot  \vn ) \delta \id_s - (\grad_s \vn) \delta + (\grad_s \delta) \vn $, as reported in \cite{Rivero2018,Rivero2018Corr}.
By virtue of \eqref{Dthetas}, it writes as
\begin{align}
r \mathcal{D}_{s} \cdot \id = \mathcal{D}_{s,{rz}} r  - \ve_r \,, \qquad
r \mathcal{D}_{s} \cdot \Psi = \mathcal{D}_{s,{rz}} \cdot r \Psi  + \partial_\theta \vpsi_{\theta} \,,
\end{align}
where $\vpsi_{\theta} = \ve_\theta \cdot \Psi$. The terms $\mathcal{D}_{s,{rz}} \cdot r \Psi$ and $ \partial_\theta \vpsi_{\theta}$ are written after the expansion \eqref{expdelta} up to first-order $ \Psi = \varepsilon \exp{i \theta} \Psi_1 + \O(\varepsilon^2) $, as
\begin{subequations}
\label{pertST}
\begin{align}
\label{pertSTRZ}
 \mathcal{D}_{s,{rz}} \cdot r \Psi_1   &=  \mathcal{D}_{s,{rz}} \delta_1 n_r + \mathcal{D}_{s,{rz}} \cdot [(\grad_{s,{rz}} \delta_1) r\vn ] \,, \\
\label{pertSTTH}
 \partial_\theta  \vpsi_{\theta1} &= (-\ve_r + i \ve_\theta) \delta_1 (\grad_{s,{rz}} \cdot \vn)  + (-\vn + i \ve_{\theta} \ve_r \cdot \vn) \frac{\delta_1}{r} \,.
\end{align}
\end{subequations}
The first terms represent a change of length whereas the second ones represent a rotation of the surface.

To perturb the flux term, integration of \eqref{NS} over the volume generated by the displacement of an arbitrary subset of revolution of the boundary is carried out. Conveniently using the Green theorem, and substituting the boundary conditions \eqref{velcontYL} and the perturbation of the interfacial tension term \eqref{pertPsi}, it writes at the unperturbed interphase
\begin{subequations}
\begin{align}
\vn \cdot \boldsymbol{v}_c - \mathcal{D}_{ s}  \cdot \left( \delta \boldsymbol{v}_c \right)  &=0  \,,
\\
\vn \cdot  \dbrackets{\T} - \mathcal{D}_s \cdot  \left( \delta \dbrackets{ \T } \right)  &= -\delta \varphi \Re \dbrackets{  \boldsymbol{v} \cdot \grad \boldsymbol{v}} + \vf \cdot \vx \vn +  \Ca^{-1} [ \mathcal{D}_s \cdot \id + \mathcal{D}_s \cdot \Psi ] \,,
\end{align}
\end{subequations}
at $\Gamma$. 
Considering the axisymmetry of the unperturbed geometry, it can be rewritten using \eqref{gradtheta} as
\begin{subequations}
\begin{align}
r \vn \cdot  \boldsymbol{v}_c - \mathcal{D}_{s,{rz}}  \cdot \left( r \delta \boldsymbol{v}_c \right) -  \partial_{\theta} \left( \delta v_{c \theta} \right) &=0  \,,
\\
r \vn \cdot  \dbrackets{\T} - \mathcal{D}_{s,{rz}} \cdot \left( r \delta \dbrackets{\T} \right) -  \partial_{\theta} \left( \delta \dbrackets{\vT_{\theta}} \right)  &= 
-r   \delta \varphi \Re \dbrackets{ \boldsymbol{v} \cdot \grad \boldsymbol{v} } + \\ \nonumber
+ r\vf \cdot \vx \vn &
 + \Ca^{-1} [ \mathcal{D}_{s,{rz}} r + \mathcal{D}_{s,{rz}} \cdot \left( r \Psi \right) - \ve_r+ \partial_{\theta} \vpsi_\theta ]
  \,,
\end{align}
\end{subequations}
at $\Gamma$ where $\vT_\theta = \ve_\theta \cdot \T$.


Finally, the equations (\ref{global}a,b) can be perturbed by using the Reynolds transport theorem as

\begin{align} \label{pertvolpos}
\int_{\mathcal{V}_{d0}}        \dd \V + \int_{\Sigma_{d0}}       \delta  \dd \Sigma = \V_d \,, \qquad 
\int_{\mathcal{V}_{d0}}  \vx \dd \V + \int_{\Sigma_{d0}}  \vx \delta \dd \Sigma = \V_d \veps \,,
\end{align}
which can be simplified by carrying out the integrals in $\theta$,
\begin{align} \label{thetaintegral}
\int_{0}^{2 \pi}  \dd \theta=  2 \pi \,, \qquad 
\int_{0}^{2 \pi}  \exp{i \theta}   \dd \theta=  0 \,, \qquad 
\int_{0}^{2 \pi}  \exp{i \theta} \ve_r   \dd \theta=  \pi \left( \ve_x+i\ve_y \right) \,,
\end{align}
leading to \eqref{integralesthetahechas}.


\section{Decomposition of differential operators} \label{Decomp}

Since the Finite Element Method is used to solve the partial differential equations, it is convenient to write the differential operators in \eqref{defdifop} in order to lead to partial differential equations written in conservation form. To do so, it is considered its integral over a volume generated by the revolution of $S$ and $\Gamma$ along a differential angle $\dd \theta$. On the one hand, applying the mean value theorem leads to
\begin{subequations}
\label{decompLHS}
\begin{align}
 \lim_{\dd \theta \rightarrow 0} \frac{1}{\dd \theta} \int_S \int_\theta^{\theta+\dd \theta}  r \grad \star \dd \theta  \dd \Sigma &=
\int_S r  \grad \star \dd \Sigma
 \,,\\
 \lim_{\dd \theta \rightarrow 0} \frac{1}{\dd \theta} \int_\Gamma \int_\theta^{\theta+\dd \theta}  r \D_s \star \dd \theta  \dd \Gamma &= 
 \int_\Gamma r  \D_s \star \dd \Gamma  
\,.
\end{align}
\end{subequations}
On the other hand, after application of the Green theorem it can be written as
\begin{subequations}
\label{decompRHS}
\begin{align}
 \lim_{\dd \theta \rightarrow 0} \frac{1}{\dd \theta} \int_S \int_\theta^{\theta+\dd \theta}  r \grad \star \dd \theta  \dd \Sigma &=
\int_S \left(  \grad_{rz} r \star + \partial_{\theta} \ve_{\theta} \star \right) \dd \Sigma
 \,,\\
 \lim_{\dd \theta \rightarrow 0} \frac{1}{\dd \theta} \int_\Gamma \int_\theta^{\theta+\dd \theta}  r \D_s \star \dd \theta  \dd \Gamma &= 
\int_\Gamma \left( \D_{s,{rz}}  r \star + \partial_{\theta} \ve_{\theta} \star \right) \dd \Gamma  
\,,
\end{align}
\end{subequations}
Then, equating the RHS of \eqref{decompLHS} and \eqref{decompRHS} leads to
\begin{subequations}
\begin{align}
 \int_S r  \grad \star \dd \Sigma  &=
 \int_S \left(  \grad_{rz} r \star + \partial_{\theta} \ve_{\theta} \star \right) \dd \Sigma \,
 \,,\\
 \int_\Gamma r  \D_s \star \dd \Gamma  
 &= 
 \int_\Gamma \left( \D_{s,{rz}}  r \star + \partial_{\theta} \ve_{\theta} \star \right) \dd \Gamma  \,,
\end{align}
\end{subequations}
which for any arbitrary $S$ and $\Gamma$ reduces to the searched form
\begin{subequations}
\label{gradtheta}
\begin{align}
\label{gradthetas}
  r  \grad \star  &=  \grad_{{rz}} r \star + \partial_{\theta} \ve_{\theta} \star \,,
\\
\label{Dthetas}
  r  \mathcal{D}_s \star  &=  \mathcal{D}_{s,{rz}} r \star + \partial_{\theta} \ve_{\theta} \star \,.
\end{align}
\end{subequations}

\section{Fittings}
\label{app:velocity}

\begin{figure}
\input{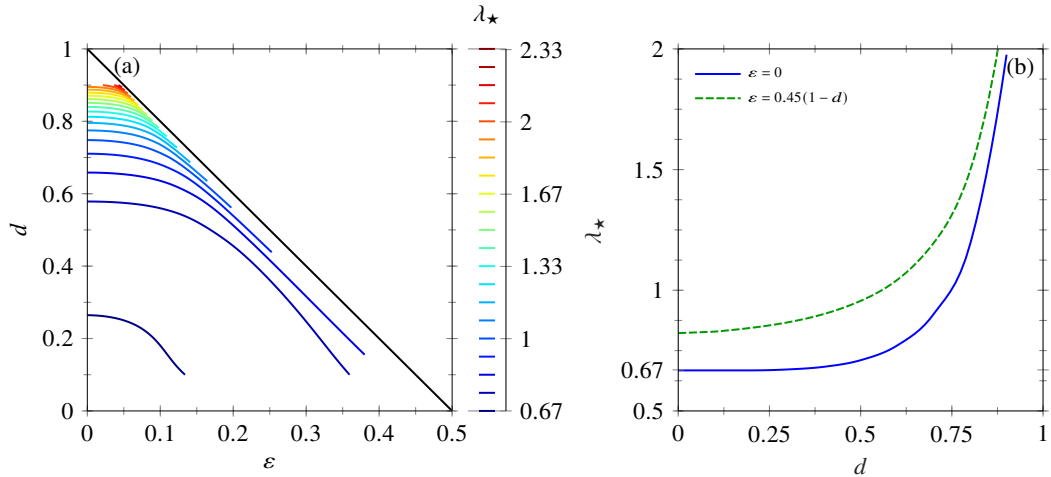}
\input{./texfigures/Velocity/inset_paper.tex}
\caption{
(a) $\lambda_\star$ as functions of $d$ and $\varepsilon$, and (b)  $\lambda_\star$ as a function of $d$ for the two given eccentricities involved in \figref{fig:velocity_evo_numerics}: centered ($\varepsilon=0$) and close to the wall ($\varepsilon=0.45(1-d)$). $\lambda_\star$ is involved in equation \eqref{Vdropecc} and is here numerically computed in the $\Re=0$ and $\Ca=0$ limit.}
\label{mustar}
\end{figure}

In this appendix, we provide the polynomial fittings, with less than 1$\%$ error, of the quantities  $\eta~=~(\Vb,~\Vp,\lambda_\star)$ in the form 
\begin{equation}
\label{fit}
\eta \left( \varepsilon,d \right) = \sum_{i,j} \eta_{ij} \left( \frac{\varepsilon}{\varepsilon_*} \right) ^i  d  ^j\,
\end{equation}
in which the respective coefficients $\eta_{ij}$ are provided in \tabref{tab:coef} and $\varepsilon_*=0.5(1-d)$. Note that the functions $\Vb$ and $\Vp$ correspond to the bubble and bead limits previously studied and derived in~\cite{Rivero2018}, i.e. for $\lambda \to 0$ and $\lambda \to \infty$, respectively. Hence, only the numerically computed function $\lambda_{\star}$ is plotted as functions of $d$ and $\varepsilon$ in \figref{mustar}. In the latter, we can in particular observe how $\lambda_{\star}$ increases sightly with $\varepsilon$ and strongly with $d$. 

%
%

\begin{table}
  \begin{center}
\def~{\hphantom{0}}
  \begin{tabular}{lrrrrr}
     $\Vb$  & $j=0$         &   $j=1$ & $j=2$ & $j=3$ & $j=4$\\
     $i=0$   & $2.0180$   &   $-0.4310$   & $2.2886$    & $-4.0375$ & $1.1047$\\
     $i=2$   & $-2.0370$   &   $4.4183$   & $-2.3396$    & $-1.0086$ & $1.0093$\\
     $i=4$   & $0.0405$   &   $-0.1170$    & $-4.2568$ & $8.5856$ & $-4.2313$\\[4pt]
     $\Vp$ & $j=0$         &   $j=1$ & $j=2$ & $j=3$ & $j=4$\\
     $i=0$   & $1.9764$   &   $0.2213$   & $-2.0961$    & $0.9186$ & $-0.0192$\\
     $i=2$   & $-2.0424$   &   $4.7348$   & $-4.3428$    & $2.8893$ & $-1.2567$\\
     $i=4$   & $0.1284$   &   $-1.7981$    & $1.6230$ & $1.8201$ & $-1.7819$\\[4pt]
$\lambda_\star$  & $j=0$         &   $j=2$ & $j=4$ & $j=6$ & $j=8$\\
     $i=0$   & $0.6728$   &   $-0.2613$   & $2.8477$    & $-6.1694$ & $6.7940$\\
     $i=2$   & $0.0217$   &   $-0.4273$   & $1.2292$    & $-0.3968$ & $-0.3587$\\
     $i=4$   & $0.1659$   &   $4.1323$    & $-15.1319$ & $22.7114$ & $-12.5874$\\
     $i=6$   & $-0.1048$  &   $-13.5725$ & $49.6521$  & $-73.0672$ & $39.2562$\\
     $i=8$   & $0.1995$   &   $12.4285$  & $-42.7485$ & $60.2374$ & $-31.7364$\\
  \end{tabular}
  \caption{Coefficients of the polynomial fitting $\eta_{ij}$ of $\Vb$, $\Vp$ and $\lambda_\star$ involved in \eqref{fit}.}
 \label{tab:coef}
  \end{center}
\end{table}

%
%
%
%
%
%

%
%
%

%% file: texfigures/Velocity/inset_paper.tex
%
\begin{tikzpicture}[%
baseline
]

\begin{axis}[%
width=.3\textwidth,
height=.3\textwidth,
scale only axis,
xmin=0,
xmax=1,
xlabel style={font=\color{white!15!black}},
xlabel={$d$},
ymin=0.5,
ymax=2,
ylabel style={font=\color{white!15!black}},
ylabel={$\lambda_\star$},
axis background/.style={fill=white},
legend style={legend cell align=left, align=left, draw=white!15!black},
xtick={0,.25,...,1},
minor xtick={0,0.125,...,1},
ytick={0.5,0.67,1,1.5,2},
minor ytick={0,0.125,...,2},
legend style={at={(0.02,0.98)},anchor=north west,draw=none,fill=none}
]

\node[anchor=north east,xshift=0.0cm,yshift=0.0cm] at (rel axis cs:1,1) {(b)};

\addplot [color=blue]
  table[row sep=crcr]{%
0	0.668010883592571\\
0.2	0.668010883592571\\
0.214	0.668123838442436\\
0.228	0.668377804262236\\
0.242	0.668772781051971\\
0.256	0.669308768811642\\
0.27	0.669985767541248\\
0.284	0.670803777240789\\
0.298	0.671762797910265\\
0.312	0.672806077404802\\
0.326	0.673879837334592\\
0.34	0.675055985246834\\
0.354	0.676408255935814\\
0.368	0.678010384195821\\
0.382	0.67993610482114\\
0.396	0.682259152606059\\
0.41	0.684925907800606\\
0.424	0.68769246077016\\
0.438	0.690680348004442\\
0.452	0.694032423112288\\
0.466	0.697891539702536\\
0.48	0.702400551384022\\
0.494	0.707702311765583\\
0.508	0.713774420031816\\
0.522	0.72006812184783\\
0.536	0.726780208841537\\
0.55	0.734205067000777\\
0.564	0.742637082313392\\
0.578	0.752370640767224\\
0.592	0.763700128350114\\
0.606	0.776700178148485\\
0.62	0.790033347939035\\
0.634	0.804022053214952\\
0.648	0.819403479212446\\
0.662	0.836914811167728\\
0.676	0.857293234317006\\
0.69	0.881275933896492\\
0.704	0.909265695761641\\
0.718	0.936795709720957\\
0.732	0.964387510022802\\
0.746	0.994826822690407\\
0.76	1.030899373747\\
0.774	1.07539088921582\\
0.788	1.13108709512008\\
0.802	1.20070468345621\\
0.816	1.28221188806246\\
0.83	1.37346459328247\\
0.844	1.47446279911624\\
0.858	1.58520650556376\\
0.872	1.70569571262505\\
0.886	1.83593042030009\\
0.9	1.97591062858889\\
};
\addlegendentry{$\varepsilon=0$}

\addplot [color=green,dashed, smooth]
  table[row sep=crcr]{%
0 0.8224 \\ 
0.1	0.828716412721808\\
0.127586206896552	0.832520940200732\\
0.155172413793103	0.836643990819635\\
0.182758620689655	0.841174384376324\\
0.210344827586207	0.846200940668607\\
0.237931034482759	0.851812479494289\\
0.26551724137931	0.858097820651177\\
0.293103448275862	0.865145783937078\\
0.320689655172414	0.873050659721373\\
0.348275862068966	0.881954352607519\\
0.375862068965517	0.892023283464179\\
0.403448275862069	0.903424069171699\\
0.431034482758621	0.916318540104754\\
0.458620689655172	0.93085530267821\\
0.486206896551724	0.947180698794592\\
0.513793103448276	0.965456899328088\\
0.541379310344828	0.986210141501128\\
0.568965517241379	1.01033072888438\\
0.596551724137931	1.03872479402017\\
0.624137931034483	1.07225406573637\\
0.651724137931034	1.11152097070317\\
0.679310344827586	1.157034079343\\
0.706896551724138	1.20931917474277\\
0.73448275862069	1.27100043667768\\
0.762068965517241	1.34877744344973\\
0.789655172413793	1.44981801063847\\
0.817241379310345	1.58128995382345\\
0.844827586206897	1.75036108858424\\
0.872413793103448	1.96419923050037\\
0.9	2.2299721951514\\
};
\addlegendentry{$\varepsilon=0.45 (1-d)$}

\end{axis}
\end{tikzpicture}%